\documentclass[11pt,a4paper]{article}
\pdfoutput=1
\usepackage{jheppub}
\usepackage[utf8]{inputenc}
\usepackage{xcolor}
\usepackage{epstopdf}
\usepackage{graphicx}
\usepackage{epsfig}
\usepackage{dcolumn}  
\usepackage{bm}    
 \usepackage{caption}
 \usepackage{subcaption}
\usepackage{amssymb} 
\usepackage{amsmath,bm}
\usepackage{amsfonts} 
\usepackage{simpler-wick}
\usepackage{amsmath}  
\usepackage{slashed}  
\usepackage{enumitem}
\usepackage[mathscr]{euscript}
\usepackage{tabu}
\usepackage{epsfig}
\makeatletter
\g@addto@macro\bfseries{\boldmath}
\makeatother
\def\l1{{{1-loop}}}

\def\o{\mathcal{O}}

\def\n1{\Bigg|_{n=1}}

\def\n{{(n)}}

\def\p{\partial}

\usepackage{bookmark}
  \title{\textbf{\textsf{Entanglement  in descendants }}}
    \author{Barsha G. Chowdhury, Justin R. David}
\affiliation{\vspace{.1cm} Centre for High Energy Physics, \\ ${}^{~\,}$Indian Institute of Science,\\
${}^{~\,}$C. V. Raman Avenue, Bangalore 560012, India.}
\emailAdd{barsha@iisc.ac.in, justin@iisc.ac.in}
\abstract{ 
We study the single interval entanglement and relative entropies 
of conformal descendants  in 2d CFT. 
Descendants contain non-trivial  entanglement,  though the   entanglement entropy  of 
the canonical primary  in the free boson CFT  contains  
no additional entanglement compared to the vacuum, we show
that the 
 entanglement entropy of the state created by its  level one descendant
is  non-trivial and is identical to that of the 
$U(1)$ current in this theory.  
We determine 
the first sub-leading corrections to the short  interval expansion of the entanglement entropy 
 of descendants in a general CFT from their 
four point function on the $n$-sheeted plane. 
We show that these corrections are determined by 
multiplying squares of appropriate dressing factors 
to the corresponding corrections of the primary. 
 Relative entropy between descendants of the same primary is proportional 
to the square of the difference of their dressing factors.
We apply our results to a class of descendants of 
generalized free fields and 
descendants of the vacuum and show that their  dressing 
factors   are universal. }

\begin{document}

\maketitle

\section{Introduction}

Ideas from information theory have played a key role in our recent understanding of 
holography,  quantum gravity and black hole physics. 
The most well studied and useful information theoretic quantity has proven to be 
entanglement entropy and relative entropy.  
In theories which admit a holographic duals, entanglement entropy can be evaluated 
from areas of extremal surfaces \cite{Ryu:2006bv,Hubeny:2007xt}. Relative entropy between 
two nearby excited states is equal to the corresponding bulk relative entropy \cite{Jafferis:2015del}
Relative entropy  has been applied to obtain a precise version 
of the Bekenstein bound and the second law of black hole thermodynamics \cite{Casini:2008cr,Wall:2010cj,Wall:2011hj}.  

Quantum field theories generally admit symmetries for example conformal field theories
in 2 dimensions  admit Virasoro symmetry.  It is interesting to study how such a symmetry 
is reflected in information theoretic quantities like entanglement entropy and relative entropy. 
In particular since  the global part of  the Virasoro algebra corresponds to isometries
of $AdS_3$, there should be simple modifications of extremal surfaces or 
relative entropies when one considers   these quantities in the bulk. 
for states excited by such a symmetry.  As far as we are aware this simple question 
has not been investigated in the literature.

In this paper we  study this question by evaluating single interval  entanglement entropy and relative 
entropies of conformal descendants of primaries 
 in 2 dimensional conformal field theories. 
Information theoretic properties of conformal descendants have been studied earlier \cite{Palmai:2014jqa,Caputa:2015tua,Taddia:2016dbm,Brehm:2020zri}, 
but it has been confined only the the 2nd R\'{e}nyi entropy and the trace square distance. 
However
to understand how information theoretic quantities of descendants 
 can be read out from 
extremal surfaces in holography  it is important to study entanglement entropies and 
relative entropies.  In fact as we will  see in this paper, there are certain simplifications 
when one considers entanglement entropy or relative entropy. 
One might expect that descendants do not contain interesting information theoretic properties. 
This is easily dismissed by first considering relative entropy.  We know that relative entropies 
is  always positive  and is a measure of the distance between states, see \cite{Witten:2018zva} for a
review of information theoretic measures in quantum field theory. 
Therefore descendants of the same primary,  and in fact 
distinct descendants  of a given primary at the same level 
should differ in their relative entropy.   

To  begin,  we show that descendants contain non-trivial entanglement 
by considering  the free boson theory. It is known that the single interval 
entanglement entropy of its canonical excitation $e^{i l X}$ is identical to that of the vacuum
\cite{Alcaraz:2011tn, Berganza:2011mh}. In this paper we evaluate the single interval entanglement entropy of the 
level one descendant $\partial e^{i l X}$ exactly. 
We use the  replica trick developed in 
\cite{Alcaraz:2011tn,Berganza:2011mh,Lashkari:2014yva,Sarosi:2016oks} 
to evaluate single interval entanglement 
entropy 
Evaluating the $2n$ point function of the descendant on the uniformized  plane we show that 
 the single interval entropy   is identical to that of the $U(1)$ current in this theory. 
 As far as we are aware this is the first exact result of entanglement entropy of a descendant.

 For a general conformal field theories it  is not easy to evaluate the single interval entanglement entropy 
 of excited states exactly since it involves knowing the  $2n$  point function of the corresponding operator
 on the uniformized plane involved in implementing the replica trick. 
 However we can study entanglement entropies or relative entropies in the short interval expansion. 
 The leading term in the short distance expansion of the entanglement entropy 
  of  any primary  in 2 dimension conformal field theory was obtained in 
 \cite{Alcaraz:2011tn}.  This contribution to the  short interval expansion 
 is obtained by factoring the $2n$  point function 
 into $n$ 2-point functions where the pairs of operators are located on each wedge of the 
 uniformized plane as in figure \ref{uniplane2}. 
 A method to obtain the first sub-leading corrections 
 to the entanglement entropy and relative entropy was developed in \cite{Sarosi:2016oks}. 
 In this paper we revisit this method and simplify the evaluation of the first sub-leading corrections. 
 We show that the sub-leading correction can be extracted from factorizing the $2n$ point function 
 into    2-point functions  on $n-2$ wedges in the uniformized plane 
   and a four point function valued on $4$ operators on the remaining pair of wedges as shown 
   in figure \ref{n2correlators}.
   Then one needs 
   to sum over   different pairs of wedges.  This view of the first sub-leading correction 
   allows us to generalise these computations to conformal descendants. 
  We obtain the  following expression for the short interval expansion   of the 
    entanglement entropy 
      of the primary \footnote{ (\ref{eeleadsubi}) is the difference of the entanglement entropy of the 
      excited state  and the vacuum state. }
    \begin{eqnarray}\label{eeleadsubi}
\hat S(\rho_{\o} ) &=&  2h ( 1- \pi x \cot\pi x)  - \frac{8 h^2}{15c}  ( \sin\pi x)^4  \\ \nonumber
&& - C_{\o\o\o_p} C^{\o_p}_{\; \o\o}  (\sin\pi x)^{2h_p} 
\frac{\Gamma(\frac{3}{2} ) \Gamma( h_p +1) }{ 2\Gamma( h_p + \frac{3}{2} ) } 
+ \cdots. 
\end{eqnarray}
Here we are  considering only the holomorphic sector of the conformal field theory,  the primary 
$\o$ has conformal dimension $(h, 0)$. $h_p$ is the conformal dimension of the 
lowest lying primary in the theory
and $C_{\o\o\o_p}$ is the corresponding OPE coefficient.  
Further more among the 
two sub-leading terms only one contributes depending on the certain conditions 
discussed after equation (\ref{lowestc4}). For example,  the term proportional to $1/c$ which 
arises from the stress tensor exchange is negligible 
for  low lying primaries 
in large $c$ conformal field theories.

We develop methods that enable the evaluation of the sub-leading corrections to short
interval expansion of  entanglement entropy of conformal descendants in 2d  conformal field theories  efficiently.  
After studying  all descendants till level 3 and the class of descendants $\partial^l \o $ , we 
find the  leading  contributions to the single interval entropy of descendants  take the form
\begin{eqnarray}\label{eeleadsubdl2i}
\hat S(\rho_{\partial\o^{[-l] } } ) &=&  2(h+l)  ( 1- \pi x \cot\pi x)  - \frac{8 h^2}{15c}  [D_{\o^{ [l]} }( h, 2) ]^2
( \sin\pi x)^4  \\ \nonumber
&& - C_{\o\o\o_p} C^{\o_p}_{\; \o\o}  [D_{\o^{[-l]} } ( h, h_p)]^2
\frac{\Gamma(\frac{3}{2} ) \Gamma( h_p +1) }{ 2\Gamma( h_p + \frac{3}{2} ) } (\sin\pi x)^{2h_p} 
+ \cdots.
\end{eqnarray}
The  leading term behaves as though the weight has shifted by $h\rightarrow h+l$. 
The sub-leading terms are modified compared to the expression for the corresponding primary 
in (\ref{eeleadsubi}) by appearance of a square of a pre-factor 
$D_{\o^{[-l]} } ( h, h_p)$ which we call the dressing factor.  This factor depends on the 
level and nature  of the descendant, weight of the primary and central charge. 
$C_{\o\o\o_p}$ is the OPE coefficient. 
We evaluate the dressing factor for all states till level $3$ and for   the class $\partial^l \o$ . 
From the computations which lead us to (\ref{eeleadsubdl2i}), we see that the dressing factor 
is the ratio of a deformed norm of the descendant, to the usual norm. 
The notion of the deformed norm is explained in section \ref{section3.3}. 

We show that the leading contributions  to the relative entropy between descendants at 
level $l, l'$ of primaries with weight $h, h'$ respectively is given by 
 \begin{eqnarray}\label{relenfdi}
 &&S( \rho_{\o^{[-l] }} | \rho_{\o^{\prime [-l']  }}) = \frac{ 8 }{15 c} 
 \Big[  hD_{\o^{ [-l]}}( h, 2)  - h' D_{\o^{\prime [-l']}}( h, 2)  \Big]^2 ( \sin\pi x)^4  \\ \nonumber
& &  + \Big|\Big| ( C_{\o\o\o_p} D_{\o^{ [-l]}}( h, h_p)   - C_{\o'\o'\o_p}
D_{\o^{\prime [-l'] } }( h, h_p)    \Big|\Big|^2
%( C^{\o_p}_{\, \o\o} D_{\o^{ [-l] } }( h, h_p)  - C^{\o_p}_{\, \o'\o'} 
%D_{\o^{\prime [-l'] } }( h, h_p)   )
 \frac{\Gamma(\frac{3}{2} ) \Gamma( h_p +1) }{ 2\Gamma( h_p + \frac{3}{2} ) } (  \sin\pi x)^{2h_p} 
+ \cdots
 \end{eqnarray}
 Here $|| \cdot ||^2$ refers to the fact that we need to take the square of its argument together 
 with the index $\o_p$ in one of the term raised by the Zamolodchikov metric.
 Again,  only one of the 2 terms is the leading contribution. 
 It is easy to see that the relative entropy of descendants of  the same primary is proportional 
 to the square of the difference of their dressing factors.  
 We test the results (\ref{eeleadsubdl2i}) and (\ref{relenfdi}) using the exact results from the free boson theory.

Finally we apply our  results to  a class of  descendants  of 
generalized free fields and  the 
vacuum.  Generalized free fields in holographic conformal field theories.
are known to be  dual to minimally coupled scalars 
The entanglement entropy of $\partial^l \o$ where $\o$ is a generalised free field is given by 
\begin{eqnarray} \label{eegenfie1i}
\hat S( \rho_{\partial^l \o} ) = 2(h+l)  ( 1- \pi x \cot\pi x)  -  
( \sin\pi x )^{4h}  \frac{\Gamma ( \frac{3}{2} ) \Gamma( 2h + 1) }{ \Gamma ( 2h + \frac{3}{2} )} 
\times \left( \frac{ {\cal N}_l}{(  l !)^2} \right)^2  .
\end{eqnarray}
Here ${\cal N}_l$ is the norm of the global descendant $\partial^l\o$ at level $l$. 
For primaries that is  $l=0$,  the sub-leading correction correction has been reproduced from the bulk 
\cite{Belin:2018juv}. 
This involved not only corrections to the Ryu-Takayanagi surface,  but also evaluation of 
the  bulk entanglement across this surface.  The operator $\partial^l\o$ corresponds 
to the state $( L_{-1})^l  |h\rangle$.  which is obtained by the action 
of an isometry in the bulk on the excitation of the minimally coupled scalar. 
Therefore it should be possible to reproduce the result in (\ref{eegenfie1i}) 
for $l\neq 0$
by generalising the methods of \cite{Belin:2018juv}. 
For  the class of descendants $L_{-n} |0\rangle$ we show that the leading 
contributions to the single interval entanglement is given by 
\begin{eqnarray} \label{eegrav2i}
\hat S( \rho_{L_{-(l+2)}|0\rangle } ) &=&  2( l+2)  ( 1- \pi x \cot\pi x)  - \frac{ 8 ( l + 2)^2 }{ 15c} \sin^4 \pi x
\\ \nonumber
&& 
 - \Big[  \frac{ ( l +3) ( l + 2) ( l + 1) }{ 3!} \Big]^2 \frac{128}{315}  \sin^8 \pi x
+ \cdots. 
\end{eqnarray}
Again only one among the last two terms contribute, 
 for holographic conformal field theories and for low lying excitations, 
  it is only the third term that contributes at the 
sub-leading order.  As expected the entanglement entropy of descendants of the vacuum
is universal.  It should be possible to verify  (\ref{eegrav2i}) using holography.

The organization of the paper is as follows. 
In section \ref{section1}.  we briefly  review of the set up involved in evaluating entanglement properties 
of low lying states 
and use it to evaluate the entanglement entropy of the level one descendant 
of $e^{il X}$  exactly in the free boson theory. 
In section \ref{section2}.  we develop a simplified method to obtain the 
entanglement entropy of low lying states in the short interval expansion and then apply it 
to conformal descendants.  We also evaluate the relative entropy between various descendants.
We apply these results to a class of descendants of generalized free fields and the vacuum in 
section \ref{section4}.   Section \ref{section5}. contains our conclusions. 
Appendix \ref{appen1}.
  contains the details for evaluating the $2n$ point function of the level one descendant 
of $e^{il X}$  on the uniformized plane. 
Appendix \ref{appen2}.
discusses the properties of conformal descendants needed to arrive at the results
in the paper.

\section{Entanglement and relative entropy  of excited states} \label{section1}

\begin{figure} %[!t]
	\centering
	\includegraphics[width=.6\textwidth]{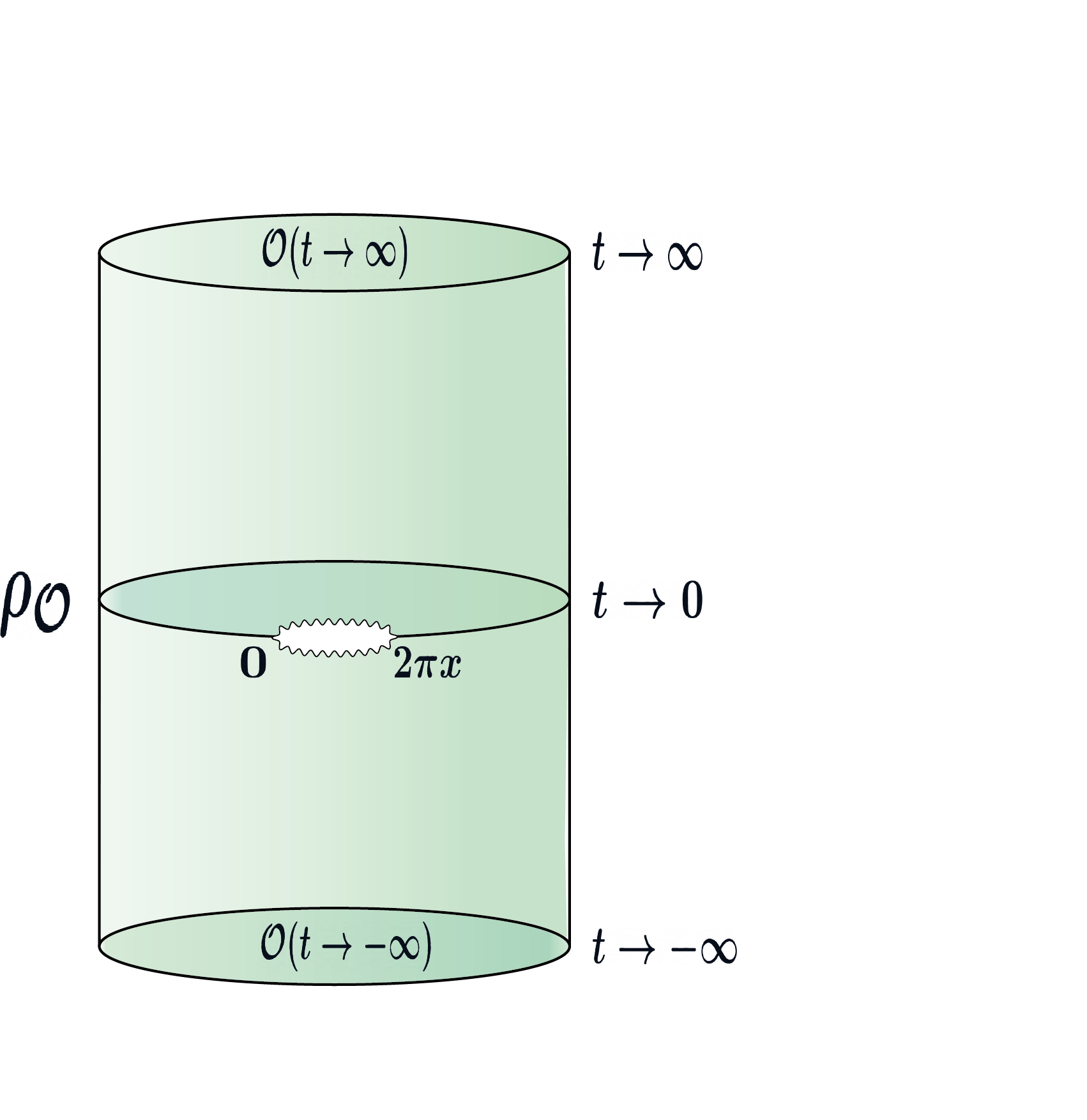}
	\caption{The path integral for the density matrix $\rho_{\o}$.}
	\label{pintegral}
\end{figure}

\begin{figure} %[!t]
	\centering
	\includegraphics[width=.6\textwidth]{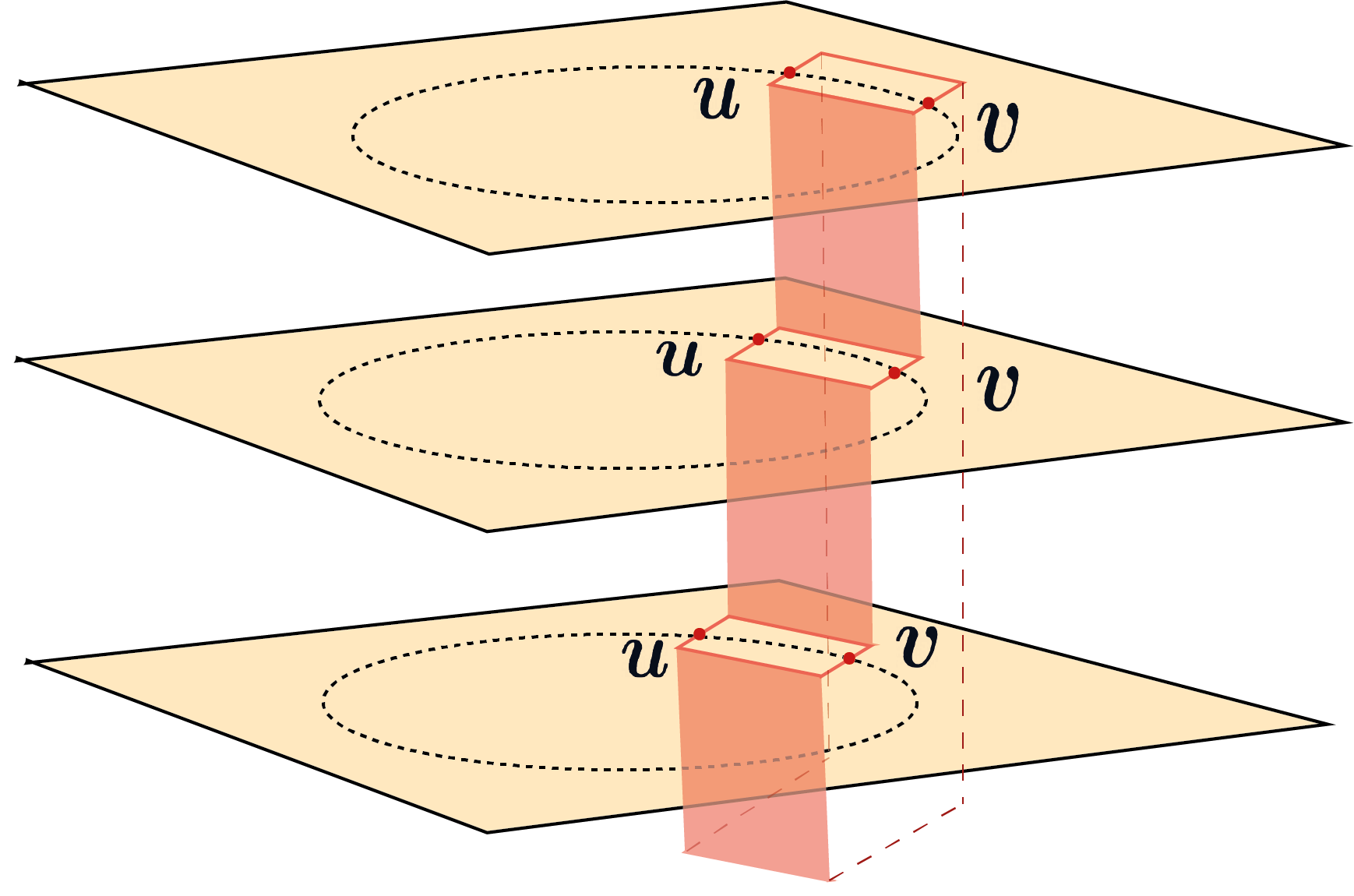}
	\caption{The path integral for $\rho_{\o}^n$ on the $n$-sheeted plane obtained using the cylinder to the plane map. The entangling interval is on the arc between $v=1$ and  $u= e^{2\pi i x}$ on the unit circle. The operators at mapped to the origin and infinity on each plane.}
	\label{nsheet}
\end{figure}

\begin{figure} %[!t]
	\centering
	\includegraphics[width=.6\textwidth]{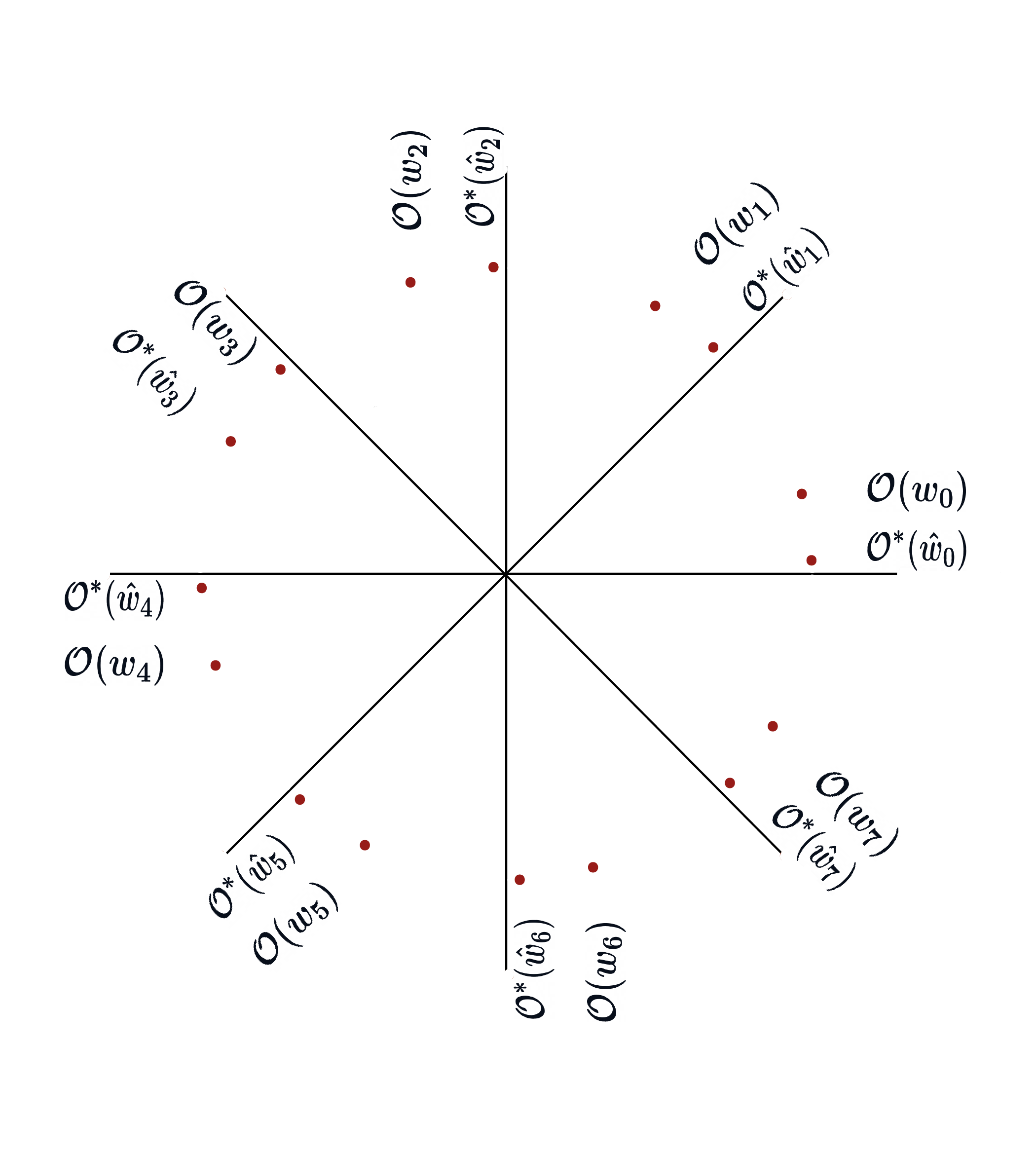}
	\caption{The uniformized plane with $2n$ operator insertions. The figure shows the example of $n=8$. Each sheet of figure \ref{nsheet} is mapped to a wedge on the uniformized plane.}
	\label{uniplane}
\end{figure}

We briefly review the set up of evaluating entanglement and relative entropy of 
excited states.
Consider  a 2d CFT on a cylinder  parametrised by $(t, \phi)$ where $t$ is the 
Euclidean time and $\phi \sim \phi + 2\pi$  is the spatial coordinate. 
We excite the CFT by 
placing an operator $\o$, not necessarily a primary at $t\rightarrow -\infty$, let us 
call  refer to this state as $|{\o} \rangle$. 
In this paper we will restrict our discussions 
only to the holomorphic sector of the CFT to keep the discussion uncluttered. 
All conclusions obtained in this paper 
can easily be generalised to include the anti-holomorphic sector. 
We wish to consider the reduced density matrix of $\rho_{\o}$. 
obtained by tracing over the complement of the interval $[0, 2\pi x]$ at 
the $t=0$  time slice. 
\begin{eqnarray}
\rho_{\o} = {\rm Tr}_{[0, 2\pi x]}  (  | \o \rangle \langle \o | ) .
\end{eqnarray}
The path integral picture of this reduced density matrix is given in 
figure  \ref{pintegral}
As we mentioned the  state  $|\o\rangle$  is obtained by placing  the operator 
$ \o $ at $t\rightarrow -\infty$, while the state $\langle \o|$ is its 
adjoint obtained  by placing the operator at $t \rightarrow +\infty$. 
The partial trace over the complement of $[0, 2\pi x]$ 
leaves a cut at this interval. 
To evaluate  the entanglement entropy of the reduced density matrix $\rho_{\o}$, we  
would need to  evaluate ${\rm Tr} (  \rho_{\o}^n) $.
The path integral then  involves  $n$ copies of the cylinder in  
figure \ref{pintegral} glued along the cut with the last cylinder glued to the first. 
It is convenient to use the  cylinder to plane map given in 
\begin{equation}\label{cyl-plane}
 z= e^{t + i \phi}, 
 \end{equation}
to equivalently regard the path integral on an  $n$ sheeted plane $\Sigma_n$. 
In this picture,  $n$ copies of the  operator $\o$  are placed at 
the origin as well  at $\infty$ on each sheet as shown in figure \ref{nsheet}. 
Using the state operator correspondence on the plane, the state 
  $|\o\rangle$  at the origin can be written as 
\begin{eqnarray}
|{ \o} \rangle = { \o} (0)  |0\rangle,
\end{eqnarray}
while the state at infinity is given by 
\begin{eqnarray} \label{adj1}
\langle \o | =  \langle 0  | \o^\dagger ( \infty) = \lim_{z\rightarrow 0} 
 \langle 0 | I\circ \o^* ( z ).
\end{eqnarray}
Here $I\circ \o$ refers to the action of the  inversion which takes the origin to 
$\infty$. 
\begin{equation} \label{inv}
I (z) =- \frac{1}{z} , 
\end{equation}
which is a map in  $SL(2, \mathbb{ Z} )$. 
If $\o$ is a primary of weight $h$. 
\begin{equation} \label{adj2}
I \circ \o ( z) =  ( \partial_z I (z) )^h  \o ( I( z) ) =  z^{-2h} \o \left (- \frac{1}{z}  \right) .
\end{equation}
However if $\o$ is a descendent one needs to use the appropriate conformal 
transformation.  
For example let $\partial \o$ be  the level 1 descendant, then  we  have
\begin{equation}
I \circ \partial  \o(z) =  \big[ \partial_z I ( z) \big]^{h+1 } \o (  I( z)  )
+  h \partial^2_z I  (z)   \big[\partial_z I(z) \big]^{h-1} \o ( I (z) ).
\end{equation}
Using the map (\ref{cyl-plane}) we see that the cut along the interval on the 
cylinder lies on the unit circle at the points
\begin{equation} \label{uvdef}
v = 1, \qquad \qquad u = e^{2\pi i x} .
\end{equation}
From the path integral 
on the $n$ sheeted plane, the trace of the reduced density matrix  $\rho_{\o}^n$ 
can be written 
as 
\begin{eqnarray}
\frac{ {\rm Tr} \rho_{\o} ^n  }{ {\rm Tr }\rho_{(0)}^n } 
= \langle \prod_{k =0}^{n-1} \o (0_k)  \prod_{{k'}=0}^{n-1} \o^*( \infty_{k'}  )  \rangle _{\Sigma_n}.
\end{eqnarray}
Here by $0_k$ we mean the origin and 
$\infty_k$ we mean $\infty$  at the $k$-th cover. 
Note to obtain $\infty_k$, we can take $z\rightarrow 0$ in the inversion map (\ref{inv}). 
The  reduced density matrix $\rho_{(0)}$ refers to the path integral on $\Sigma_n$ without
any operator insertions. 
To make the evaluation of the correlation function tractable we use the uniformization 
map which takes the branched cover $\Sigma_n$ to the plane
\begin{equation} \label{u1}
w(z) = \left( \frac{ z- u  }{ z-v }  \right)^{\frac{1}{n}}.
\end{equation}
The uniformized plane is  shown in figure \ref{uniplane}. 
For points at $\infty$ we compose this map with the inversion and use
\begin{equation} \label{u2}
\hat w(z) = \left( \frac{ - \frac{1}{ z} - u  }{- \frac{1}{z} -v  }  \right)^{\frac{1}{n}}.
\end{equation}
The maps in (\ref{u1}) and ( \ref{u2})  the operators placed on the branched planes are 
mapped as follows
\begin{equation} \label{locate}
0_k \rightarrow w_k =  e^{\frac{2\pi i ( k + x) }{n}}, \qquad \qquad \infty_k \rightarrow 
\hat w_k = e^{\frac{2\pi i k }{n} }.
\end{equation}
Using the uniformization map, the reduced density matrix is given by 
\begin{equation} \label{eecor}
\frac{ {\rm Tr} \rho_{\o} ^n  }{ {\rm Tr }\rho_{(0)}^n }  = 
\langle \prod_{k =0}^{n-1} w\circ \o ( w_k )  \prod_{k'=0}^{n-1} 
 \hat w \circ \o^*( \hat w_{k'})  \rangle.
\end{equation}
Here $w\circ \o$  and 
$\hat w \circ \o^* $ refers to the action of the  conformal transformation given in (\ref{u1}) 
and (\ref{u2}) respectively \footnote{
The complex conjugation on the operator $\o$ refers to taking complex conjugates 
of the parameters in the operator, but keeping holomorphic $z$. 
eg. $( e^ {i l X(z) } )^* = e^{ - i l X(  z) } $. This action  arises on taking the adjoint since 
we are interested in working only in the holomorphic sector.  }.
The  R\'{e}nyi entropies and entanglement entropies are then  obtained by evaluating 
\begin{eqnarray}\label{defreee}
S_n ( \rho_{\o } ) = \frac{1}{1-n} \log ( {\rm Tr} ( \rho_{\o}^n ) ), \qquad\qquad 
S( \rho_{\o}) = \lim_{n\rightarrow 1} S_n( \rho_{\o}) .
\end{eqnarray}
Thus these entropies can be obtained provided the $2n$ point function of the 
operator $\o$ on the uniformized plane can be evaluated.  Since the single interval R\'{e}nyi entropies 
of the  vacuum of the CFT  is known  and universal we will be interested in the difference of the 
entanglement entropy of the excited state from that of the vacuum.  To evaluate this difference, 
it is sufficient to examine the ratio of the traces of the density matrices given in (\ref{eecor}). 
We refer to this as the entanglement entropy of the excited state and it is given by 
\begin{eqnarray}\label{extee}
\hat S_n ( \rho_{\o} ) = \frac{1}{1-n} \log \left( \frac{ {\rm Tr} \rho_{\o} ^n  }{ {\rm Tr }\rho_{(0)}^n } \right) , 
\qquad  S  ( \rho_{\o} )  = \lim_{n\rightarrow 1}  \hat S_n ( \rho_{\o}). 
\end{eqnarray}

Let us now consider the relative entropy between the reduced density matrix $\rho_{\o}$ 
and that corresponding to another operator $\rho_{\o'}$ which is given by
\begin{eqnarray}
S( \rho_{\o} || \rho_{\o'} ) = {\rm Tr} (  \rho_\o \log \rho_{\o})   - {\rm Tr} (  \rho_\o \log \rho_{\o'} ).
\end{eqnarray}
To re-write the relative entropy in terms of correlators we use the identity 
\begin{eqnarray} \label{defsn}
S( \rho_{\o} || \rho_{\o'} )  &=& \lim_{n\rightarrow 1} S_{n} ( \rho_{\o} || \rho_{\o'} ) , \\ \nonumber
&=& \lim_{n\rightarrow 1} \frac{1}{n-1}  \left[
\log ( {\rm Tr} \rho_\o^n)  - \log ( {\rm Tr} \rho_\o \rho_{\o'}^{n-1})    \right].
\end{eqnarray}
Going through a similar analysis for the trace  ${\rm Tr}  \rho_\o \rho_{\o'}^{n-1}  $
we see that  the path integral  involves the $n$-branched plane in which one plane say the 
$0$-th plane contains the operator $\o$ and the rest $n-1$ planes contain 
the operator $\o'$. 
Therefore we have 
\begin{eqnarray}\label{mrdm}
\frac{ {\rm Tr} ( \rho_\o \rho_{\o'}^{n-1} )  }
{ {\rm Tr} ( \rho_{(0)} )^n } &=& 
\left\langle \o( 0_0) \o(\infty_0) \prod_{k =1}^{n-1} \Big[\o' (0_k)  \o^{\prime *} ( \infty_k  ) \Big]  \right\rangle _{\Sigma_n}, \\ \nonumber
&=& 
\left\langle  w\circ \o( w_0)   \hat w\circ \o^*( \hat w_0) 
\prod_{k =1}^{n-1} \Big[w\circ \o' ( w_k )   \hat w \circ \o^{\prime *} ( \hat w_{k} )  \Big] 
 \right\rangle.
\end{eqnarray}
Using the definition of $S_{n} ( \rho_{\o} || \rho_{\o'} ) $   in (\ref{defsn}) 
and (\ref{mrdm}) we obtain 
\begin{eqnarray}\label{relentropy1}
&& S_{n} ( \rho_{\o} || \rho_{\o'} ) = \frac{1}{n-1}  \log \frac
{\langle \prod_{k =0}^{n-1} w\circ \o ( w_k )  \prod_{k'=0}^{n-1} 
 \hat w \circ \o^*( \hat w_k')  \rangle}
 {\langle  w\circ \o( w_0)  \hat w \circ\o^*(\hat w_0) 
\prod_{k =1}^{n-1}  \left[ w\circ \o' ( w_k ) 
 \hat w \circ \o^{\prime *} ( \hat w_{k} )  \right] \rangle}.  \nonumber \\
 \end{eqnarray}

 We will begin by applying these formulae to the case of the free boson  
 to evaluate single interval  entanglement entropies and relative entropies 
 excited by the primary $\o = e^{i l X}$ and the descendant $\partial \o = \partial e^{i l X} $. 
 We will see that in particular the case of the  entanglement entropy of the state 
 excited by the descendant   can be exactly evaluated.

\subsection{Free boson CFT}

Consider the CFT of  a single boson $X$  and $\o = e^{i l X}, {\cal O }^* = e^{-i l X} $. 
Then  from (\ref{eecor}) we see that to evaluate the R\'{e}nyi entropy we would need to evaluate
the following  $2n$ point function for this operator
\begin{eqnarray}\label{cor1}
\langle \prod_{k =0}^{n-1} w\circ \o ( w_k )  \prod_{k'=0}^{n-1} 
 \hat w \circ \o^*( \hat w_{k'})  \rangle
 = \prod_{k=0}^{n-1} \left[  \left.  \frac{\partial w}{\partial z} \right|_{0_k}  
\left. \frac{\partial \hat w}{\partial z} \right|_{\infty_k}   \right]^{\frac{l^2}{2} } 
f( w; \hat w) , \\ \nonumber
f( w, \hat w ) = 
[ \prod_{k, k'=0; \,  k'>k}^{n-1}  ( w_k - w_{k'} )( \hat w_k - \hat w _{k'} ) ] ^{l^2} 
[ \prod_{k', k=0}^{n-1} ( w_k - \hat w_{k'} ) ]^{-l^2}. \\ \nonumber
\end{eqnarray}
Here the products over $k, k'$ run over $0, 1, \cdots n-1$. 
From the uniformization map (\ref{u1}) and ( \ref{u2}) we see that 
\begin{eqnarray}
&&w_k = e^{\frac{2\pi i (k +x) }{n} }, \qquad \qquad\qquad\qquad  \hat w_k = e^{\frac{2\pi i k }{n} }, 
\end{eqnarray} 
and 
\begin{eqnarray} \label{slope}
 \left. \frac{\partial w}{\partial z} \right|_{0_k}  &=&  \frac{1}{n} 
e^{\frac{ 2\pi i ( x+k) }{n} } ( 1- e^{-2\pi i x} ) ,    \qquad
\left. \frac{\partial \hat w}{\partial z} \right|_{\infty_k}    = - \frac{1}{n} e^{\frac{2\pi i k }{n}}  
( 1- e^{2\pi i x} ) , \\ \nonumber
&\equiv& B_k,  \qquad\qquad \qquad\qquad  \qquad\qquad \qquad \equiv \hat B_k .
\end{eqnarray}

The correlator (\ref{cor1}) have been evaluated before in 
\cite{Berganza:2011mh,Lashkari:2014yva}, but here we revaluate it  so as to develop the 
methods which will enable to to evaluate the corresponding correlator  for the  level one descendant of 
${\cal O }$. 
To proceed, we would need the following  algebraic identities 
\begin{eqnarray} \label{iden1}
\prod_{k'=0}^{n-1} ( z - e^{ \frac{ 2 \pi  ik'}{n} }  )  =   z^n- 1, 
 \\
\prod_{    k'\neq k, k' =0}^{n-1} 
 ( z- e^{\frac{ 2\pi i k'}{n} } ) =  \frac{ z^n - 1}{ z- e^{\frac{ 2\pi i k}{n} } } .
\end{eqnarray}
From these identities, it is easy to derive the relations. 
\begin{eqnarray} \label{iden4}
\prod_{k' =0}^{n-1} ( e^{\frac{ 2\pi i ( k+x) }{ n} } - e^{ \frac{ 2 \pi  ik'}{n} }  ) = e^{ 2\pi i x } - 1, \\
\label{iden2}
\prod_{ k';  k'\neq k} (  e^{ \frac{2\pi i k}{n} } - e^{\frac{ 2\pi i k'}{n} } )  = n e^{ -\frac{2\pi i k}{n} } .
\end{eqnarray}
To obtain  (\ref{iden2}) we can substitute $z= e^{\frac{2\pi i ( k +x)}{n} }$ and then take the 
$x\rightarrow 0$ limit. 
Using these identities we can simplify the first set of terms that occur in the function $f(w, \hat w)$ 
of (\ref{cor1}).  
\begin{eqnarray}
[ \prod_{k, k'=0; \,  k'>k}^{n-1} ( w_k - w_{k'} )( \hat w_k - \hat w _{k'} ) ]  &= &
 \prod_{k, k=0'; \,   k'>k}^{n-1}  (  e^{\frac{ 2\pi i ( k+x) }{ n} }  
  -  e^{ \frac{2\pi i (  k'  +x) }{n} }  )(  
 e^{\frac{ 2\pi i  k }{ n} }  
  -  e^{ \frac{2\pi i   k'   }{n} }  ) , \nonumber \\
  &=&  
  e^{  \pi i (n-1) x } \prod_{ k= 0}^{n-1} 
   \prod_{ k'; \, k'\neq k }   e^{ \frac{2\pi i k}{n} } - e^{\frac{ 2\pi i k'}{n} } ) ,  \\ \nonumber
   & =&    e^{  \pi i (n-1) x }  \prod_{ k= 0}^{n-1}   n e^{ - \frac{2\pi i k}{n} }  = e^{  \pi i (n-1) x }   n^{ n} e^{ - i \pi ( n-1) } .
\end{eqnarray}
Substituting this result in (\ref{cor1}), we obtain 
\begin{eqnarray} \label{iden3}
\langle \prod_{k =0}^{n-1} w\circ \o ( w_k )  \prod_{k'=0}^{n-1} 
 \hat w \circ \o^*( \hat w_{k'})  \rangle \nonumber
 &=&  \left( \frac{1}{n} \right)^{n l^2}   e^{ \pi i x l^2} e^{ \pi i (n-1) l^2}  ( e^{ i \pi })^{ \frac{nl^2}{2}}
( 2 \sin \pi x )^{ n l^2}  \\ \nonumber
&& \qquad\times   ( e^{  \pi i (n-1) x }   n^{ n} e^{ - i \pi ( n-1) } )^{ l^2} \\ \nonumber
& & \qquad\qquad  \times ( 2i  e^{ \pi i x }      \sin \pi x )^{- nl^2}, \\ 
&=& 1.
\end{eqnarray}
The first line arises from the conformal transformations, 
the second line from using  (\ref{iden3}) in (\ref{cor1})  and the third line 
arises from terms of (\ref{iden4}).  
Therefore we have the relation 
\begin{equation}\label{iden5}
\prod_{k= 0 }^{n-1} ( B_k \hat B_k ) ^{\frac{l^2}{2} } f( w , \hat w)  = 1.
\end{equation}
where $B_k, \hat B_k$ and  $f(w, \hat w)$  are defined in (\ref{slope}) and (\ref{cor1}) respectively. 

Using the result for the correlator  (\ref{cor1}) in (\ref{eecor}) we obtain 
\begin{equation}
\frac{ {\rm Tr} \rho_{e^{i l X}}^n  }{ {\rm Tr} \rho_{(0)}^n }  = 1.
\end{equation}
Therefore from (\ref{defreee}) we see that the single interval 
R\'{e}nyi entropies  as well as the entanglement entropy 
of the state $e^{i kX} |0\rangle$ is identical to the vacuum.  To be explicit we write 
\begin{equation} \label{eetriv}
\hat S( \rho_{e^{i l X}} )  = 0.
\end{equation}
where $\hat S$  is defined in (\ref{extee}).

\subsection{Entanglement entropy of $\partial e^{ i  l X} $} 

Let us proceed to evaluate the single interval 
entanglement entropy of the level one descendant $\partial e^{i l X}$. 
From (\ref{eecor}), we see that we need the conformal transformation of the level one descendant. 
This is given by 
\begin{equation} \label{cfttrans}
f\circ \partial {\cal O } ( z) =  \big[ \partial_z f(z) \big]^{h+1}  {\cal O } ( f(z) )  + h \partial_z^2 f(z) 
\big[ \partial_z f(z)  \big]^{h-1} {\cal O } ( f(z) ) .
\end{equation}
Thus we would need the second derivatives of the conformal transformation 
in (\ref{u1}) and which is given by 
\begin{eqnarray}  \label{secder0}
\left. \frac{\partial^2 w}{\partial z^2}  \right|_{0_k} 
&=& e^{\frac{ 2\pi i ( k + x) }{ n} } \frac{1}{n} ( 1- e^{-2\pi i x} ) 
\left[ ( 1+ e^{-2\pi i x} ) + \frac{1}{n} ( 1 - e^{-2\pi i x} ) \right] , \\ \nonumber
&=& B_k A, 
\end{eqnarray}
where
\begin{eqnarray} \label{defA}
A &\equiv&  \left[ ( 1+ e^{-2\pi i x} ) + \frac{1}{n} ( 1 - e^{-2\pi i x} ) \right].
\end{eqnarray}
Similarly the derivative of the conformal transformation (\ref{u2})   is given by 
\begin{eqnarray} \label{secder1}
\left. \frac{\partial^2 \hat w}{\partial z^2}  \right|_{\infty_k} 
&=&  \frac{1}{n} e^{\frac{2 \pi i k}{n} } ( 1- e^{2\pi i x} ) 
\left[ ( 1+ e^{2\pi i x} ) + \frac{1}{n} ( 1- e^{2\pi i x})  \right], \\ \nonumber
&=& \hat B_k  \hat A ,
\end{eqnarray}
where 
\begin{eqnarray}\label{defhatA}
\hat A &\equiv& - \left[ ( 1+ e^{2\pi i x} ) + \frac{1}{n} ( 1- e^{2\pi i x})  \right].
\end{eqnarray}
Using this input into the conformal transformation of the 
descendant  $\partial e^{i l X}$ given in (\ref{cfttrans}) we see that the  relevant 
$2n$ point correlator can be written as 
\begin{eqnarray} \label{expcor}
{\cal C}_{2n}  =  \left. \prod_{k =0}^{n-1} ( B_k \hat B_k )^{\frac{l^2}{2} }  \left[   \frac{l^2}{2} A + B_k \partial_{u_k}  \right]
\left[  \frac{l^2}{2} \hat A + \hat B_k \partial_{\hat u_k}  \right] f( u, \hat u ) \right|_{ (u, \hat u) = (w , \hat w )}.
\end{eqnarray} 
In (\ref{expcor})   we first act the derivatives on the function 
\begin{eqnarray} \label{taccor}
f( u , \hat u) = 
[ \prod_{k, k'=0; \,  k'>k}^{n-1}  ( u_k - u _{k'} )( \hat u_k - \hat u _{k'} ) ] ^{l^2} 
[ \prod_{k', k=0}^{n-1} ( u_k - \hat u_{k'} ) ]^{-l^2}, 
\end{eqnarray}
and then set  $( u, \hat u )$ to the point  $(w, \hat w)$, which are the points on the 
uniformized plane. 
The operator $\partial e^{il X}$ is not normalized. 
Evaluating the norm  using the adjoint defined in (\ref{adj1}), (\ref{adj2})  we obtain 
\begin{eqnarray}\label{normcalc}
\langle \partial e^{i l X}| \partial e^{i l X} \rangle &=& 
\lim_{z\rightarrow 0}  \langle I \circ \partial e^{i l X (z) }   e^{i l X(0)}  \rangle, \\ \nonumber
&=& - l^2.
\end{eqnarray}
Note that the reason we obtain a negative sign is due to our definition of the adjoint.  It is easy to 
see that according to this definition, the adjoint of the Virasoro generator $( L_{n})^\dagger = (-1)^n L_{-n}$
\footnote{ This can be seen from performing the inversion transformation on 
$ T(z)  = \sum_{n}\frac{L_{n}}{z^{n+2}}$ }, 
therefore  we get 
\begin{eqnarray}
\langle h  | L_{1}^\dagger L_{1} |h \rangle =  - \langle h  | L_{-1}  L_{1} |h \rangle  = - 2h .
\end{eqnarray}
Here $|h\rangle$ is a primary of weight $h$.  The above equation agrees with the explicit 
calculation in (\ref{normcalc}).

Before we proceed we will need the following identities. From (\ref{iden1}) we can obtain
\begin{eqnarray} \label{sumiden1}
 \sum_{k =0}^{n-1} \frac{1}{z - e^{\frac{2\pi i k}{n} } } =   \frac{ n z^{n-1}}{z^n -1}.
\end{eqnarray}
On taking the limit that $z$ is one of the $n$-th root of unity we obtain 
\begin{equation}
\label{sumiden2}
\sum_{j =0, j \neq k}^{n-1} \frac{1}{e^{\frac{ 2\pi i  ( k +x) }{n}} -  e^{\frac{2\pi i (j +x) }{n} } }
= \frac{n-1}{2} e^{-  \frac{2\pi i( k+x) }{n} } .
\end{equation}
Then from  the obvious substitutions in (\ref{sumiden1}) we obtain 
\begin{eqnarray} \label{sumiden3}
& & \sum_{j=0  }^{ n-1} \frac{1}{ e^{  \frac{ 2\pi i ( k+x) }{n} } - e^{ \frac{2\pi i j }{n}} } = 
\frac{ n e^{ - \frac{2\pi i (k+x) }{n} } }{ 1 - e^{-2\pi i x}} ,  \\ 
& & \sum_{j =0}^{n-1} \frac{1}{  e^{ \frac{ 2\pi i  k }{n} }  - e^{ \frac{ 2\pi i ( j +x) }{n } } }
= \frac{ n e^{ - \frac{2\pi i k}{n} }}{ 1 - e^{2\pi i  x}  }.
\end{eqnarray}
Now examining (\ref{expcor}) we see that there are  derivatives acting on the function $f(u, \hat u)$.
Consider a single derivative
\begin{eqnarray} \label{definigh}
\partial_{u_k } f(u, \hat u) &=& f(u, \hat u ) g_k( u, \hat u ) , \\ \nonumber
g_k(u, \hat u )  &=& l^2 \left[ \sum_{j =0 j\neq k }^{n-1}   \frac{1}{ u_k - u_j } - \sum_{j =0 }^{n-1} \frac{1}{ u_k - \hat u_j} 
\right], \\ \nonumber
\partial_{\hat u_k} f(u, \hat u) &=& f(u, \hat u ) h_k( u, \hat u ) , \\ \nonumber
h_k(u, \hat u ) &=&  l^2\left[ \sum_{j =0 j\neq k }^{n-1}   \frac{1}{ \hat u_k - \hat u_j } - 
\sum_{j =0 }^{n-1} \frac{1}{ \hat u_k -  u_j} 
\right].
\end{eqnarray}
Now setting $(u, \hat u) = ( w, \hat w)$ we obtain 
\begin{eqnarray}
\partial_{u_k } f(u, \hat u) |_{(u, \hat u) = ( w, \hat w)} &=& 
l^2 f(w, \hat w) \left[ \sum_{k \neq j }  \frac{ 1}{ e^{\frac{2\pi i ( x+j )}{ n } } -
e^{\frac{2\pi i ( x+k )}{ n } }  } - 
\sum_k \frac{1}{  e^{\frac{2\pi i ( x+j )}{ n } }  -  e^{\frac{2\p i  k }{ n } } } \right],  \nonumber \\
&=& l^2 f( w, \hat w ) \left[  \frac{n-1}{2} e^{ - \frac{ 2\pi i ( x+ j) }{n}} - 
\frac{n}{ 1 - e^{-2\pi i x} } e^{ - \frac{ 2\pi i ( x+ j }{n}} \right].
\end{eqnarray}
After substituting the points on the uniformized plane, in the second line we have used the
identities (\ref{sumiden2}) and (\ref{sumiden3}). 
In (\ref{expcor}), the derivative at $u_k$ occurs with $B_k$. 
Therefore let us evaluate 
\begin{eqnarray} \label{derrel1}
B_k\partial_{u_k } f(u, \hat u) |_{(u, \hat u)  = ( w, \hat w ) } 
&=& g_k (u, \hat u) f( u, \hat u)   |_{(u, \hat u)  = ( w, \hat w ) } , \\ \nonumber
&=& -  l^2 f( w, \hat w ) \left(  \frac{1}{2} ( 1+ e^{-2\pi i x} ) + \frac{1}{2n} ( 1- e^{ -2\pi i x}) \right) ,
\\ \nonumber
&=& -  \frac{l^2}{2}  f ( w, \hat w )  A.
\end{eqnarray}
Note that this relation does not involve a sum over $k$. 
This relation helps us simplify the various terms which involve the derivatives. 
Similarly we have the following relation for  the derivative  with respect to $\hat u_k$. 
\begin{eqnarray}\label{derrel2}
\hat B_k\partial_{\hat u_k } f(u, \hat u) |_{(u, \hat u) = ( w, \hat w) } &=&  
g_k(u, \hat u) f( u, \hat u)  |_{(u, \hat u) = ( w, \hat  w) } , \\ \nonumber
&= &- \frac{l^2}{2} f( w, \hat w) \hat A.
\end{eqnarray}

We are now ready to evaluate the correlator (\ref{expcor}). 
To illustrate the  manipulations involved we first consider the case $n=2$. 
Let us organise the terms according to the number of derivatives. 
The term involving  no derivatives is given by 
\begin{eqnarray} \label{2n0}
{\cal C}_2^{(0)} &=&\left(  \prod_{k=0}^1 (B_k \hat B_k) ^{\frac{l^2}{2}} ( \frac{l^2}{2} )^2 A \hat A  \right) f(w, \hat w) , 
\\ \nonumber
&=&  ( \frac{l^2}{2} )^4 ( A \hat A)^2 .
\end{eqnarray}
The superscript indicates the number of derivatives. 
To arrive at the second line in the above equation we have used the relation (\ref{iden5}) with $n=2$
All the terms with a single derivative  results in 
\begin{eqnarray}\label{2n1}
{\cal C}_2^{(1)} &=&  \left. \prod_{k =0}^1  (B_k \hat B_k) ^{\frac{l^2}{2}}
\left( A \hat A^2 B_j  \sum_{j =0}^1 \partial_{u_j}  + A^2 \hat A 
\sum_{j = 0}^1 \hat B_j \partial_{\hat u_j} \right) f( u, \hat u ) \right|_{( u, \hat u ) = ( w, \hat w) }, \\ \nonumber
&=& - 4  ( \frac{l^2}{2} )^4   ( A \hat A)^2 .
\end{eqnarray}
Here we have used (\ref{derrel1}) , (\ref{derrel2})  to simplify the action of 
the derivatives  and (\ref{iden5}) to arrive at the last line. 
The terms involving 2 derivatives  can be simplified using  the following relations
\begin{eqnarray}
\partial_{u_j} \partial_{u_l} f(u, \hat u ) &=&  f(u, \hat u ) g_j ( u, \hat u ) g_l ( u, \hat u )  + 
{l^2} \frac{ f( u, \hat u ) }{ ( u_j - u_l )^2 }, \\ \nonumber
\partial_{\hat u_j} \partial_{\hat u_l} f(u, \hat u ) &=&  f(u, \hat u ) h_j ( u, \hat u ) h_l ( u, \hat u )  + 
{l^2} \frac{ f( u, \hat u ) }{ ( \hat u_j - \hat u_l )^2 }, \\ \nonumber
\partial_{\hat u_j} \partial_{u_l} f(u, \hat u ) &=&  f(u, \hat u ) g_j ( u, \hat u ) h_j ( u, \hat u )  -
{l^2} \frac{ f( u, \hat u ) }{ ( u_l - \hat u_j )^2 },
\end{eqnarray}
where $j\neq l $. 
Then using these formulae for the derivatives together with the relations (\ref{derrel1}) ,  (\ref{derrel2}) 
and (\ref{iden5}) 
we obtain 
 \begin{eqnarray}\label{2n2}
 && {\cal C}_{2}^{(2)}  = ( \frac{l^2}{2} )^4  A^2 \hat A+    ( \frac{l^2}{2} )^2  \hat A^2 
 \frac{ l^2  }{  (w_0 - w_1)^2} B_0B_1   \\ \nonumber
 && +  ( \frac{l^2}{2} )^4  A^2 A^{\prime 2}  +   ( \frac{l^2}{2} )^2 A^2  
 \frac{ l^2 }{  (\hat w_0 - \hat w_1)^2}  \hat B_0 \hat B_1  \\ \nonumber
 && +4 ( \frac{l^2}{2} )^4  A^2 \hat A^{2} - ( \frac{l^2}{2})^2    A \hat A \left( 
 \frac{  l^2 B_0 \hat B_0}{ (w_0 - \hat w_0)^2}  
 + \frac{ l^2 B_0 \hat B_1}{ (w_0 - \hat w_1)^2}  
  + \frac{ l^2 B_1 \hat B_0}{ (w_1 - \hat w_0)^2 }  + 
  \frac{l^2 B_1 \hat B_1}{ (w_1 -\hat w_1)^2}  \right).
 \end{eqnarray} 
 The last line in the above equation is due to the contribution of the mixed derivative 
 $\partial_{u_j} \partial_{u_l} f( u, \hat u ) $.
Similarly using the  expressions for $3$ derivatives in  (\ref{3derv})  we obtain 
 \begin{eqnarray}\label{2n3}
  {\cal C}_2^{(3)}&=& - 2 ( \frac{l^2}{2} )^4 A^2 A^{\prime 2} - 2 ( \frac{l^2}{2} )^2 \hat A^{ 2} 
   \frac{ l^2 B_0 B_1} {(w_0 -w_1)^2 }
  \\ \nonumber
& & +  ( \frac{l^2}{2} )^2  A \hat A   \left( 
 \frac{ l^2 B_0 \hat B_0}{ (w_0 - \hat w_0)^2}  
 + \frac{ l^2 B_0  \hat B_1 }{ (w_0 - \hat w_1)^2}  
  + \frac{ l^2 B_1 \hat B_0 }{ (w_1 - \hat w _0)^2 }  + 
  \frac{B_1 \hat B_ 1}{ (w_1 -\hat w_1)^2}  \right) \\ \nonumber
  && - 2   ( \frac{l^2}{2} )^4  A^2 \hat A^{2} - 2 ( \frac{l^2}{2} )^2  A^2 
   \frac{  l^2 \hat B_0 \hat B_1}{ (\hat w_0 - \hat w_1)^2}  \\ \nonumber
  && + ( \frac{l^2}{2} )^2  A \hat A   \left( 
 \frac{ l^2 B_0 \hat B_0}{ (w_0 - \hat w_0)^2}  
 + \frac{ l^2 B_0 \hat B_1}{ (w_0 - \hat w_1)^2}  
  + \frac{ l^2 B_1 \hat B_0}{ (w_1 - \hat z_0)^2 }  + 
  \frac{B_1 \hat B_1}{ (w_1 -\hat w_1)^2}  \right).
  \end{eqnarray}
  The first two lines arise from from derivatives of the type $\partial_{\hat u_i} \partial_{u_j} \partial_{u_k} $,
  while the last two lines arise from derivatives of the type $\partial_{ u_i} \partial_{\hat u_j} \partial_{\hat u_k} $.
  Finally using (\ref{4derv}) the terms arising  from the action of   $4$ derivatives are given by 
  \begin{eqnarray}\label{2n4}
   &&{\cal C}_2^{(4)}= ( \frac{l^2}{2} )^4 
   A^2 A^{\prime 2}  +  ( \frac{l^2}{2} )^2 A^2  \frac{  l^2 \hat B_0 \hat B_1 }{ ( \hat w_0 - \hat w_1)^2} 
 + ( \frac{l^2}{2} )^2  \hat A^{ 2} \frac{ l^2  B_0 B_1}{ ( w_0 - w_1)^2} \\ \nonumber
 && -  ( \frac{l^2}{2} )^2 A \hat A   \left( 
 \frac{  l^2 B_0 \hat B_0 }{ (w_0 - \hat w _0)^2}  
 + \frac{ l^2  B_0 \hat B_1 }{ (w_0 - \hat w_1)^2}  
  + \frac{  l^2 B_1 \hat  B_0 }{ (w_1 - \hat w_0)^2 }  + 
  \frac{l^2 B_1 \hat B_1}{ (w_1 -\hat w_1)^2}  \right)
  \\ \nonumber
  && + l^4B_0 B_1\hat  B_0 \hat B_1  \left( 
  \frac{1}{( w_0 - \hat w_0)^2 ( w_1 - \hat w_1)^2 } 
  +  \frac{1}{( w_0 - \hat w_1)^2 ( w_1 - \hat w_0)^2 } 
  +  \frac{1}{( w_0 - w_1)^2 ( \hat w_0  - \hat w_1)^2 } 
  \right).
  \end{eqnarray}
  Adding (\ref{2n0}), (\ref{2n1}), (\ref{2n2}), (\ref{2n3},  (\ref{2n4}), we 
  obtain  the relevant $4$ point function of the descendant 
  to be 
  \begin{eqnarray}
  &&{\cal C}_2 = \sum_{a= 0}^{4}  {\cal C}_{2}^{(a)}, \\ \nonumber
  &=&   l^4B_0 B_1\hat  B_0 \hat B_1  \left( 
  \frac{1}{( w_0 - \hat w_0)^2 ( w_1 - \hat w_1)^2 } 
  +  \frac{1}{( w_0 - \hat w_1)^2 ( w_1 - \hat w_0)^2 } 
  +  \frac{1}{( w_0 - w_1)^2 ( \hat w_0  - \hat w_1)^2 } 
  \right) .
  \end{eqnarray}
  Note that all powers of $l^2$ greater than $2$ cancel, or in another words all terms containing $A$or $\hat A$ 
  cancel. 
  Also observe that the final result is identical to the four point function of the $U(1)$ current $i\partial X$ 
  on the uniformized plane together with the slopes of the uniformization maps (\ref{u1}), (\ref{u2}). 
 Let us finally evaluate the normalised correaltor by dividing by the norm evaluated in (\ref{normcalc}). 
 This removes the factor $l^4$ from the correlator and we obtain 
 \begin{eqnarray}\label{2ndfreed}
& &\frac{  {\rm Tr} ( \rho_{\partial e^{il X}}^2)  }{{\rm Tr} ( \rho_{(0)}^n) } 
= B_0 B_1\hat  B_0 \hat B_1   \\ \nonumber 
  &&\qquad \times  \left(  \frac{1}{( w_0 - \hat w_0)^2 ( w_1 - \hat w_1)^2 } 
  +  \frac{1}{( w_0 - \hat w_1)^2 ( w_1 - \hat w_0)^2 } 
  +  \frac{1}{( w_0 - w_1)^2 ( \hat w_0  - \hat w_1)^2 } 
  \right).
  \end{eqnarray}

\begin{figure} %[!t]
	\centering
	\includegraphics[width=.6\textwidth]{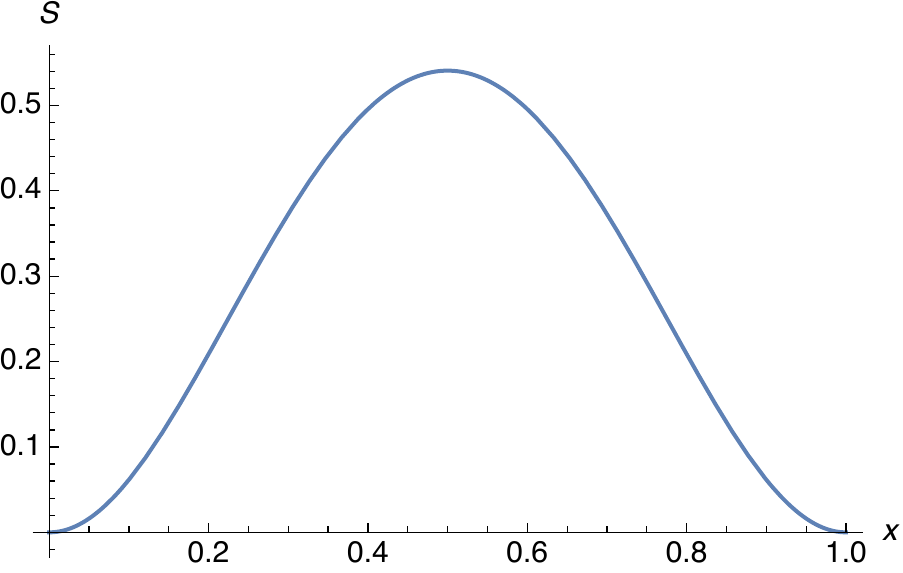}
	\caption{Single interval entanglement entropy of the descendant $\partial e^{ilX}$. Note that 
	as expected the entanglement entropy interval $x$ is the same as its complement $1-x$. }
	\label{eedeikx}
\end{figure}

The result  in (\ref{2ndfreed}) 
implies that the single interval  $2$nd R\'{e}nyi entropy of the descendant $\partial e^{il X}$ 
is identical to that of the holomorphic current $\partial X$. 
This observation can be generalized to the $n$-th  R\'{e}nyi entropy. The details of this proof is given in 
appendix \ref{appen1}.   
The result of evaluating (\ref{expcor}) for arbitrary $n$ is given by 
\begin{eqnarray}\label{expcorf}
  {\cal C}_{2n} &=& l^{2n} (-1)^n \left( \prod_{k=0}^{n-1}  ( B_k \hat B_k) \right) \left( 
  \prod_{k = 0}^{n-1} \frac{1}{ ( w_k - \hat w_k) } + {\rm distinct \;  permutations} \right) .
  \end{eqnarray}
  Here $w_k, \hat w_k$ are given in (\ref{locate}) and the slopes of the conformal transformations 
  are given in (\ref{slope}). 
  As remarked for the $n=2$ case, this correlator is identical to the $2n$ point function of 
  the primary $i\partial X$ 
  on the  uniformized plane.  To evaluate the entanglement entropy we need to consider the normalized 
  operator for which we divide (\ref{expcorf}) by the norm. 
  Therefore using (\ref{extee}) we obtain that the  $n$-th R\'{e}nyi entropy 
   of the level one descendant $\partial e^{i l X}$ is given by 
  \begin{eqnarray}
  \hat S_n( \rho_{\partial e^{i l X}}) = \frac{1}{1-n} \log \Big(\frac{  {\cal C}_{2n} }{ (-l^2)^n} \Big). 
  \end{eqnarray}
  Since we have shown that ${\cal C}_{2n}$ is identical to that of the $U(1)$ current, we can use the earlier 
  results of both the R\'{e}nyi and entanglement entropy for the state excited by the  $U(1)$ current
  obtained in \cite{Calabrese_2014,Ruggiero:2016khg}. 
  After substituting for $w_k, \hat w_k$ from  (\ref{locate}) and the slopes of the conformal transformations
  form (\ref{slope}) we can write the correlator as 
  \begin{eqnarray}
  \frac{{\cal C}_{2n} }{(-1)^n l^{2n}} = \left(  \frac{ \sin \pi x}{n } \right)^{2n} {\rm Hf} ( M ) .
  \end{eqnarray}
  The Hafnian is defined by 
\begin{eqnarray}
{\rm Hf} (M) \equiv \frac{1}{2^n n!} \sum_{\sigma \in S_{2n} }  \prod_{i =0}^{n-1} 
\frac{1} { \left( \sin \frac{ \theta_{\sigma(2 i) }- \theta_{ \sigma( 2i +1) }  } {2} 
\right)^2} , \\ \nonumber
\end{eqnarray}
where 
\begin{eqnarray}
\theta_{ k} - \theta_{ l } =
\begin{cases}
\frac{2\pi}{n}( k - l ), \qquad \qquad k, l \leq n-1 \\
\frac{2\pi }{n} ( k + x - l ), \qquad \qquad  k\leq  n-1, l > n-1, \\
\frac{2\pi }{n} ( k -  x - l ), \qquad \qquad  l \leq n-1, k > n-1, \\
\frac{2\pi }{n} ( k -l ), \qquad \qquad  k, l > n-1,
\end{cases}
\end{eqnarray}
and $M$ is the matrix whose entries are given by 
\begin{equation}
M_{kl}= \frac{1}{\left( \sin\frac{ \theta_k   -\theta_l}{2} \right)^2} .
\end{equation}
The Hafnian  can be simplified using the methods of \cite{Calabrese_2014,Ruggiero:2016khg} and 
can be obtained in terms of known functions 
\begin{equation}
{\rm Hf} (M)  
= 4^n 
\left( \frac{ \Gamma( \frac{ 1 +n + n \csc \pi x}{2} ) }{\Gamma(  \frac{ 1 -n + n \csc \pi x}{2} )  } \right)^2.
\end{equation}
Substituting this in the equation of the R\'{e}nyi entropies  of the excited state in (\ref{extee}), we obtain 
\begin{equation}
\hat S_n (\rho_{\partial e^{i l X}} ) = \frac{1}{ 1- n}  \log\left[  ( 2 \frac{\sin\pi x}{n} )^{2n} 
\left( \frac{ \Gamma( \frac{ 1 +n + n \csc \pi x}{2} ) }{\Gamma(  \frac{ 1 -n + n \csc \pi x}{2} )  } \right)^2
\right].
\end{equation}
Taking the $n\rightarrow 1$ limit we obtain the entanglement entropy 
\begin{eqnarray} \label{exacl1}
\hat S(\rho_{\partial e^{i l X}} )  &=& -2 \left( \log( 2 \sin \pi x)   + \psi ( \frac{\csc \pi x}{2} )  + \sin \pi x \right) , \\ \nonumber
&=& \frac{2 }{3} \pi^2 x^2 + O( x^4) . 
\end{eqnarray}
where 
\begin{equation}
\psi( x) = \frac{d}{dx} \Gamma(x) .
\end{equation}
As we mentioned earlier, this result is identical to then entanglement entropy of the $U(1)$ current since
the $2n$ point function of the descendant $\partial e^{il X}$ on the uniformized plane coincides with that 
of the $U(1)$ current $i \partial X$. 
The plot of this entanglement entropy as a function of $x$ is given in figure \ref{eedeikx}.
At the point we can also emphasise that the entanglement entropy of the   primary $e^{il X}$  vanishes 
(\ref{eetriv}), however its level one descendant shows interesting entanglement structure. 
In the next subsection we will extract some of the universal relations which determine the 
entanglement entropy  of the descendant given the entanglement entropy of the primary 
in the  short distance $x<<1$ limit.

We can proceed and evaluate various relative entropies associated with the level one descendant with 
respect to the vacuum and the primary.

\subsubsection*{$S(\rho_{\partial e^{i l X} } |\rho_{(0)} )$}

 Consider the relative entropy $S(\rho_{\partial e^{i l X} } |\rho_{(0)} )$
where $\rho_{(0)}$ is the reduced density matrix of a single interval in vacuum. 
From (\ref{relentropy1}), we see that the correlator in the numerator is  given by ${\cal C}_{2n}$. 
The correlator in the denominator consists of $n-1$ insertions of the identity and one insertion of the descendant. 
Therefore we need to evaluate the following 
\begin{eqnarray}
{\cal D} &=& \langle w \circ \partial e^{i l X ( w_0) } \,  \hat w \circ \partial  e^{ - i l X ( \hat w_0) } \rangle.
\end{eqnarray}
This is just the two point function on the uniformized plane in which both the operators are on a single wedge. 
It is easily evaluated and  the normalised correlator is given by 
\begin{eqnarray}
{\cal D} &=& \left( \frac{\sin \pi x}{ n \sin \frac{\pi x}{n} } \right)^{ 2(  \frac{l^2}{2} + 1) }
\left[ 1+ l^2  \big( \cos \frac{\pi x}{n}  - n \cot \pi x \sin \frac{\pi x }{n } \big)^2 \right].
\end{eqnarray}
Substituting this and ${\cal C}_{2n}$ into the expression for the relative R\'{e}nyi entropies in (\ref{relentropy1}) 
we obtain 
\begin{eqnarray}
&& S_n(\rho_{\partial e^{i l X} } |\rho_{(0)} ) =
\frac{1}{n-1} \Bigg\{  \log\bigg[  \Big( 2 \frac{\sin\pi x}{n} \Big)^{2n} 
\left( \frac{ \Gamma( \frac{ 1 +n + n \csc \pi x}{2} ) }{\Gamma(  \frac{ 1 -n + n \csc \pi x}{2} )  } \right)^2
\bigg] \\ \nonumber
&& \qquad  - 2 \Big( \frac{l^2}{2} +1 \Big) \log \left( \frac{\sin\pi x}{ n \sin \frac{\pi x}{n} } \right)
- \log\left[ 1+ l^2  \Big( \cos \frac{\pi x}{n}  - n \cot \pi x \sin \frac{\pi x }{n } \Big)^2  \right]
\Bigg\} .
\end{eqnarray}
The $n\rightarrow 1$ limit results in the relative entropy of the primary and the descendent
\begin{eqnarray}
S (\rho_{\partial e^{i l X} } |\rho_{(0)} )&=& 2 \left( \log( 2 \sin \pi x)   + \psi \Big( \frac{\csc \pi x}{2} \Big)  + \sin \pi x \right)  \\ \nonumber
 & & +  2 \Big( \frac{l^2}{2} +1\Big) ( 1- \pi x \cot \pi x) .
\end{eqnarray}

\subsubsection*{$S(\rho_{\partial e^{i l X} } |\rho_{e^{il X} } )$}

\begin{figure} %[!t]
	\centering
	\includegraphics[width=.6\textwidth]{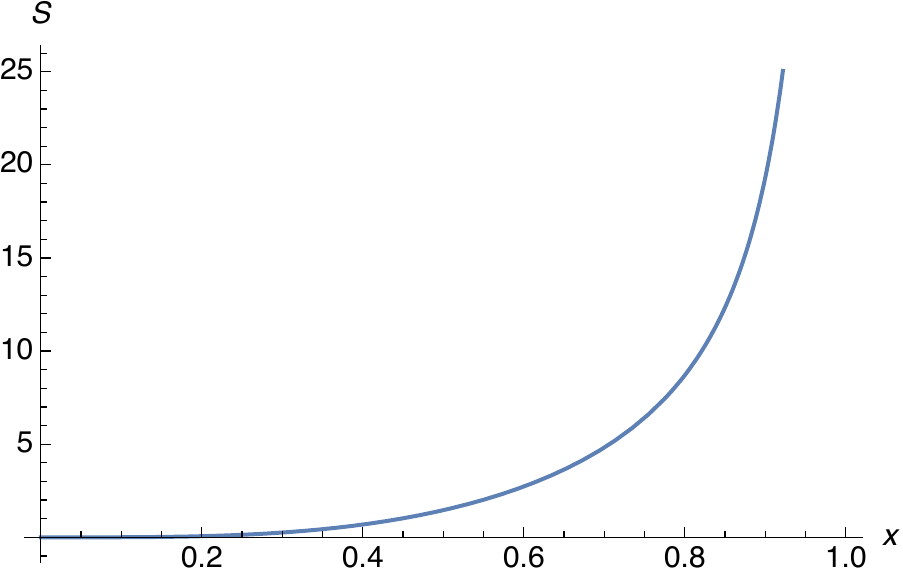}
	\caption{ Relative entropy of the descendant $\partial e^{il X} $ with respect to the primary $e^{ilX}$. }
	\label{redeikx}
\end{figure}

From (\ref{relentropy1}) we see that to evaluate the relative entropy of the descendant with the primary we need the 
following correlator which occurs in the denominator
\begin{eqnarray}
{\cal D} = \langle w\circ \partial e^{i l X ( w_0) }\,  \hat w \circ  \partial e^{i l X(\hat w_0) }  
\prod_{k=1}^{n-1}  w\circ  \partial e^{i l X ( w_k) }\,  \hat w  \circ \partial e^{i l X(\hat w_k) } \rangle .
\end{eqnarray}
From the conformal transformation of the descendant we see that this correlator is given by 
\begin{equation}
{\cal D} = 
 \left. \left( \prod_{k =0}^{n-1} ( B_k \hat B_k )^{\frac{l^2}{2} } \right)  \left[   \frac{l^2}{2} A + B_0 \partial_{u_0}  \right]
\left[  \frac{l^2}{2} \hat A + \hat B_0 \partial_{\hat u_0}  \right] f( u, \hat u ) \right|_{ (u, \hat u) = (w , \hat w )}.
\end{equation}
We can evaluate this correlator 
using the methods developed in appendix \ref{appen1}, this results in 
\begin{equation}
{\cal D} = \frac{ -B_0 \hat B_0 l^2}{ ( w_0 - \hat w_0)^2} .
\end{equation}
As expected this is the two point function of $U(1)$ currents on one wedge of the uniformised plane. 
Substituting the normalised correlators in 
(\ref{relentropy1})  we obtain 
 \begin{eqnarray}
  S_n ( \rho_{\partial e^{ikX} }|\rho_ {e^{ikX}} )  &=& 
 \frac{1}{n-1} \Bigg\{  \log\Big[  \Big( 2 \frac{\sin\pi x}{n} \Big)^{2n} 
\left( \frac{ \Gamma( \frac{ 1 +n + n \csc \pi x}{2} ) }{\Gamma(  \frac{ 1 -n + n \csc \pi x}{2} )  } \right)^2
\Big]  \\ \nonumber
&& \qquad\qquad 
-2  \log \left( \frac{\sin\pi x}{ n \sin \frac{\pi x}{n} } \right) \Bigg\}.
 \end{eqnarray}
 We obtain the relative entropy by taking the $n\rightarrow 1$ limit, which is given by 
 \begin{equation}
  S ( \rho_{\partial e^{ikX} }|\rho_ {e^{ikX}} ) = 2 \left( \log( 2 \sin \pi x)   + \psi ( \frac{\csc \pi x}{2} )  + \sin \pi x \right)  + 2  ( 1- \pi x \cot \pi x) .
 \end{equation}
 Figure \ref{redeikx}
  plots this relative entropy, as expected it is positive and it increases monotonically with respect
 to the interval length $x$. 

\section{Short distance expansion} \label{section2}
 
 In the previous section we have seen the example of  level one descendant of the operator 
 $e^{ilX}$ in the free boson theory  for which the entanglement entropies and certain relative 
 entropies  could be evaluated exactly. In general it is not possible to evaluate  the $2n$-point functions 
 involved in evaluating the entanglement entropies or the relative entropies (\ref{eecor}), (\ref{relentropy1}). 
 In this section we develop a systematic approximation that enables us to obtain these quantities 
 in the short interval approximation. 
 This approximation was developed in \cite{Sarosi:2016oks}, here we first revisit it and formulate it 
 so that it can be used easily when the excitations involve descendants. 
 
 \subsection{Primaries} \label{primary}
 
\begin{figure} %[!t]
	\centering
	\includegraphics[width=.6\textwidth]{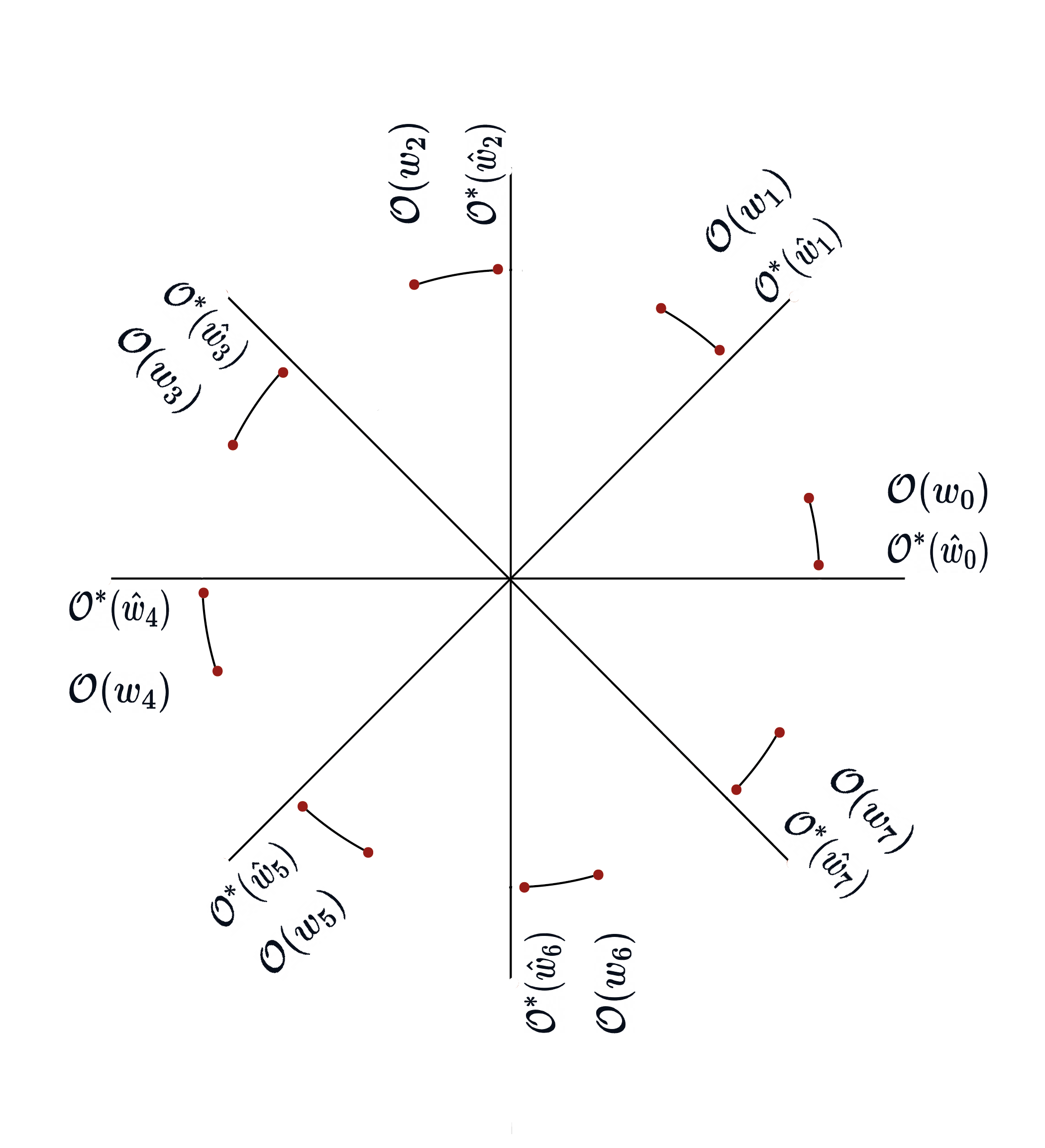}
	\caption{Factorization of the $2n$-point functions into $n$   2-pt functions on the same wedge.
	The figure shows the example of $n=8$.}
	\label{uniplane2}
\end{figure}

 Before we proceed let us discuss the case of primaries in the $2n$ point function (\ref{eecor}). 
 The positions of the operators in the uniformised plane is shown in figure \ref{uniplane}. 
 In the short distance limit, the distance between the 2 operators on the same wedge is small and 
 therefore the leading contribution is obtained by considering the factorisation of the $2n$-point 
 function into $n$ two point functions as shown in figure \ref{uniplane2}
 We write this leading term as 
 \begin{eqnarray}\label{factorapprox}
 {\cal C}_{2n} &=& 
 \langle \prod_{k =0}^{n-1}  w\circ \o ( w_k )  
 \hat w \circ \o^*( \hat w_{k})  \rangle, \\ \nonumber
 &\simeq& \prod_{k=0}^{n-1} \langle w\circ \o ( w_k )    \hat w \circ \o^*( \hat w_{k})  \rangle + \cdots.
 \\ \nonumber
 \end{eqnarray}
When the operator  $\o$ is a primary, the two point function is easily evaluated and is given by 
\begin{eqnarray}\label{samewedge}
\langle w\circ \o ( w_k )    \hat w \circ \o^*( \hat w_{k})  \rangle =
\left(  \frac{ \sin \pi x}{ n \sin \frac{\pi x}{n} } \right)^{2h}.
\end{eqnarray}
Here we have used the conformal transformation (\ref{u1}) and (\ref{u2}) and the corresponding 
slopes (\ref{slope}) at $w_k$ and $\hat w_{k}$. Note that the $2$-point function on the same 
wedge is independent of the wedge. 
Now substituting (\ref{samewedge}) into the factorized approximation (\ref{factorapprox}) of the $2n$ point function, 
we can evaluate the  R\'{e}nyi entropies and the entanglement entropy.
\begin{eqnarray}
\hat S_n (\rho_\o)  &=&  \frac{1}{1-n}\log\left(\frac{\sin\pi x}{n\sin \frac{\pi x}{n}}\right)^{2nh} + 
\cdots,  \\ \nonumber
\hat S (\rho_\o)  &=& \lim_{n\rightarrow 1} \hat S_n(\rho_\o) =  2h(1-\pi x\cot\pi x)+\cdots.
\end{eqnarray}
The leading short interval contribution is universal, it just depends on the dimension of the operator
and is independent of the details of the conformal field theory.
This contribution has been 
studied earlier both in CFT and in holography \cite{Alcaraz:2011tn,Bhattacharya:2012mi, Blanco:2013joa,Lashkari:2014yva,Sarosi:2016oks,Belin:2018juv}

Let us proceed to evaluate the sub-leading corrections to the short interval limit. 
The approach used in  \cite{Sarosi:2016oks} involves using the  description of the 
$2n$ point function as a 4-point function between $\o, \o^*$ and two twist operators. 
This 4-point function is rewritten as a sum over states in the  $\mathbb{Z}_n$ orbifold 
theory. 

\begin{figure} %[!t]
	\centering
	\includegraphics[width=.6\textwidth]{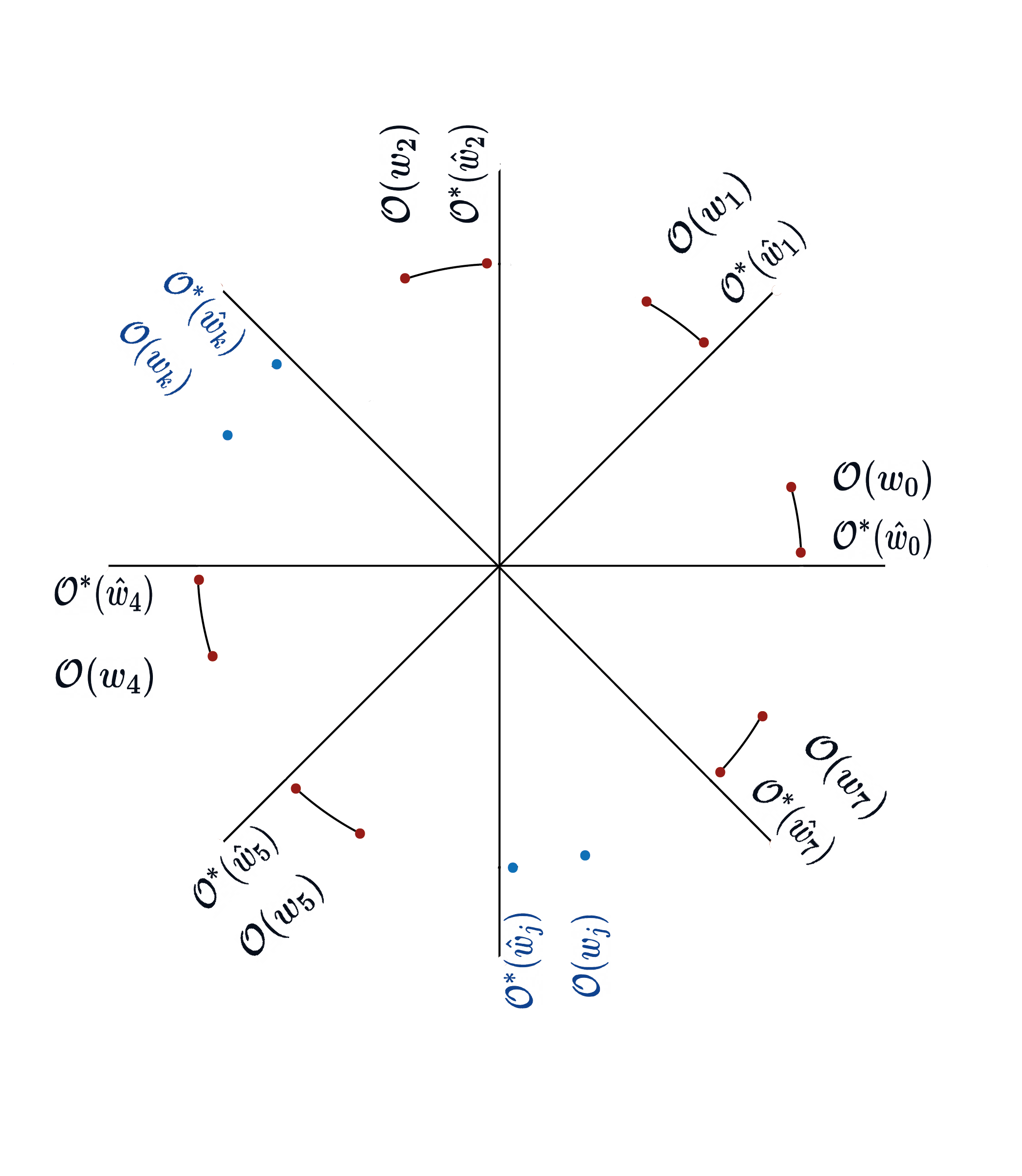}
	\caption{Subleading contribution to the short distance expansion from the factorization of the $2n$-point function to $(n-2)$ two point functions on the same wedge and a $4$-point function of operators 
	on a pair of wedges. These operators are shaded in blue. 
	One needs to sum all  pairs of wedges which contains the $4$-point function. The figure shows the example of $n=8$. }
	\label{n2correlators}
\end{figure}

Here we will obtain these corrections directly from the $2n$ point function. 
The leading approximation in (\ref{factorapprox}) involved the factorization of the $2n$ point 
functions into $n$ two point functions on the same wedge. 
The sub-leading  correction found by \cite{Sarosi:2016oks} arises 
by examining the lightest states formed by primaries 
in $2$ copies of the orbifold theory.  
Thinking of the $2n$-point function on the uniformized plane, this correction 
arises from the factorization into  $n-2$ pairs of $2$-point functions
on the same wedge. The remaining $4$ operators are contracted using the 
connected 4-point function \footnote{The connected four point function 
does not contain the pair wise contraction of these operators on the same wedge.
This contribution is already accounted.}. 
The figure \ref{n2correlators} illustrates this contraction. 
The operators shaded in blue 
are contracted using the connected four point function. 
One can chose any $2$ wedges among the $n$ wedges  in which these $4$ operators 
lie. Therefore we should sum over all such pairs. 
This discussion leads to the following approximation of the $2n$ point function
\begin{eqnarray} \label{subleadap}
&& {\cal C}_{2n} \simeq  \prod_{k=0}^{n-1} \langle w\circ \o ( w_k )    \hat w \circ \o^*( \hat w_{k})  \rangle
\\ \nonumber
&+&  \left(\frac{\sin\pi x}{n\sin \frac{\pi x}{n}}\right)^{2(n-2) h}  \times \sum_{j, k=0, j\neq k }^{n-1}
 \langle w\circ \o ( w_j )     \hat w \circ \o^*( \hat w_{j})  w\circ \o ( w_k )     \hat w \circ \o^*( \hat w_{k}) 
 \rangle_c + \cdots.
\end{eqnarray}
The subscript `$c$'   refers to the connected $4$ point function. 

We can evaluate the connected $4$-point function using its conformal block decomposition
which is given by, see for example in \cite{Perlmutter:2015iya}. 
\begin{eqnarray} \label{blockc}
&& \langle w\circ \o ( w_j )     \hat w \circ \o^*( \hat w_{j})  w\circ \o ( w_k )   \hat w \circ \o^*( \hat w_{k}) 
 \rangle_c  =   ( B_j \hat B_j B_k \hat B_k )^h  \times \\ \nonumber
&& \frac{1}{ (w_j - \hat w_j)^{2h} ( w_k - \hat w_k)^{2h} }   \left( \sum_{q =1}^{\infty} \chi_{{\rm vac},\,  q } w^2 \, {}_2F_1( q, q, 2q, w) + 
 \sum_{p}C_{\o\o\o_p} C^{\o_p}_{\; \o\o} w^{h_p} {\cal F}( c, h, h_p; w) \right),
\end{eqnarray}
where  the cross ratio $w$ is given by 
\begin{eqnarray} \label{crossratio}
w &=& \frac{(  w_j - \hat w_j)( w_k - \hat w_k) }{ (w_j - w_k) ( \hat w_j - \hat w_k) }
= \left( \frac{\sin \frac{\pi x}{n}}{ \sin \frac{\pi}{n} ( j- k ) } \right)^2 .
\end{eqnarray}
The hypergeometric functions ${}_2F_1( q, q, 2q, w) $ are the global $SL(2, C)$ 
 blocks and the sum over $q$ expresses the 
Virasoro block corresponding to the exchange of the stress tensor and its descendants in terms of these
global blocks.
Note that this is the vacuum block without the $q=0$ term since we  are interested in the 
connected $4$ point function. 
${\cal F}( c, h, h_p; w) $ is the Virasoro block corresponding to the exchange of the primary $\o_p$  with 
weight $h_p$ and 
its descendants.    The  leading term in the expansion of the Virasoro block  is unity
\begin{eqnarray} \label{blockc1}
{\cal F}( c, h, h_p; w)  = 1+  O( w) , 
\end{eqnarray}
$C_{\o\o\o_p} C^{\o_p}_{\; \o\o} $  is the  product of the structure constants, 
the raised superscript $\o$ refers to the fact that this index in the structure constants are raised by the 
Zamolodchikov metric.  The coefficients $\chi$ have been evaluated in \cite{Perlmutter:2015iya}. 
The first two coefficients are of importance to us
\begin{eqnarray} \label{blockc2}
\chi_1 =0, \qquad\qquad \chi_2 = \frac{2h^2}{c}.
\end{eqnarray}
where $c$ is the central charge of the CFT. 
It is clear from the conformal block expansion in (\ref{blockc}), the expression for the 
crossratio (\ref{crossratio}) and the property (\ref{blockc1}), (\ref{blockc2}) that the leading  terms in short 
interval contribution to the four point function is given by 
\begin{eqnarray} \label{4ptconformalb}
&& \langle w\circ \o ( w_j )     \hat w \circ \o^*( \hat w_{j})  w\circ \o ( w_k )     \hat w \circ \o^*( \hat w_{k}) 
 \rangle_c  = ( B_j \hat B_j B_k \hat B_k )^h  \times \\ \nonumber
 && \frac{1}{ (w_j - \hat w_j)^{2h} ( w_k - \hat w_k)^{2h} } 
 \left\{ \chi_2  w^2 \Big[ 1+ O(w)  \Big]   + C_{\o\o\o_p} C^{\o_p}_{\; \o\o} w^p \Big[ 1+ O(w) \Big]
 \right\}.
 \end{eqnarray}
 Here we have kept the leading term from the vacuum block and the exchange corresponding to the 
 primary with the lowest weight.   We will simply refer to  this primary as ${\cal O}_p$ with weight $h_p$. 
 Substituting for $w$ we obtain
 \begin{eqnarray} \label{lowestc4}
 &&  \langle w\circ \o ( w_j )     \hat w \circ \o^*( \hat w_{j})  w\circ \o ( w_k )     \hat w \circ \o^*( \hat w_{k}) 
 \rangle_c  =
  ( B_j \hat B_j B_k \hat B_k )^h \times  \\ \nonumber
 &&   \frac{1}{ (w_j - \hat w_j)^{2h} ( w_k - \hat w_k)^{2h} }
 \left[ \chi_2  \left( \frac{\sin \frac{\pi x}{n}}{ \sin \frac{\pi}{n} ( j- k ) } \right) ^4
 + C_{\o\o\o_p} C^{\o_p}_{\; \o\o}    \left( \frac{\sin \frac{\pi x}{n}}{ \sin \frac{\pi}{n} ( j- k ) } \right)^{2h_p} 
 \right] + \cdots. 
\end{eqnarray}
Though we have kept leading terms from stress tensor exchange together with that
of the lowest primary, only one of the terms contribute depending on the weight $h_p$ and the 
central charge $c$. 
We note the following
\begin{enumerate} 
\item
For  conformal field theories at finite central charge $c$, 
when $h_p<2$, the contribution is from the lowest primary and the $4$ point function is sensitive to the 
details of the CFT through the structure constants. However when $h_p>2$, the contribution from the 
stress tensor contribution is  dominant. 
\item
Finally for large $c$ CFT's and when $h\sim O(c)$, then  $\chi_2  \sim O(c)$. 
Then the leading term arises from the stress tensor exchange. 
\item
For large $c$ CFT's  and $h\sim O(1)$ we have $\chi_2 \rightarrow 0$, therefore the contribution 
due to the stress tensor exchange is negligible and the  leading  term is 
due to exchange of  the lowest primary $h_p$. 
An example in which this situation occurs is in generalised free field theory. 
Then $h_p =2h$, we will discuss this  case in some detail  in section \ref{section4.1}. 
\end{enumerate}
%Though we have 
%written  the contribution  of both the stress tensor exchange and the lowest primary   in 
%(\ref{lowestc4}) 
%it is understood that only one of the terms contribute. 

Substituting the conformal block decomposition of (\ref{lowestc4})  in (\ref{subleadap}) we obtain
\begin{eqnarray}
 {\cal C}_{2n} &=& \left(\frac{\sin\pi x}{n\sin \frac{\pi x}{n}}\right)^{2hn}
\times \left[  1 + \chi_2 \Big( \sin\frac{\pi x}{n} \Big)^4 \sum_{j, k =0, j\neq k}^{n-1} 
\left(  \frac{1}{ \sin \frac{\pi}{n} ( j-k) }  \right)^4 
 \right. \\ \nonumber
&& \left. 
% 1 + \chi_2 \Big( \sin\frac{\pi x}{n} \Big)^4 \sum_{j, k =0, j\neq k}^{n-1}
%\left(  \frac{1}{ \sin \frac{\pi}{n} ( j-k) }  \right)^4 
\qquad + 
C_{\o\o\o_p} C^{\o_p}_{\; \o\o}  \Big( \sin\frac{\pi x}{n} \Big)^{2h_p} 
 \sum_{j, k =0, j\neq k}^{n-1}
\left(  \frac{1}{ \sin \frac{\pi}{n} ( j-k) }  \right)^{2h_p} \right] ,
\end{eqnarray}
where  it is understood this result for the $2n$ point function includes only the leading and the first 
 sub-leading 
corrections. 
After changing variables, one of the sums can be performed and we obtain 
\begin{eqnarray} \label{sublead1}
{\cal C}_{2n} &=& \left(\frac{\sin\pi x}{n\sin \frac{\pi x}{n}}\right)^{2hn}
\left[ 1 + \chi_2 \Big( \sin\frac{\pi x}{n} \Big)^4 \sum_{l=1}^{n-1}
 \frac{n-l}{ (\sin \frac{\pi l }{n} )^4 } \right.  \\ \nonumber
& & \qquad\qquad\qquad\left. + 
C_{\o\o\o_p} C^{\o_p}_{\; \o\o}  
\Big( \sin\frac{\pi x}{n} \Big)^{2h_p} 
 \sum_{l=1}^{n-1}
\frac{n-l}{ (\sin \frac{\pi l}{n} )^{2h_p} }  \right] .
\end{eqnarray}
To perform the last sum we use  \cite{Calabrese:2010he}
\begin{eqnarray}\label{cardysum}
f( \alpha, n ) &=& \sum_{l=1}^{n-1} 
\frac{n-l}{(  \sin \frac{\pi l  }{n} )^{2\alpha } } , \\ \nonumber
& =& (n-1) \frac{\Gamma( \frac{3}{2} ) \Gamma( \alpha +1) }{  2 \Gamma( \alpha + \frac{3}{2} ) }
+ O( (n-1)^2) .
\end{eqnarray}
Using the above approximation in  (\ref{sublead1}) and then substituting the 
expansion of the $2n$ point function  in the expression for the entanglement entropy
(\ref{extee}), we obtain
\begin{eqnarray}\label{eeleadsub}
\hat S(\rho_{\o} ) &=&  2h ( 1- \pi x \cot\pi x)  - \frac{8 h^2}{15c}  ( \sin\pi x)^4  \\ \nonumber
&& - C_{\o\o\o_p} C^{\o_p}_{\; \o\o}  (\sin\pi x)^{2h_p} 
\frac{\Gamma(\frac{3}{2} ) \Gamma( h_p +1) }{ 2\Gamma( h_p + \frac{3}{2} ) } 
+ \cdots.
\end{eqnarray}
Let us recall that $h_p$ is the lowest dimension primary in the CFT and among the two sub-leading terms
only one contributes depending on the conditions discussed after equation (\ref{lowestc4}). 

For completeness we discuss 2 applications of the result (\ref{eeleadsub}). 
Let us consider the free boson theory for which $c=1$ with $\o = e^{i l X}$ and   $h = \frac{l^2}{2}$.  The lightest primary is the 
$U(1)$ current $\o_p = i \partial X$   with 
$h_p =1 <2$.   Therefore the 3rd term in (\ref{eeleadsub}) is the dominant contribution and we should 
trust this equation to $O(x^2)$.  Expanding the first term in(\ref{eeleadsub}) to $O(x^2)$ together with the 
3rd term we obtain 
\begin{equation}
\hat S( \rho_{e^{il X}}) = \frac{1}{3} ( l^2 - C_{\o \o\o_p}  C^{\o}_{\, \o\o})  + O(x^4) .
\end{equation}
The OPE coefficient  can be evaluated easily and is given by $C_{\o \o\o_p}C^{\o_p}_{\, \o\o}  = l ^2$.
Therefore we obtain $\hat S( \rho_{e^{il X}}) = 0 + O(x^4)$ which is consistent with 
the result in  (\ref{eetriv}).

For the second application we consider a 
CFT with large central charge  $c$ CFT with $h\sim O(c)$, then the 
single interval entanglement entropy of 
the excited state can be obtained by considering the heavy-heavy-light-light correlator 
\cite{Asplund:2014coa}. 
\begin{eqnarray}\label{coneee}
S( \rho_{\o} )  = \frac{c}{6} \log \left( \frac{2 \alpha  }{ \epsilon}  \sin \frac{\pi x  }{ \alpha} \right) , 
\qquad \alpha = \frac{1 }{ \sqrt{ \frac{24 h}{c} - 1} }.
\end{eqnarray}
Note that we have the factor $\frac{c}{6} $ rather than $\frac{c}{3}$ 
since we are in the holomorphic sector of the CFT
\footnote{The $\frac{c}{6}$ arises from the conformal dimension of twist operator 
$\frac{c}{24} ( n - \frac{1}{n}) $ in the holomorphic sector.}. 
The entanglement entropy of the vacuum is given by  taking the $h = 0$ limit in 
(\ref{coneee})
\begin{equation}
S({\rho_{0}}) = \frac{c}{6} \log\left(  \frac{2}{\epsilon} \sin \pi  x  \right) . 
\end{equation}
Evaluating the entanglement entropy of the excited state  we obtain 
\begin{eqnarray}\label{conexpee}
\hat S_{\rho_{\o}} &=&  S( {\rho_{\o}} )  -  S( {\rho_{0}})  , \\ \nonumber
%&=& \frac{h y^2}{ 6}  + \Big( \frac{ h}{360} - \frac{h^2}{30 c} \Big) y^6 +  O( y^6),   \\
&=& \frac{2h }{3}  \pi^2 x^2 + \Big( \frac{2 h}{45}  - \frac{ 8 h^2}{15 c} \Big) \pi^4 x^4 + O ( x^6) . 
\end{eqnarray}
From (\ref{eeleadsub}) see see that 
when $c$ is large and $h\sim O(c)$ it is only the first two terms which contribute
\begin{eqnarray}
\hat S(\rho_{\o} ) &=&  2h ( 1- \pi x \cot\pi x)  - \frac{8 h^2}{15c}  ( \sin\pi x)^4 , \\ \nonumber
&=& \frac{2h }{3}  \pi^2 x^2 + \Big( \frac{2 h}{45}  - \frac{ 8 h^2}{15 c} \Big) \pi^4 x^4 + O ( x^6).
\end{eqnarray}
As expected the  above equation precisely agrees with (\ref{conexpee}). 

Our aim in the next subsections is to derive a similar expression to that in 
(\ref{eeleadsub})  for the entanglement entropy 
when the excited state is a descendant of $\o$ in a generic CFT. 

\subsubsection*{Relative entropy between primaries}

Let us use the short distance expansion developed to evaluate the relative entropy between 
a primary $\o, \o'$  of weight $h$ and  of weight $h'$ respectively. 
From (\ref{relentropy1}), we see that we would need to evaluate the following correlator
in the short distance expansion
\begin{eqnarray}
\widehat{\cal C}_{2n} =\Big\langle  w\circ \o( w_0)  \hat w \circ \o ^* (\hat w_0) 
\prod_{k =1}^{n-1} \Big[ w\circ \o' ( w_k ) \hat w\circ \o^{\prime *}( w_k)  
 \Big]  \Big\rangle.
\end{eqnarray}
Again the leading contribution  to this correlator  in the short distance expansion arises from 
factorization of the $2n$  point function into $n$ two point functions, 
\begin{eqnarray}
\widehat{\cal C}_{2n} &\simeq&  \langle  w\circ \o' ( w_k )  \,  \hat w\circ \o^*( w_0)  \rangle
\prod_{k =1}^{n-1} \Big\langle  w\circ \o' ( w_k ) \, \hat w\circ \o^{\prime *}( w_k)   \Big\rangle +
\cdots , 
\\ \nonumber
&=& \left(  \frac{\sin\pi x}{ n \sin\frac{\pi x}{n}} \right)^{2 ( h + ( n-1) h') } + \cdots.
\end{eqnarray}
Following the same arguments as in the case of the correlator ${\cal C}_{2n}$, we see that 
that the leading corrections arise from the following 
\begin{eqnarray} \label{hatC}
&& \widehat{\cal C}_{2n} \simeq    \left(  \frac{\sin\pi x}{ n \sin\frac{\pi x}{n} }\right)^{2 ( h + ( n-1) h') }
\\ \nonumber
& &+   \left(  \frac{\sin\pi x}{ n \sin\frac{\pi x}{n}} \right)^{2 ( h + ( n-3) h') }
\times \sum_{j, k=1, j\neq k }^{n-1}
 \langle w\circ \o' ( w_j )     \hat w \circ \o^{\prime *}( \hat w_{j})  w\circ \o' ( w_k )     
 \hat w \circ \o^{\prime *}( \hat w_{k}) 
 \rangle_c  \\ \nonumber
 & & +   \left(  \frac{\sin\pi x}{ n \sin\frac{\pi x}{n} }\right)^{2  ( n-2) h' }
 \times \sum_{j =1  }^{n-1}
 \langle w\circ \o ( w_0 )     \hat w \circ \o^*( \hat w_{0})  w\circ \o' ( w_j )     \hat w \circ \o^{\prime *}
 ( \hat w_{j}) 
 \rangle_c + \cdots.
\end{eqnarray}
The second line contains the sub-leading correction that arises from the 
factorization into the connected  4 point functions 
of only  $\o'$ and  $n-2$ pairs of 2-point functions which includes the pair $\o$. 
The third line contains the sub-leading correction that arises from the factorization into 
the connected 4 points functions containing a pair of $\o'$ and $\o$, while the remaining 
operators are contracted pairwise on the same wedge. 
Using the conformal block decomposition for these four point functions we obtain
 \begin{eqnarray} \label{lowestcp4} \nonumber
 &&  \langle w\circ \o' ( w_j )     \hat w \circ \o^{\prime *}( \hat w_{j})  w\circ \o' ( w_k )     \hat w \circ \o^{\prime *} ( \hat w_{k}) 
 \rangle_c  =
  \frac{ ( B_j \hat B_j B_k \hat B_k )^{h' } }{  (w_j - \hat w_j)^{2h'} ( w_k - \hat w_k)^{2h'} } \\ 
 && \quad\quad\times  
 %\frac{1}{ (w_j - \hat w_j)^{2h'} ( w_k - \hat w_k)^{2h'} }
 \left[ \frac{ 2 h^{\prime \, 2}  }{c}  \Big( \frac{\sin \frac{\pi x}{n}}{ \sin \frac{\pi}{n} ( j- k ) } \Big) ^4
 + C_{\o'\o'\o_p} C^{\o_p}_{\; \o'\o'}    \Big( \frac{\sin \frac{\pi x}{n}}{ \sin \frac{\pi}{n} ( j- k ) } \Big)^{2h_p} 
 \right] + \cdots. 
\end{eqnarray}
The analysis leading upto this equation is the same as that of (\ref{lowestc4}) except that 
the operator $\o$ is replaced by $\o'$, here $h_p$ is the lowest  weight of the  primary 
that appears in the OPE of $\o'$. 
Similarly the leading contributions from the 
 four point function involving a pair of $\o$ and a pair of $\o'$ is given by 
\begin{eqnarray}\label{lowetcp4} \nonumber
 &&  \langle w\circ \o ( w_0 )     \hat w \circ \o^{*}( \hat w_{0})  w\circ \o' ( w_j )     \hat w \circ \o^{\prime *} ( \hat w_{j}) 
 \rangle_c  =
 \frac{  ( B_0 \hat B_0 )^h (B_j \hat B_j )^{h' } }{ (w_0 - \hat w_0)^{2h} ( w_j - \hat w_j)^{2h'} } \\ 
 &&\qquad\qquad\qquad \times  
 %\frac{1}{ (w_0 - \hat w_0)^{2h} ( w_j - \hat w_j)^{2h'} }
 \left[ \frac{ 2 h h' }{c}  \Big( \frac{\sin \frac{\pi x}{n}}{ \sin \frac{\pi j}{n}  } \Big) ^4
 + C_{\o\o \o_p} C^{\o_p}_{\; \o'\o'}    \Big( \frac{\sin \frac{\pi x}{n}}{ \sin \frac{\pi j}{n} } \Big)^{2 h_p} 
 \right] + \cdots.
\end{eqnarray}
Here we have assumed that the  lowest weight of the  primary that appears in the OPE of 
$\o$ as well as $\o'$ is  the same operator $\o_p$ with weight $ h_p$. 
Now we substitute the expressions for the four point functions  (\ref{lowestcp4}) and (\ref{lowetcp4})
 into (\ref{hatC})  to obtain 
 \begin{eqnarray} \label{hatC1} 
 &&\widehat{\cal C}_{2n} \simeq    \left(  \frac{\sin\pi x}{ n \sin\frac{\pi x}{n}}\right)^{2 ( h + ( n-1) h') }
 \times  \\ \nonumber
 & & 
 \left[ 1 + \big( \sin\frac{\pi x}{n} \big)^4  \Big( 
   \frac{ 2 h^{\prime\,  2}  }{c} \sum_{ j, k  =1,  j\neq k}^{n-1} \frac{1}{ \big(  \sin\frac{\pi}{n} ( j - k)  \big)^4} 
     + \frac{ 2 h h'}{c} 
   \sum_{j = 1}^{n-1}   \frac{1}{ \big(\sin\frac{\pi j }{n}   \big)^4} \Big) \right. \\ \nonumber
 &&  \left.  +  \big( \sin \frac{\pi x}{n} \big)^{2h_p}  \Big(  C_{\o'\o'\o_p} C^{\o_p}_{\; \o'\o'} 
 \sum_{j, k =1, j\neq k}^{n-1}
  \frac{1}{ \big( \sin \frac{\pi}{n} ( j-k)\big)^{2h_p} }  
+   C_{\o\o \o_p} C^{\o_p}_{\; \o'\o'} \sum_{j = 1}^{n-1}   
\frac{1}{ \big( \sin \frac{\pi j }{n} \big)^{2h_p} }  \Big) \right] .
 \end{eqnarray}
 Using the result for the sum in (\ref{cardysum}),  the two sums that occur in (\ref{hatC1}) can be performed
 and they are given by 
 \begin{eqnarray}
 \sum_{j, k =1, j\neq k}^{n-1}
\left(  \frac{1}{ \sin \frac{\pi}{n} ( j-k) }  \right)^{2\alpha }  
& =&- (n-1) \frac{\Gamma( \frac{3}{2} ) \Gamma( \alpha +1) }{  2 \Gamma( \alpha + \frac{3}{2} ) }
+ O( (n-1)^2),  \\ \nonumber
\sum_{j=1}^{n-1} \left( \frac{1}{\sin\frac{\pi j } {n} } \right)^{2\alpha} 
&= & 2  ( n-1) \frac{\Gamma( \frac{3}{2} ) \Gamma( \alpha +1) }{  2 \Gamma( \alpha + \frac{3}{2} ) }
+ O( (n-1)^2) .
 \end{eqnarray}
 Substituting these expressions  for the sums into (\ref{hatC1}) and using  the similar approximation  for
 the correlator ${\cal C}_{2n}$ given in  (\ref{sublead1})  
 in the expression for the relative entropy in (\ref{relentropy1}) we obtain 
 \begin{eqnarray}\label{relenf}
 S( \rho_{\o} | \rho_{\o'}) &=& \frac{ 8 }{15 c} ( h- h')^2 ( \sin\pi x)^4  \\ \nonumber
& &  + ( C_{\o\o\o_p} - C_{\o'\o'\o_p} )( C^{\o_p}_{\, \o\o}  - C^{\o_p}_{\, \o'\o'} ) 
 \frac{\Gamma(\frac{3}{2} ) \Gamma( h_p +1) }{ 2\Gamma( h_p + \frac{3}{2} ) } (  \sin\pi x)^{ 2h_p} 
+ \cdots.
 \end{eqnarray}
 Here we have assumed that $h_p$ is the lowest weight primary that occurs in the 
 OPE's. 
 Again we mention that depending on the conditions discussed after equation (\ref{lowestc4})
 it is only one of the terms in (\ref{relenf}) which contributes to the leading approximation 
 for the relative entropy.  Note that the expression for relative entropy is manifestly 
 positive.

\subsection{Descendants: leading contribution to entanglement entropy} \label{dessubl}

As discussed in section \ref{primary} the leading contribution is obtained by approximating the 
$2n$ point function as $n$ factorised 2-point functions on the uniformised plane. 
Therefore we begin by evaluate the two point function of the level-1 descendant $\partial \o$. 
Using the conformal transformation of the level  descendant given in (\ref{transdelO}) 
 we obtain 
\begin{eqnarray} \label{level1d}
\langle w\circ \partial \o(w_k) 
\hat w\circ \partial \o (\hat w_k)
\rangle = (-2h)  \left( \frac{\sin \pi x}{ n \sin\frac{\pi x}{n}} \right)^{2(h+1)} 
\left[ 1+ 2h \Big( \cos \frac{\pi x}{n} - n \cot\pi x\sin \frac{\pi x}{n} \Big)^2 \right].
\nonumber \\
\end{eqnarray}
The important point to notice is that the prefactor is identical to that of a primary of 
weight $h+1$, however there is a correction given by the second  factor. 
This second factor has the property 
\begin{equation}
\left[ 1+ 2h \Big( \cos \frac{\pi x}{n} - n \cot\pi x\sin \frac{\pi x}{n} \Big)^2 \right] = 
1+ O( (n-1)^2). 
\end{equation}
Indeed due to this property,  the leading contribution  
to the  entanglement entropy in the short distant limit is given by 
\begin{eqnarray}
S(\rho_{\partial \o} ) = 2 (h+1) (  1- \pi x \cot\pi x) + \cdots.
\end{eqnarray} 
To obtain this we have to normalise the correlator in (\ref{level1d}) and use (\ref{extee}). 

Now this property we have seen for the level one descendant can be shown to be true for all descendants. 
It simplifies both the computation of entanglement and relative entropies at the leading order as
well as the first sub-leading order. 
Let us first demonstrate this property for all global descendants of ${\cal O}$  of the 
form $\partial^l \o$. 

\subsubsection*{Global descendants}
The two point function for these descendants on the same wedge can be obtained by 
\begin{eqnarray}
\langle w\circ \partial^l \o (w_k) \, \partial^l \o^*( \hat w_k) \rangle
&=&  \partial_{z}^l \partial_{\hat z}^l G( z, \hat z, n) |_{(z, \hat z) = ( 0_k,  \hat 0_k) }, \\ \nonumber
G(z, \hat z, n ) &=& (  \partial_z w(z) \partial_{\hat z} \hat w( \hat z) )^h 
\left( \frac{1}{ w( z) - \hat w( \hat z) } \right)^{2h}.
\end{eqnarray}
In this expression we take the the derivatives first and then $(z, \hat z) = (0_k, \hat 0_k)$ which 
ensures that the  $w(z) \rightarrow 0_k, \hat w(z) \rightarrow \hat w_k$ as can seen from the 
maps (\ref{u1}) and (\ref{u2}). 
The strategy to evaluate the derivatives with respect to $z$ and $\hat z$ is to first 
expand the function $G(z, \hat z, n )$ around $n=1$. 
This results in 
\begin{eqnarray}
&& G(z, \hat z, n )  = \frac{1}{ (1 + z\hat z)^{2h} }  
-   (n-1) \times \frac{ h}{ (1 + z\hat z)^{2h} }   \left\{   2 + \log\Big(\frac{ z- u}{ z-v} \Big) +
 \log\Big( \frac{1+ u \hat z}{1+ v\hat z}\Big)  \right.  \\ \nonumber 
 && \qquad\qquad \left. + 
 \frac{ 2 }{(u-v)( 1+ z\hat z)}  \Big[
 ( z-u) ( 1+ v\hat z) \log\Big(\frac{ z- u}{ z-v} \Big)  
 + ( v-z) ( 1+ u \hat z)  \log\Big( \frac{1+ u \hat z}{1+ v\hat z}\Big) \Big]
 \right\}  \\ \nonumber
 &&\qquad\qquad \qquad\qquad\qquad\qquad  + O ((n-1)^2) .
\end{eqnarray}
Since we need to differentiate with respect to $z$ and $\hat z$ equal number of times 
and then set   $(z, \hat z) = ( 0_k,  \hat 0_k) $, we just need to examine  terms in 
$G(z, \hat z, n)$ dependent on the product $z\hat z$. 
This is given by 
\begin{eqnarray}
G(z, \hat z, n) |_{z\hat z} &=& \frac{1}{ ( 1+ z\hat z)^{2h} }
- ( n-1) \times  \frac{h}{ ( 1+ z\hat z)^{2h}} \Bigg\{ 2 + \log\Big(\frac{u}{v} \Big)  
\\ \nonumber
&& 
+ \frac{2}{ ( u-v) ( 1+ z\hat z) } \left[  ( - u  + z\hat z v) \log\Big(\frac{u}{v} \Big)    + 2 z\hat z( v- u) 
\right] \Bigg\}+ O( ( n-1)^2). 
\end{eqnarray}
Differentiating with respect to $z$ and $\hat z$, $l$-times  and then setting 
$(z, \hat z) = ( 0_k,  \hat 0_k) $. This differentiation is easy to carry out, we just have 
to group the terms proportional to $( z\hat z)^l $. 
We obtain
\begin{eqnarray} \label{lder}
\partial_z^l \partial_{\hat z}^l   G(z, \hat z, n )  |_{ (z, \hat z) = ( 0_k,  \hat 0_k)}
&=& \frac{  (-1)^l  \Gamma( 2h +l )\, l!  }{\Gamma( 2h) } 
\left[ 1 -  (n-1)2 (  ( h  + l )  (  1 + \pi x \cot \pi x)   \right]  \nonumber  \\&& + O( ( n-1)^2 ) .
\end{eqnarray}
where we have substituted the values of $u =e^{2\pi i x}, v=1$ from (\ref{uvdef}). 
Note that prefactor $\frac{ (-1)^l \Gamma( 2h +l )\, l!  }{\Gamma( 2h) } $ is the norm of the 
global descendant $\partial^l \o$.  
Therefore the  leading contribution to the entanglement entropy in the 
short distance limit  is given by
\begin{eqnarray}
S( \rho_{\partial^l\o} ) = 2 ( h+l ) ( 1-\pi x \cot\pi x) + \cdots .
\end{eqnarray}
The result in (\ref{lder}) also implies that the two point function of the descendent 
is given by 
\begin{eqnarray}
\langle w\circ \partial^l \o (w_k) \, \hat w \circ \partial^l \o^*( \hat w_k) \rangle= 
 \frac{ (-1)^l  \Gamma( 2h +l )  \, l! }{\Gamma( 2h) }  
 \left( \frac{\sin \pi x}{ n \sin\frac{\pi x}{n}} \right)^{2( h+l ) }  ( 1+ O ( ( n-1)^2 ). \nonumber 
 \\
\end{eqnarray}
Here is it understood that the pre-factor  should be expanded to $O(n-1)$ only. 
Thus the 2-point function of global descendants at level $l$  of a primary of weight $h$ 
in  the same wedge on the 
uniformized plane is identical to that of a primary of weight $h+1$ to order $(n-1)$. 

\subsubsection*{Virasoro descendants}

A class of Virasoro descendants descendants  of  the primary $\o$ can be defined by the OPE 
\begin{eqnarray}\label{TOope}
T(z) \o (w) = \sum_{k\geq 0} \frac{\o^{(-k) } (w)}{(z-w)^{2-k} } .
\end{eqnarray}
Form this definition, we see that 
\begin{equation}
\o^{(0)} (w) = h \o( w) , \qquad  \o^{(-1)} = \partial_z \o.
\end{equation}
Indeed another definition of these descendants is $\o^{-k} = L_{-k} \o$. 
From (\ref{TOope}), the inverse relation which directly defines the descendants 
is given by 
\begin{equation} \label{defvo}
\o^{(-k)} (z) = \oint_z \frac{d\tilde z }{ ( \tilde z-z)^{k-1} } T(\tilde z) \o(z) .
\end{equation}
where the contour  in $\tilde z$ is a small circle around $z$. 

Let us now examine the descendants at level 2. 
We  have 2 states,  the global descendant  $\partial^2 \o = L_{-1}^2 \o$ and  the 
Virasoro descendant $\o^{-2} = L_{-2} \o$. 
We have already evaluated the two point function of global descendants of the
same wedge, therefore we proceed to 
compute the two point function of the Virasoro descendant $
\langle  w\circ \o^{(-2)} ( w_k)  \, \hat w \circ \o^{(-2)} ( \hat w_k) \rangle $. 
To compute this we would need the conformal transformation of this descendant, 
evaluated in (\ref{O2transformation}). 
We see that the conformal transformation takes $\o^{-2}$ to 
a linear combination of $\o^{-2}, \o^{-1}$ and $\o$. 
This implies that we  require the two point function between 
any $2$ of these operators. These two point functions can be found in \ref{2ptlist}. 
There are 9 terms in all and finally we can expand the correlator  to $O(n-1)^2$. 
We have performed this computation in Mathematica and we find
\begin{eqnarray}  \label{2ptominuso}
\langle  w\circ \o^{(-2)} ( w_k)  \, \hat w \circ \o^{(-2)} ( \hat w_k) \rangle = \big(  \frac{c}{2} + 4h \big) 
 \left( \frac{\sin \pi x}{ n \sin\frac{\pi x}{n}} \right)^{2( h+2 ) }  ( 1+ O ( ( n-1)^2 ).
\end{eqnarray}
The calculation is straight forward though tedious. 
 To demonstrate a simple consistency  check let us 
examine the term proportional to  $O(c^2)$ in this correlator. 
From  the conformal transformation of $\o^{(-2)}$ given in (\ref{O2transformation}) 
we see that this term is given by
\begin{eqnarray}
&& \left. \langle  w\circ \o^{(-2)} ( w_k)  \, \hat w \circ \o^{(-2)} ( \hat w_k) \rangle \right|_{c^2}
\\ \nonumber
&& \qquad\qquad\qquad
= c^2 \left( \frac{\sin \pi x}{ n \sin\frac{\pi x}{n}} \right)^{2( h+2 ) }
\left(  -\frac{ A^2}{2} + F \right) \left( -\frac{ \hat A^2}{2}  + \hat F \right) 
\frac{ ( w_k - \hat w )^4  }{ (B_k \hat B_k)^2} , \\ \nonumber
&& \qquad\qquad\qquad
=  c^2 \left( \frac{\sin \pi x}{ n \sin\frac{\pi x}{n} } \right)^{2( h+2 ) }
\left [ 16 (n-1)^2  \sin^4 ( \pi x)   + O((  n-1)^3 ) \right] .
\end{eqnarray}
The equation  (\ref{2ptominuso})   therefore implies that the  leading contribution to the entanglement entropy 
in the short distance expansion is given by 
\begin{eqnarray}
S( \rho_{\o^{(-2)} } ) = 2 ( h+2 ) ( 1-\pi x \cot\pi x) + \cdots.
\end{eqnarray}

Similarly  let us evaluate the two point function of $\o^{(-3)}$ on the same wedge. 
For this we would us the
conformal transformation of $\o^{(-3)}$ given in (\ref{ominus3trans}). This  involves a linear 
combination of $\o^{(-3)}, \o^{(-2)}, \o^{(-1)}, \o$.  We can evaluate the 16 two point functions
that arise using the list given in (\ref{2ptlist}). 
The end result is 
\begin{eqnarray}\label{2ptominus3o}
\langle  w\circ \o^{(-3)} ( w_k)  \, \hat w \circ \o^{(-3)} ( \hat w_k) \rangle =-2 (  c + 3h ) 
 \left( \frac{\sin \pi x}{ n \sin\frac{\pi x}{n}} \right)^{2( h+3) }  [ 1+ O ( ( n-1)^2 ].
\end{eqnarray}
Again to perform a simple consistency check we can look at the term proportional to $O(c^2)$ 
in this correlator. 
\begin{eqnarray}
&&\left. \langle  w\circ \o^{(-3)} ( w_k)  \, \hat w \circ \o^{(-3)} ( \hat w_k) \rangle \right|_{c^2}
\\ && \qquad\qquad\qquad
= \big( \frac{c}{12} \big)^2  \left( \frac{\sin \pi x}{ n \sin\frac{\pi x}{n}} \right)^{2( h+3 ) }
( {-AF + 2G}) (  -\hat A \hat F + 2 \hat G) \frac{ ( w_k - \hat w_k )^6}{ ( B_k \hat B_k)^3},
 \nonumber \\
&&  \qquad\qquad\qquad \nonumber
 =\big( \frac{c}{12} \big)^2  \left( \frac{\sin \pi x}{ n \sin\frac{\pi x}{n}} \right)^{2( h+3 ) }
\Big[  -256 \cos ^2 \pi x  \sin^4 \pi  (n-1)^2 + O ( ( n-1)^3)  \Big].
\end{eqnarray}
The result in (\ref{2ptominus3o}) implies that the leading contribution to the entanglement entropy 
is given by 
\begin{eqnarray}
\hat S( \rho_{\o^{(-3)} } ) = 2 ( h+3 ) ( 1-\pi x \cot\pi x) + \cdots.
\end{eqnarray}
We have also repeated this calculation for the descendant $\partial \o^{(-2)}$ and have seen 
that again the leading contribution just depends on the level of the descendant. 
\begin{eqnarray}
\hat S( \rho_{\partial \o^{(-2)} } ) = 2 ( h+3 ) ( 1-\pi x \cot\pi x) + \cdots.
\end{eqnarray}

Based on the explicit calculations with the global descendants and all descendants till 
level $3$ we expect the two point function for descendants at level $l$ on same 
wedge in the uniformized plane is given by
\begin{equation}
\langle  w\circ \o^{[-l]} ( w_k)  \, \hat w \circ \o^{[-l]} ( \hat w_k) \rangle=
(-1)^l {\cal N}_{(l)}  \left( \frac{\sin \pi x}{ n \sin\frac{\pi x}{n}} \right)^{2( h+l ) }
\left[ 1+  O ( ( n-1)^2)  ) \right].
\end{equation}
Here  the superscript $l$ refers to any descendant at level $l$ and ${\cal N}_{(l)} $ is 
its norm. 
It will be interesting to provide a proof of the observation that the two point function on 
the same wedge of the uniformized plane 
of descendants of a primary of weight $h$ at level $l$ is same as that of a primary of weight
$h+l$ to order $O(n-1)$.

\subsection{Descendants: sub-leading contribution to entanglement entropy} \label{section3.3}

We now proceed to evaluate the sub-leading contribution to the short interval expansion 
of the entanglement entropy of descendants. 
Just as we have discussed for the primaries, the  leading and the sub-leading contribution 
to the short interval expansion 
of the $2n$-point function on the uniformized  plane is given by 
(\ref{subleadap}) where now ${\cal O}$ is a descendant of a primary of weight $h$. 
We have seen  in section \ref{dessubl}
 that the two point function of descendant at level $l$  on the same wedge 
behave as a primary of weight  $h+l$. 
Therefore we get
\begin{eqnarray}\label{2nlevel1d}
&&{\cal C} _{2n}  =  \Big\langle \prod_{k=0}^{n-1}
\big( w\circ \o^{[l]} (w_k)\,  w\circ \o^{*\, [l]} ( \hat w_k) \big) \Big\rangle, \\ \nonumber
&\simeq&  \Big( \frac{\sin \pi x}{ n \sin\frac{\pi x}{n} } \Big)^{2n ( h + l ) }  \\ \nonumber
& & +  \Big( \frac{\sin \pi x}{ n \sin\frac{\pi x}{n} } \Big)^{2( n-2)  ( h + l ) } 
\times \sum_{i, j = 0, i\neq j}^{n-1} 
 \Big\langle w\circ \o^{[l]}  ( w_j )     \hat w \circ \o^{*\, [l]} ( \hat w_{j})  
 w\circ \o^{[l]}  ( w_k )     \hat w \circ \o^{*\, [l]} ( \hat w_{k}) 
 \Big\rangle_c  \\ \nonumber
 & & + O( ( n-1)^2) + \cdots.
\end{eqnarray}
Here $\o^{[l]}$  refers to a descendant of level $l$.  Note that this relation is true only to 
 $O( (n-1)^2)$ since we have replaced the two point function of descendants 
 at level one on the same wedge with that of the primary of weight $h+l$.  
 Here are working with the normalised descendant at level $l$. 
 What remains now is to evaluate the connected   $4$-point function on pairs of wedges 
 in the conformal block approximation.  We need this only to $O(n-1)$ and 
 only the leading term in the short distance approximation. 
 
 \subsubsection{Level 1}
 
 We discuss the level one descendant $\partial O$  in detail. This will serve to illustrate the methods
 and approximations involved in obtain the leading contribution from the $4$-point function 
 from the operators on pairs of wedges. The relevant correlator is given by 
 \begin{eqnarray} \label{level14pt}
 &&{\cal F}_4 =  \frac{1}{(-2h)^2}  \Big\langle w\circ \partial\o ( w_j )     \hat w \circ \partial\o^{*} ( \hat w_{j})  
 w\circ \partial\o ( w_k )     \hat w \circ \partial\o^* ( \hat w_{k}) 
 \Big\rangle_c,  \\ \nonumber
 &=&\frac{1}{(2h)^2} 
 (B_j\hat B_jB_k \hat B_k)^h\left(h {A}+B_j\partial_{w_j}\right)\left(h {\hat A }+{\hat B_j}\partial_{\hat{w}_j}\right)\left(h {A}+{B_k}\partial_{w_k}\right)\left(h {\hat A}+{\hat B_k}\partial_{\hat{w}_k}\right)\\ 
 \nonumber
  && \times  \Big[\chi_{2}\frac{(w_j-w_k)^{-2}(\hat{w}_j-\hat{w}_k)^{-2}}{(w_j-\hat{w}_j)^{2h-2}(w_k-\hat{w}_k)^{2h-2}} + C_{\mathcal{O}\mathcal{O}\mathcal{O}_p}C_{\;\mathcal{O}\mathcal{O}}^{\mathcal{O}_p}\frac{(w_j-w_k)^{-h_p}(\hat{w}_j-\hat{w}_k)^{-h_p}}{(w_j-\hat{w}_j)^{2h-h_p}(w_k-\hat{w}_k)^{2h-h_p}}\Big].
\end{eqnarray}
Here we have substituted for the leading contribution to the connected $4$-point function 
from (\ref{4ptconformalb}), used the definition of the cross ratio in (\ref{crossratio}) 
 and used the action of the conformal transformation (\ref{u1}) and (\ref{u2}) on 
the level-1 descendants.  We have also normalized the  descendant operator. 
Expanding the action of the  $4$ terms   of the kind $( h A + B\partial_w)$ one obtains  $16$ terms. 
However it is reasonably  easy to isolate among these the leading terms which 
contribute in the short distance approximation. From (\ref{slope}) note that $B$'s vanish 
when $x\rightarrow 0$ as 
\begin{eqnarray}
B_k \sim \sin\pi x , \qquad \hat B_k \sim - \sin\pi x \qquad \hbox{as} \quad  x\rightarrow 0,
\end{eqnarray}
while $A$'s defined in (\ref{defA}) and (\ref{defhatA}) remain finite as $x\rightarrow 0$. 
\begin{eqnarray}
A \sim  2 , \qquad \hat A \sim - 2 , \qquad \hbox{as} \quad  x\rightarrow 0.
\end{eqnarray}
This implies that  for instance in  $(h A + B _j\partial_{w_j}  )$, 
the term involving $B\partial_{w_j} $ is generally suppressed 
compared to $hA$ in the short distance limit, unless the derivative acts on 
$( w_j - \hat w_j)^{-(2h -h_p)} $ since  form (\ref{locate})  we see that 
\begin{eqnarray}
(w_j - \hat w_j) \sim \sin\frac{\pi x}{n} , \qquad  \hbox{as} \qquad x\rightarrow 0. 
\end{eqnarray}
Thus the  action of $ B _j\partial_{w_j}  $  or the multiplication 
of $A$ 
on $( w_j - \hat w_j)^{-(2h -h_p)} $   
have identical behaviour as
$x\rightarrow 0$ limit. 
Similar arguments can be applied for all the terms of the form $(h A + B \partial_w)$. 
This leads us to conclude that the leading term in the short distance limit of 
the $4$-point function in (\ref{level14pt}) is given by 
\begin{eqnarray}\label{f4pt}
 {\cal F}_4 &=&  \frac{1}{(-2h)^2}   (B_j\hat B_jB_k \hat B_k)^h \\ \nonumber
 && \times   \left[ 
\chi_2   ( w_ j- w_k)^{-2} ( \hat w_j - \hat w_k )^{-2}
 \left(h {A}+B_j\partial_{w_j}\right)\left(h {\hat A }+{\hat B_j}\partial_{\hat{w}_j}\right)
\frac{1}{ ( w_j - \hat w_j)^{ 2h - 2 }  }   \right. \\  \nonumber
&& \qquad\qquad\qquad\qquad\times 
\left(h {A}+ B_k\partial_{w_k}\right)\left(h {\hat A }+{\hat B_k}\partial_{\hat{w}_k}\right)
\frac{1}{ ( w_k - \hat w_k)^{ 2h - 2 }  }   + O ( x^5 )   \\ \nonumber
& & + C_{\o\o\o_p}C^{\o_p}_{\, \o\o}  
( w_ j- w_k)^{-h_p} ( \hat w_j - \hat w_k )^{-h_p}
 \left(h {A}+B_j\partial_{w_j}\right)\left(h {\hat A }+{\hat B_j}\partial_{\hat{w}_j}\right)
\frac{1}{ ( w_j - \hat w_j)^{ 2h - h_p }  }   \\ \nonumber 
&&  \qquad\qquad\qquad\qquad\times  \left. 
\left(h {A}+ B_k\partial_{w_k}\right)\left(h {\hat A }+{\hat B_k}\partial_{\hat{w}_k}\right)
\frac{1}{ ( w_k - \hat w_k)^{ 2h - 2h_p}  }  + O( x^{2h_p +1})  \right]. \\ \nonumber
\end{eqnarray}

To illustrate a further simplification, let us consider the following action of the derivatives
\begin{eqnarray}\label{deform2ptfn}
 && \left(h {A}+B_j\partial_{w_j}\right)\left(h {\hat A }+{\hat B_j}\partial_{\hat{w}_j}\right)
\frac{1}{ ( w_j - \hat w_j)^{ 2h - h_p}  } 
= \frac{B_j\hat B_j }{ ( w_j - \hat w_j)^{ 2(h+1) - h_p }  } \times  \\ \nonumber
&& \left(h^2 A\hat A  \frac{ ( w_j - \hat w_j)^2}{B_j \hat B_j} 
 - (2h - h_p) h \hat A   \frac{w_j - \hat w_j  }{ \hat B_j} 
+ ( 2h - h_p) h A \frac{ w_j - \hat w_j }{B_j}  
- ( 2h - q) ( 2h - q +1)  \right).
\end{eqnarray}
Substituting this result in (\ref{f4pt}), we see that the term proportional to $C_{\o\o\o_p}$ reduces to 
\begin{eqnarray}
&& {\cal F}_4|_{C_{\o\o\o_p}} = \frac{C_{\o\o\o_p} C^{\o_p}_{\, \o\o} }{(-2h)^2} \frac{(B_j\hat B_jB_k \hat B_k)^{h+1} }
 {(w_j - \hat w_j)^{2(h+1)}  ( w_k - \hat w_k)^{2(h+1) } } \times w^{h_p} \times   \\ \nonumber
 && \Big[ h^2 A\hat A \frac{ (w_j - \hat w_j )^2}{B_j \hat B_j}
  - (2h - h_p) h \hat A   \frac{( w_j - \hat w_j) }{  \hat B_j} 
+ ( 2h - h_p) h A \frac{ ( w_j - \hat w_j) }{ B_j } 
- ( 2h - q) ( 2h - q+ 1) \Big]  \\ \nonumber
&& \qquad\qquad \times 
\Big[  j \rightarrow k\Big].
\end{eqnarray}
Here we have reassembled the dependences  in $w$'s back to the cross ratio defined in (\ref{crossratio}). 
Substituting for  the $w$'s from (\ref{locate}) and the $B$'s from (\ref{slope}) 
we obtain
\begin{eqnarray}\label{l1ddressf1}
&& {\cal F}_4|_{C_{\o\o\o_p}} = \frac{1 }{(-2h)^2}
\Big( \frac{ \sin\pi x}{ n \sin \frac{\pi x}{n} } \Big)^{4( h+1)} \times
\Big(  \frac{\sin \frac{\pi x}{n}}{ \frac{\pi}{n} ( j-k) } \Big)^{2h_p}  \times  \\ \nonumber
&&\Big[ h^2 A\hat A \frac{ (w_j - \hat w_j )^2}{B_j \hat B_j}
  - (2h - h_p) h \hat A   \frac{( w_j - \hat w_j) }{  \hat B_j}
+ ( 2h - h_p) h A \frac{ ( w_j - \hat w_j) }{ B_j } 
- ( 2h - q) ( 2h - q+1) \Big]  \\ \nonumber
&& \qquad\qquad \times 
\Big[ j \rightarrow k \Big].
\end{eqnarray}
From the definition of $w_j, \hat w_j$ in (\ref{locate}) and $B_j$ from (\ref{slope}) 
we see that the terms in bracket of the second line or 
the third line are independent of $j$ and $k$ respectively. 
Let us define
\begin{eqnarray}
 D( h, h_p, n ) &=&  \frac{1}{(-2h) } \times  \\ \nonumber
& &    \left( h^2 A\hat A \frac{ (w_j - \hat w_j )^2}{B_j \hat B_j}
  - (2h - h_p) h \hat A   \frac{( w_j - \hat w_j) }{  \hat B_j}  \right. \\ \nonumber
 && \left. \left. 
+ ( 2h - h_p) h A \frac{ ( w_j - \hat w_j) }{ B_j } 
- ( 2h - q) ( 2h - q+1) \right) \right|_{x\rightarrow 0}.
\end{eqnarray}
The quantity $D(h, h_p, n )$ also depends on $n$, but is 
independent of $j$.    We have taken the $x\rightarrow 0$ limit in this quantity since all higher
orders in $x$ contribute as $O( x^{2h_p + 1}) $. 
A  equation similar to (\ref{l1ddressf1}) results on performing the same analysis 
for the term proportional to $\chi_2$ in (\ref{f4pt}) with $h_p\rightarrow 2$. 
Therefore the 4-point function can be written as 
\begin{eqnarray}
{\cal F}_4  &=&  \Big( \frac{ \sin\pi x}{ n \sin \frac{\pi x}{n} } \Big)^{2( h+1)} 
\left[ \chi_2  [ D( h, 2, n) ]^2
\times \Big(  \frac{\sin \frac{\pi x}{n}}{ \frac{\pi}{n} ( j-k) } \Big)^{4}  + O ( x^{5})  \right. \\ \nonumber
& & \left. 
\qquad\qquad\qquad + C_{\o\o\o_p} C^{\o_p}_{\, \o\o} ( D( h, h_p, n) )^2
\Big( \frac{\sin \frac{\pi x}{n}}{ \frac{\pi}{n} ( j-k) } \Big)^{2h_p}  + O( x^{2h_p+1})   \right] .
\end{eqnarray}
Substituting this result for the 4-point function into (\ref{2nlevel1d}) we obtain 
\begin{eqnarray}
& & {\cal} C_{2n} =  \Big( \frac{ \sin\pi x}{ n \sin \frac{\pi x}{n} } \Big)^{2( h+1)} \times 
\Bigg[ 1 + \\ \nonumber
& &
  \quad (n-1)  \left\{
\frac{ 8 h^2}{ 15 c} [ D( h, 2, n) ]^2  \big( \sin \frac{\pi x}{n}  \big)^4  
+ C_{\o\o\o_p} C^{\o_p}_{\, \o\o} ( D( h, h_p, n) )^2
\frac{ \Gamma( \frac{3}{2} ) \Gamma( h_p+ 1) }{\Gamma( h_p+ \frac{3}{2} ) }
\right\} \Bigg] \\ \nonumber
& & \quad  + ( (n-1)^2) .
\end{eqnarray}
Here we have performed the sums over pairs of wedges using (\ref{cardysum}). 
Since we are interested in the entanglement entropy we can take the $n\rightarrow 1$ limit 
on the second term in the curved bracket. 
Thus the factor $D(h, 2, n)$ can be evaluated in the $n \rightarrow 1$ limit which results in 
\begin{eqnarray}\label{dressfac}
D_{\partial \o}( h, h_p)  &=& \lim_{n\rightarrow 1} D( h, h_p, n) , \\ \nonumber
&=&  \frac{  2h - h_p + h_p^2}{2h}.
\end{eqnarray}
Substituting  this result and evaluating the entanglement entropy we obtain 
\begin{eqnarray}\label{eeleadsubd}
\hat S(\rho_{\partial\o} ) &=&  2(h+1)  ( 1- \pi x \cot\pi x)  - \frac{8 (h+1)^2}{15c}  ( \sin\pi x)^4  \\ \nonumber
&& - C_{\o\o\o_p} C^{\o_p}_{\; \o\o} \Big(\frac{  2h - h_p + h_p^2}{2h}\Big)^2
\frac{\Gamma(\frac{3}{2} ) \Gamma( h_p +1) }{ 2\Gamma( h_p + \frac{3}{2} ) } (\sin\pi x)^{2h_p} 
+ \cdots.
\end{eqnarray}
Note that $D_{\partial \o}( h, 2) = \frac{h+1}{h}$ results in the enhancement of the contribution 
 due to the stress tensor exchange.   This factor is also expected, since the structure constant
 for the vacuum block should be proportional to the square of the weight of the external states.
 Comparing the sub-leading corrections to that of the  primary in (\ref{eeleadsub})
 owe see that the position dependence of these corrections remain the same, 
 but the pre-factors are modified by a dressing factor (\ref{dressfac}) which depends only 
 on the weight of the primary and the dimension of the lowest lying primary. 
 
 The derivation of the sub-leading corrections in the short interval expansion of the 
 entanglement entropy to the level 1 descendent makes it clear that the 
 dressing factor is the  ratio of the deformation of the norm of the level one primary to that 
 of the original norm.  This is evident from (\ref{deform2ptfn}) from which the dressing factor emerges. 
 The deformation is in which the two point function of the primary 
  is changed to that of an operator of weight $h - h_p/2$ while the conformal transformation 
  of the descendant is that of an operator of weight $h$ as it is clear from the left hand side 
  of  (\ref{deform2ptfn}). 
 Note that the terms in the brackets are finally evaluated in the $n\rightarrow 1$ limit and observe 
 that when $h_p=0$ one is essentially evaluating the two point function of the descendants. 
 These observations are useful in evaluating the dressing factors at higher levels. 
 
 Let us  perform a  check of the result in (\ref{eeleadsubd})  using the exact result for the 
 entanglement entropy of the descendant $\partial e^{il X}$ in (\ref{exacl1}). 
 Substituting $h = \frac{l^2}{2}, h_p = 1, C_{\o\o\o_p} C^{\o_p}_{\; \o\o} = l^2$ 
 in  (\ref{eeleadsubd}) we obtain 
 \begin{eqnarray}
 \hat S(\rho_{\partial e^{il X}}) = \frac{2}{3} \pi^2 x^2 + O( x^4) , 
 \end{eqnarray}
 which precisely agrees with  the leading term in the expansion of (\ref{exacl1}). 

\subsubsection{Level 2}

At level 2 we have two descendants $\partial^2 \o$ and $\o^{(-2)}$. 
The analysis for $\partial^2 \o$ proceeds identically to that discussed for the level one descendant. 
The final result for the entanglement entropy is given by 
\begin{equation} \label{level2des}
    \begin{split}
    \hat S( \rho_{\partial^2 \o} ) =&2(h+2)(1-\pi x\cot \pi x)- \frac{8 (h+2)^2}{15c} 
(\sin\pi x)^4\\
&-   C_{\o\o\o_p}C_{\, \o\o}^{\o_p} [ D_{\partial^2\o} ( h, h_p) ]^2 
\frac{\Gamma(3/2)\Gamma(h_p+1)}{2\Gamma(h_p+3/2)} (\sin \pi x)^{2h_p} + \cdots,
    \end{split}
\end{equation}
\begin{eqnarray}
D_{\partial^2 \o} ( h, h_p) &=&
\frac{8 h^2 + h (8 h_{p}^{2} - 8 h_p + 4) + h_p (h_p - 1) (h_{p}^{2} - h_p + 2)}{4h(2h+1)} \\ \nonumber
&=&  \frac{2! \Gamma(2h) }{ \Gamma( 2h - h_p)  \Gamma( 2h + 2) } \sum_{k = 0}^2
\frac{  \Gamma( 2h - h_p + k ) }{ k!  } \left( 
\frac{ \Gamma ( h_p + 2 - k ) }{ (2-k) ! \Gamma( h_p)  } \right)^2 .
\end{eqnarray}
As a consistency check note that 
\begin{eqnarray} \label{dconsti} 
D_{\partial^2 \o} ( h, 0 ) = 1,  \qquad D_{\partial^2 \o} ( h, 2) = \frac{h+2}{h}.
\end{eqnarray}
The first equation in (\ref{dconsti}) 
 results due to the fact explained in the level 1 case. 
The dressing factor is a ratio of the deformed norm of the descendant to the undeformed norm and it 
should reduce to unity when $h_p=0$. 
The second equation ensures that  the stress tensor contribution in (\ref{level2des}) 
is proportional to $( h+2)^2$. 

%\begin{eqnarray}
% {\cal F}_4|_{C_{\o\o\o_p}}&=& \left( \frac{\sin\pi x}{ \sin \frac{\pi x}{n} }  \right)^{

We will discuss the Virasoro descendant  $\o^{(-2)}$ in detail as it involves using conformal 
Ward identities to evaluate the relevant correlator. 
From (\ref{2nlevel1d}) we see  that we need the 4-point function 
\begin{eqnarray}\label{leve2f4}
{\cal F}_4 = \frac{1}{ \big( \frac{c}{2} + 4h\big)^2 } 
 \Big\langle w\circ \o^{(-2)}    \hat w \circ \o^{*\, (-2)} ( \hat w_{j})  
 w\circ \o^{(-2)} ( w_k )     \hat w \circ \o^{*\, (-2)} ( \hat w_{k}) 
  \Big\rangle_c .
\end{eqnarray}
The conformal transformation of  $\o^{(-2)}$ is given by 
\begin{eqnarray}{\label{O2transformationm}}
       w\circ \o^{(-2)}(w_k)&=& B_{k}^{h}\Big[ B_k^2 \o^{(-2)}(w_k)+\frac{3A B_k }{2}\o^{(-1)}(w_k)
       \\ \nonumber & & \qquad 
       +\frac{1}{12}\Big((5h-\frac{c}{2})A^2 +2(4h+\frac{c}{2})F \Big)\o(w_k)\Big], \\ \nonumber
         \hat w \circ \o^{(-2)} (\hat w_k)&= &\hat B_{k}^{h}\Big[ \hat B_k^2 
         \o^{(-2)}(\hat w_k)+\frac{3\hat A \hat B_k }{2 
        }\o^{(-1)}( \hat w_k) \\ \nonumber
        && \qquad +\frac{1}{12}\Big((5h-\frac{c}{2})\hat A^2 +2(4h+\frac{c}{2})
        \hat F \Big)\o ( \hat w_k)\Big].\\
\end{eqnarray}
Substituting the conformal transformation of $\o^{(-2)}$  we see that we need to 
evaluate 81 correlators  involving $\o^{(-2)}, \partial\o , \o$. 
However as we have seen earlier for the level 1 descendant,  we can simplify the calculation
since we are interested in obtaining only the leading correction at short distance 
and also the $O((n-1))$ term. 
To illustrate the simplification consider the  following  term which occurs in the evaluation 
of ${\cal F}_4$ in (\ref{leve2f4}).  
\begin{eqnarray} \label{f4onecom}
{\cal F}_4^{(1)} &=& (  B_j \hat B_j B_k \hat B_k )^h  \langle B_k^2 \o^{(-2)} ( w_j) 
\o^*( w_j) \o( w_k ) \o^*( w_k)  \rangle \\ \nonumber
& & \times \frac{1}{12}\Big((5h-\frac{c}{2})A^2 +2(4h+\frac{c}{2})F \Big) \times
\left[  \frac{1}{12}(5h-\frac{c}{2})\hat A^2 +2(4h+\frac{c}{2})
        \hat F \right]^2.
\end{eqnarray}
The superscript in ${\cal F}_4^{(1)}$ just refers to the fact  that we are looking at one term
among the $81$ terms.  
Due to the presence of $B_k^2$ along with $\o^{(-2)}$  which vanishes as $x\rightarrow 0$ limit
we need to look only for the  contributions which remain finite in this limit. 
The correlator we need  can be evaluated using conformal Ward identities and the definition of
$\o^{(-2)}$ in (\ref{defvo}). 
\begin{eqnarray}\label{f4onecom1}
&& \langle \o^{(-2)} ( w_j) \o^* (\hat w_j ) \o ( w_k) \o^* ( \hat w_k)  \rangle_c 
= \left[ \frac{h}{ ( w_j - \hat w_j)^2 } + \frac{1}{( w_j - \hat w_j} \partial_{\hat w_j }   +
\frac{h}{ ( w_j -  w_k)^2 }
\right. 
 \nonumber \\ \nonumber
&&\quad \left.  + \frac{1}{( w_j - w_k} \partial_{\hat w_k } 
+ \frac{h}{ ( w_j - \hat w_k)^2 } + \frac{1}{( w_j - \hat w_k} \partial_{\hat w_k } 
\right]  \langle \o  ( w_j) \o^*  (\hat w_j ) \o ( w_k) \o^* ( \hat w_k)  \rangle_c . \\
\end{eqnarray}
From (\ref{f4onecom}) we see that the correlator in (\ref{f4onecom1}) occurs  with a factor of $B_j^2$, 
therefore $x\rightarrow 0$ limit  only from the first two terms contribute at the leading order.
The denominators in the second line of (\ref{f4onecom1}) are finite, therefore the factor
$B_k^2 \sim x^2$ will render them sub-leading compared to the first two terms which come with an 
 an additional $1/x^2$. 
In conclusion, the leading terms can be obtained by restricting the Ward identity to 
 \begin{eqnarray} \label{wiapp}
&& \langle \o^{(-2)} ( w_j) \o^* (\hat w_j ) \o ( w_k) \o^* ( \hat w_k)  \rangle_c 
\simeq   \\ \nonumber
& &\qquad\qquad
\left[ \frac{h}{ ( w_j - \hat w_j)^2 } + \frac{1}{( w_j - \hat w_j} \partial_{\hat w_j }  \right]
\langle \o  ( w_j) \o^*  (\hat w_j ) \o ( w_k) \o^* ( \hat w_k)  \rangle_c.   \nonumber 
\end{eqnarray}
This conclusion is true for all of the $81$ correlators in (\ref{leve2f4}).  The leading contributions
can be obtained from terms contain only  powers of $(w_j- \hat w_j) $ or $( w_k - \hat w_k)$ 
in the denominator. 
We illustrate this with one more correlator whose leading terms are given by 
\begin{eqnarray}
& & \langle \o(w_j) \o^* ( \hat w_j) \o^{(-2)} (w_k) \o^{* ( -2) } ( \hat w_k) \rangle  \\ \nonumber
& & \qquad\qquad\qquad \simeq  
\left( \frac{\frac{c}{2} + 4h }{ ( w_ k - \hat w_k)^4}  +  3 \frac{1}{(  w_k - \hat w_k)^3} \partial_{ \hat w_k } 
\right) \langle \o(w_j) \o^* ( \hat w_j) \o(w_k) \o^{*} ( \hat w_k) \rangle   \\ \nonumber
& & \qquad\qquad\qquad+ \left(  
\frac{h+2}{ ( w_k - \hat w_k)^2}  +  \frac{1}{( w_k - \hat w_k) }  \right) 
\langle \o(w_j) \o^* ( \hat w_j) \o(w_k) \o^{* (-2) } ( \hat w_k) \rangle .
\end{eqnarray}
Here we have used the OPE in (\ref{tominus2}) and kept only the leading terms in the Ward identity. 

Let us proceed with the 
 approximation (\ref{wiapp})  for the Ward identity in the conformal block expansion of the 
four point function  given  in (\ref{4ptconformalb}). 
The term proportional to the OPE coefficient  in the conformal block expansion  is given by 
\begin{eqnarray}
&& \langle \o^{(-2)} ( w_j) \o^* (\hat w_j ) \o ( w_k) \o^* ( \hat w_k)  \rangle_c |_{C_{\o\o\o_p} }
\\ \nonumber
&\simeq&  \frac{ C_{\o\o\o_p} C^{\o_p}_{\, \o\o} }{( \frac{c}{2} + 4h)^2}
 \left[ \frac{h}{ ( w_j - \hat w_j)^2 } + \frac{1}{( w_j - \hat w_j} \partial_{\hat w_j }  \right]
  \frac{1}{ ( w_j - \hat w_j)^{2h - h_p} }  \times  \frac{1}{ ( w_k - \hat w_k)^{2h - h_p} }  
  \\ \nonumber
  &\simeq&   \frac{ C_{\o\o\o_p} C^{\o_p}_{\, \o\o} }{( \frac{c}{2} + 4h)^2}
    \frac{ ( 3h - h_p) }{  ( w_j - \hat w_j)^{2h - h_p} } 
   \frac{1}{ ( w_k - \hat w_k)^{2h - h_p} } .
\end{eqnarray}
Substituting this approximation in (\ref{leve2f4}) and re-grouping the cross ratio we obtain 
\begin{eqnarray}
&& {\cal F}_4^{(1)}|_{C_{\o\o\o_p} } =  \frac{ C_{\o\o\o_p} C^{\o_p}_{\, \o\o} }{ ( \frac{c}{2} + 4h)^2}
\frac{ ( B_j \hat B_j B_k \hat B_k)^{h+1} }{ ( w_j - \hat w_j)^{2( h+2) } 
( w_k - \hat w_k)^{2( h+2) } } \times w^{h_p} \\ \nonumber
&& \qquad\quad \times ( 3h- h_p)  \frac{1}{12}\Big((5h-\frac{c}{2})A^2 +2(4h+\frac{c}{2})F \Big)
 \left[ \frac{1}{12}((5h-\frac{c}{2})\hat A^2 +2(4h+\frac{c}{2})
        \hat F \right]^2 \\ \nonumber
& &\qquad \quad 
 \times  \frac{ ( w_j - \hat w_j)^2}{\hat B_j^2  } \frac{ ( w_k - \hat w_k)^4}{\hat B_k^4  }.
\end{eqnarray}
Substituting for the $w$'s  using (\ref{locate})  in the first line we obtain for the leading term
\begin{eqnarray}\label{f41coo}
&& {\cal F}_4^{(1)}|_{C_{\o\o\o_p} } =  \frac{ C_{\o\o\o_p} C^{\o_p}_{\, \o\o} }{ ( \frac{c}{2} + 4h)^2}
\Big( \frac{\sin\pi x}{ n \sin\frac{\pi x}{n} }    \Big)^{ 2( h + 2) } 
\times \Big ( \frac{\sin\frac{ \pi x}{n} }{ \sin \frac{\pi}{n} ( j - k ) } \Big)^{2h_p} 
\\ \nonumber
&& \qquad \quad  \times ( 3h- h_p)  \left. \frac{1}{12}\Big((5h-\frac{c}{2})A^2 +2(4h+\frac{c}{2})F \Big)
\left[ \frac{1}{12} (5h-\frac{c}{2})\hat A^2 +2(4h+\frac{c}{2})
        \hat F \right]^2\right|_{x\rightarrow 0}  \\ \nonumber
& &\qquad \quad  \times  \left. \left( \frac{ ( w_j - \hat w_j)^2}{\hat B_j^2  } \frac{ ( w_k - \hat w_k)^4}{\hat B_k^4  }
\right) \right|_{x\rightarrow 0 } + O ( x^{2h_p +1}) .
\end{eqnarray}
The last two lines of (\ref{f41coo}) are independent of $j, k$ and reduce to constants in 
$x\rightarrow 0$ limit. 
Now from the previous analysis  for the level 1 descendent we know that since we have the 
factor $ \big[ \sin\frac{\pi }{n} ( j-k) \big]^{-(2h_p)}$ 
and we need to sum over $j, k$, the result is proportional 
to $(n-1)$. Therefore we can set $n\rightarrow 1$ in the last 2 lines of (\ref{f41coo}) \footnote{
We can set $n=1$ in the factor 
$\Big( \frac{\sin\pi x}{ n \sin\frac{\pi x}{n} }    \Big)^{ 2( h + 2) }  ( \sin\frac{ \pi x}{n} )^{h_p} $
but it is convenient to retain this till the last step.}.
These manipulations lead us to  
\begin{eqnarray}
{\cal F}_4^{(1)}|_{C_{\o\o\o_p} } &=&  \frac{ C_{\o\o\o_p} C^{\o_p}_{\, \o\o} }{ ( \frac{c}{2} + 4h)^2}
\Big( \frac{\sin\pi x}{ n \sin\frac{\pi x}{n} }    \Big)^{ 2( h + 2) } 
\times \Big ( \frac{\sin\frac{ \pi x}{n} }{ \sin \frac{\pi}{n} ( j - k ) } \Big)^{h_p} 
\\ \nonumber
&& \times ( 3h- h_p)  \Big( \frac{ 28 h - c}{12}  \Big)^3.
\end{eqnarray} 

These  simplifications illustrate that  the Ward identities involved in evaluating each of the 
81 correators in (\ref{leve2f4}) factorize. We also note that it is sufficient to evaluate the coefficients 
involved in the conformal transformation by  grouping them as in last 2 lines of 
 (\ref{f41coo}), 
take $x\rightarrow 0$ limit and then take $n\rightarrow 1$ limit. 
These observations help us to evaluate the leading contribution to the 4-point function
without much difficulty. 
This is given by 
\begin{eqnarray} \label{f4om2}
{\cal F}_4 &=&    \Big( \frac{\sin\pi x}{ n \sin\frac{\pi x}{n} }    \Big)^{ 2( h + 2) } 
\left[ \frac{ 2h^2}{c} [D_{\o^{(-2)} }( h, 2)]^2  \Big( \frac{ \sin\frac{\pi x}{n}  }{ \sin\frac{\pi (j-k) }{n}     }
\Big)^4
+ O( x^5, (n-1)^2 )   \right. \\ \nonumber
&=&  \left.  C_{\o\o\o_p} C^{\o_p}_{\, \o\o}[ D_{\o^{(-2)} } ( h, h_p)]^2   \Big( \frac{ \sin\frac{\pi x}{n} }{ \sin \frac{\pi}{n} ( j- k ) }\Big)^{2h_p} + O ( x^{2h_p+1},  ( n-1)^2 ) \right],
\end{eqnarray}
where
\begin{eqnarray} \label{dominus2}
D_{\o^{(-2)}} ( h, h_p) = \frac{ \frac{c}{2} + 4( h + h_p( h_p-1) ) }{ \frac{c}{2} + 4h} , \qquad
D_{\o^{(-2)} }( h, 2)  = 1+ \frac{16}{ c+ 8h}.
\end{eqnarray}
This approximation of the 4-point function is sufficient to obtain the leading short distance corrections
to the 
entanglement entropy  which is given by 
\begin{eqnarray}\label{eeleadsubdl2}
\hat S(\rho_{\o^{(-2)}} ) &=&  2(h+2)  ( 1- \pi x \cot\pi x)  - \frac{8 h^2}{15c}  D_{\o^{(-2)} }( h, 2) ^2
( \sin\pi x)^4  \\ \nonumber
&& - C_{\o\o\o_p} C^{\o_p}_{\; \o\o}  D_{\o^{(-2)}} ( h, h_p)^2
\frac{\Gamma(\frac{3}{2} ) \Gamma( h_p +1) }{ 2\Gamma( h_p + \frac{3}{2} ) } (\sin\pi x)^{2h_p} 
+ \cdots.
\end{eqnarray}
The dressing factor $D_{\o^{(-2)}} ( h, h_p)$ reduces to unity also when $c>>h, h_p$. 

\subsubsection{Level 3}

Using the same techniques we can evaluate the entanglement entropy of the states at level 3. 
We obtain the following results.
\begin{eqnarray}
S( \rho_{\partial^3 O } ) 
&=&  2(h+3)  ( 1- \pi x \cot\pi x)  - \frac{8 (h+3) ^2}{15c}  D_{\partial^3 \o}( h, 2) ^2
( \sin\pi x)^4  \\ \nonumber
&& - C_{\o\o\o_p} C^{\o_p}_{\; \o\o}  [D_{\partial^3 \o } ( h, h_p)]^2
\frac{\Gamma(\frac{3}{2} ) \Gamma( h_p +1) }{ 2\Gamma( h_p + \frac{3}{2} ) } (\sin\pi x)^{2h_p} 
+ \cdots , \\ 
\end{eqnarray}
where 
\begin{eqnarray}
D_{\partial^3 \o} ( h, h_p) =  \frac{3! \Gamma( 2h) }{ \Gamma( 2h - h_p) \Gamma( 2h + 3)  } \sum_{k = 0}^3 
\frac{  \Gamma( 2h - h_p + k ) }{ k!  } \left( 
\frac{ \Gamma ( h_p + 3 - k ) }{ \Gamma( h_p)  ( 3 - k ) ! } \right)^2 .
\end{eqnarray}
This dressing factor reduces to the following as expected for the stress tensor exchange
\begin{eqnarray}
D_{\partial^3 \o} ( h, 2) = \frac{ h+3}{h}.
\end{eqnarray}
Another consistency check is that for $h_p=0$, the dressing factor reduces to unity. 
This is expected since from the evaluation of the dressing factor one can see that 
it is the ratio of a deformed norm, that is for which the conformal transformation acts
as though the state is a descendant of weight $h$ while its two point function of the primary 
is deformed as though it  has weight $h\rightarrow h- h_p/2$. 
We say this clearly for the state $\partial o$ and to some extent this property was illustrated 
for the state $\o^{(-2)}$. 

The  entanglement entropy for the other states at level 3 obey the same expression 
as in (\ref{eeleadsubdl2}),  but with different dressing factors. 
Here are the dressing factors for the remaining states at level 3
\begin{eqnarray}\label{dominus3}
&&D_{\o^{(-3)} } ( h, h_p) = \frac{ c + 4h_p( h_p + 4) + h ( 4h_p+3) }{ c + 3h },  \\ \nonumber
&&D_{\o^{(-3)}} ( h, 2) =  \frac{ c+ 11h + 48}{ c + 3h}, \\ \nonumber
  && D_{\partial\o^{(-2)} } (h,h_p)=\frac{1}{2\big[c (h + 2) + 2 h (17 + 4 h)\big]} \times \\ \nonumber
    &&\Big[16 h^2 + c \big(4 + 2 h + h_p( h_p-1) \big) + 4 h \big(17 + 6h_p ( h_p-1) \big)
   + 
 2 h_p( h_p-1)  \big(19 + 4 h_p (2 + h_p) \big)\Big],\\ \nonumber
 &&  D_{\partial\o^{(-2)} } (h, 2)=\frac{(h+3)(8h+c+34)}{c (h + 2) + 2 h (17 + 4 h)}.
\end{eqnarray}
Note  $D_{\o^{(-3)}} ( h, 0) = 1$ and  we observe as in the case of the 
Virasoro descendent  $\o^{-2}$, 
 the dressing factor reduces to unity when $c>>h, h_p$.

\subsubsection{Global descendants $\partial^l \o$}

From the explicit calculations till level $3$ we saw that the dressing factor $D_{\partial^l \o}( h, h_p)$ 
can be obtained by evaluating the deformed norm. 
That is we think of the $\o$ transforming as a primary of weight $h$, but its two point function 
is given a deformation in which the weight is shifted to $h\rightarrow h - h_p/2$. 
To evaluate this deformed norm, we can consider the $n=1$ limit of the transformations
(\ref{u1}) and (\ref{u2}) and choose $u =0, v=1$ for convenience. 
This results in the following $Sl(2, C)$ transformations
\begin{equation}
s(z) = \frac{ z }{z-1} , \qquad    \hat s(\hat z )  = \frac{ 1  }{  1+ \hat z } .
\end{equation}
We then evaluate the `deformed' two point function  to obtain the norm. 
\begin{eqnarray}
D_{\partial^l \o} ( h, h_p) &=& \frac{ (-1)^l \Gamma( 2h) }{ l! \Gamma( 2h + l ) }  \times \, 
\partial_z^l \partial^l_{\hat z} H( z, \hat z) |_{( z, \hat z) = (z, 0 ) } , 
\\ \nonumber 
H( z, \hat z) &=&  ( \partial_z s(z) \partial_{\hat z}  s( \hat z) )^h \Big( \frac{1}{ s(z) - \hat s (\hat z) } \Big)^{2h - h_p}.
\end{eqnarray}
Here we have implemented the fact that the primary transforms with the weight $h$, while 
the correlator is deformed as though it is obtained as a two point function of an operator 
of weight $h\rightarrow h - {h_p}/2$.  Also it is easy to see by construction that 
$D_{\partial^l \o} ( h, 0 )  =0$ since we have taken the ratio  of the deformed norm by the usual norm
of the state $\partial^l \o$. 
Simplifying $H(z, \hat z) $ we obtain 
\begin{equation}
H( z, \hat z) = \frac{1}{ ( 1+ z \hat z )^{2h - h_p}   ( 1-z)^{h_p}  ( 1+ \hat z)^{h_p} } .
\end{equation}
Now we can evaluate the $l$ derivatives with respective to both $z$ and $\hat z$ and then 
set $(z, \hat z) = ( 0, 0)$. This results in  the following  general expression for the dressing factor for 
global descendants. 
\begin{eqnarray} \label{gendressf}
D_{\partial^l \o} ( h, h_p) &=& 
\frac{l!\,  \Gamma (2h) }{ \Gamma( 2h - h_p) \Gamma( 2h + l )  } \sum_{k = 0}^l
\frac{  \Gamma( 2h - h_p + k ) }{ k!  } \left[ 
\frac{ \Gamma ( h_p + l - k ) }{ (l-k) ! \Gamma( h_p)  } \right]^2 .
\end{eqnarray}
We can show that as expected 
\begin{eqnarray}
D_{\partial^l \o} ( h, 2) = \frac{h +l }{h}, \qquad D_{\partial^l \o} ( h, 0 ) = 1.
\end{eqnarray}

\subsection{Relative entropy between descendants}

In this section we apply the results of  the previous section to evaluate the relative entropy 
between two states not necessary only primaries.  Consider $\o^{[-l]}$ and $\o^{\prime\, [-l'] }$ be
descendants of primaries of weight $h, h'$ and levels $l, l'$ respectively.  Recall the superscript 
refers $[l]$ refers to the level of the descendant, it can include both Virasoro or global descendants.
To obtain the relative entropy, 
from (\ref{relentropy1}), we see that we would need to evaluate the following correlator
in the short distance expansion
\begin{eqnarray} \label{hatCdd}
\widehat{\cal C}_{2n} =\Big\langle  w\circ \o^{[-l]} ( w_0)  \hat w \circ \o ^{*\, [-l]}  (\hat w_0) 
\prod_{k =1}^{n-1} \Big[ w\circ \o{\prime\, [-l']} ( w_k ) \hat w\circ \o^{\prime *\, [-l'] }( w_k)  
 \Big]  \Big\rangle.
\end{eqnarray}
Using the results of the previous sections the leading corrections to this correlator in the short
interval arises from the following terms 
\begin{eqnarray} \label{hatCd}
&& \widehat{\cal C}_{2n} \simeq    \left(  \frac{\sin\pi x}{ n \sin\frac{\pi x}{n} }\right)^
{2 ( h + l + ( n-1) ( h' +l') ) } + O( (n-1)^2 ) 
\\ \nonumber
& &+ \left(  \frac{\sin\pi x}{ n \sin\frac{\pi x}{n}} \right)^{2 ( h +l  + ( n-3) ( h' +l') ) }
  \sum_{j, k=1, j\neq k }^{n-1}
 \Big\langle w\circ \o^{\prime\, [-l']}  ( w_j )     \hat w \circ \o^{\prime *\, [-l'] }( \hat w_{j})  w\circ \o^{\prime \, [-l'] }( w_k )     
 \hat w \circ \o^{\prime * \, [-l'] }( \hat w_{k}) 
 \Big\rangle_c  \\ \nonumber
 & &   \left(  \frac{\sin\pi x}{ n \sin\frac{\pi x}{n} }\right)^{2  ( n-2) (h' +l')  }
 \sum_{j =1  }^{n-1}
 \Big\langle w\circ \o^{[-l]}  ( w_0 )     \hat w \circ \o^{*\, [-l]}
 ( \hat w_{0})  w\circ \o{\prime \, [-l']}  ( w_j )     \hat w \circ \o^{\prime * \, [-l'] }
 ( \hat w_{j}) 
 \Big\rangle_c + \cdots.
\end{eqnarray}
Where the first line contains the leading contribution which arises from the factorisation 
of the correaltor to $2n$ operators on the $n$ wedges of the uniformized plane. 
Here we have used the result in section  which showed that the result to order 
$O(n-1)$ is equivalent to that of two point functions of operators $h+l$ and $h'+l'$. 
The second and third line contains the terms which contribute at the sub-leading order 
similar to that in equation (\ref{lowestcp4}) for the case of the primaries. 

We can now use the methods developed in section to conclude that the  leading 
corrections to the four point functions of interest are given by 
\begin{eqnarray} \label{lowestcp4d}
 &&  \Big\langle w\circ \o^{\prime \, [-l]} ( w_j )     \hat w \circ \o^{\prime *\, [-l]}( \hat w_{j})  w\circ 
 \o^{\prime \, \, [-l]}( w_k )     \hat w \circ \o^{\prime *\, [-l] } ( \hat w_{k}) 
\Big\rangle_c  = \\ \nonumber
 % \frac{ ( B_j \hat B_j B_k \hat B_k )^{h' } }{  (w_j - \hat w_j)^{2h'} ( w_k - \hat w_k)^{2h'} } \\ 
 && \qquad  \Big( \frac{\sin \pi x}{ n \sin \frac{\pi x}{n} } \Big)^{ 4 ( h' + l ') }  \times  
 %\frac{1}{ (w_j - \hat w_j)^{2h'} ( w_k - \hat w_k)^{2h'} }
 \left[ \frac{ 2 h^{\prime \, 2}  }{c}  [D_{\o^{\prime [-l']}}( h, 2) ]^2 
  \Big( \frac{\sin \frac{\pi x}{n}}{ \sin \frac{\pi}{n} ( j- k ) } \Big) ^4 \right. \\  
\nonumber
&&  \qquad\qquad\qquad\qquad \left. 
 + C_{\o'\o'\o_p} C^{\o_p}_{\; \o'\o'}    [D_{\o^{\prime [-l'] }}( h, h_p) ]^2 
   \Big( \frac{\sin \frac{\pi x}{n}}{ \sin \frac{\pi}{n} ( j- k ) } \Big)^{2h_p} 
 \right] + \cdots .
\end{eqnarray}
The analysis leading upto this equation is identical to  that of  say of (\ref{f4om2}) with the 
same approximations involved. 
Similarly the leading contributions from the 
 four point function involving a pair of $\o^{ [-l]} $ and a pair of $\o^{\prime\, [-l']}$  is given by 
\begin{eqnarray}\label{lowetcp4d} 
 &&  \langle w\circ \o^{-[l]}  ( w_0 )     \hat w \circ \o^{*\, [-l] }( \hat w_{0})  w\circ \o^{\prime \, [-l]} 
 ( w_j )     \hat w \circ \o^{\prime *\, [-l] } ( \hat w_{j}) 
 \rangle_c  =  \\ \nonumber
 %\frac{  ( B_0 \hat B_0 )^h (B_j \hat B_j )^{h' } }{ (w_0 - \hat w_0)^{2h} ( w_j - \hat w_j)^{2h'} } \\ 
 && \qquad \Big( \frac{ \sin\pi x}{ n \sin\frac{\pi x}{n} } \Big)^{2 ( h + h' + l +l ') } \times  
 %\frac{1}{ (w_0 - \hat w_0)^{2h} ( w_j - \hat w_j)^{2h'} }
 \left[ \frac{ 2 h h' }{c}  D_{\o^{ [-l]}}( h, 2)  D_{\o^{\prime [-l']}}( h, 2)  
  \Big( \frac{\sin \frac{\pi x}{n}}{ \sin \frac{\pi j}{n}  } \Big) ^4 \right. \\ \nonumber
  & &\qquad\qquad\qquad  \left. 
 + C_{\o\o \o_p} C^{\o_p}_{\; \o'\o'}     D_{\o^{ [-l]}}( h, h_p)  D_{\o^{\prime [-l']}}( h, h_p)  
 \Big( \frac{\sin \frac{\pi x}{n}}{ \sin \frac{\pi j}{n} } \Big)^{2 h_p} 
 \right] + \cdots .
\end{eqnarray}
A simple way to see that the above result is obtained is to realise that  the dressing factors in 
 the four point function  of descendants in the conformal block approximation to $O((n-1))$
to the leading order in the short interval approximation  can be evaluated by thinking 
of them as a generalized norm  as we have discussed earlier. 
The generalized norm  is one that we need to compute for the operators 
at $(0, \hat 0)$ times that for the operators at $(j, \hat j)$. 
This  leads to the expression in (\ref{lowetcp4d}).  We have also evaluated these correaltors 
explicitly for various descendants till level $3$ and seen that the result 
agrees with $(\ref{lowetcp4d})$. 
We have 
assumed that the  lowest weight of the  primary that appears in the OPE of 
$\o$ as well as $\o'$ is  the same operator $\o_p$ with weight $ h_p$. 

Now we substitute the expressions for the four point functions  (\ref{lowestcp4d}) and (\ref{lowetcp4d})
 into (\ref{hatCd})  to obtain  and use the same steps as in equations  (\ref{hatC1})  to (\ref{relenf}). 
 This leads to the following result for the 
 relative entropy of descendants 
  \begin{eqnarray}\label{relenfd}
 &&S( \rho_{\o^{[-l] }} | \rho_{\o^{\prime [-l']  }}) = \frac{ 8 }{15 c} 
 \Big[  hD_{\o^{ [-l]}}( h, 2)  - h' D_{\o^{\prime [-l']}}( h, 2)  \Big]^2 ( \sin\pi x)^4  \\ \nonumber
& &  + \Big|\Big| ( C_{\o\o\o_p} D_{\o^{ [-l]}}( h, h_p)   - C_{\o'\o'\o_p}
D_{\o^{\prime [-l'] } }( h, h_p)    \Big|\Big|^2
%( C^{\o_p}_{\, \o\o} D_{\o^{ [-l] } }( h, h_p)  - C^{\o_p}_{\, \o'\o'} 
%D_{\o^{\prime [-l'] } }( h, h_p)   )
 \frac{\Gamma(\frac{3}{2} ) \Gamma( h_p +1) }{ 2\Gamma( h_p + \frac{3}{2} ) } (  \sin\pi x)^{2 h_p} 
+ \cdots.
 \end{eqnarray}
 Here $|| \cdot ||^2$ refers to the fact that we need to take the square of its argument together 
 with the index $\o_p$ in one of the term raised by the Zamolodchikov metric. 
It is useful to write down the relative entropy of descendants of the same primary
  \begin{eqnarray}\label{relenfd1}
&& S( \rho_{\o^{[-l] }} | \rho_{\o^{[-l'] }}) = \frac{ 8 h^2  }{15 c} 
 \Big[ D_{\o^{ [-l]}}( h, 2)  - D_{\o^{[-l']}}( h, 2)  \Big]^2 ( \sin\pi x)^4  \\ \nonumber
& &  +C_{\o\o\o_p}  C^{\o_p}_{\, \o\o}
 \Big[  D_{\o^{ [-l]}}( h, h_p)   - 
D_{\o^{\prime [-l']}}( h, h_p)    \Big]^2 \frac{ \Gamma(\frac{3}{2} ) 
\Gamma( h_p +1) }{ 2\Gamma( h_p + \frac{3}{2} ) } (  \sin\pi x)^{2 h_p} 
+ \cdots.
 \end{eqnarray}
 Here $D_{\o^{[-l]}},  D_{\o^{[-l']}}$  refer to the dressing factors between descendants of  the primary $\o$. 
 The  level can in fact be same but the descendant distinct  for example $\partial^2\o$ and $\o^{(-2)}$. 
 Thus  it is only the dressing factor which determines the distance between two descendants 
 in this situation.

\section{Applications} \label{section4}

In this section we apply the results of  entanglement entropy of descendants to generalised 
free fields and  descendants of the vacuum. 
The reason we focus on these applications is that they can in principle be verified using 
holography. 
Generalised  free fields are dual to minimally coupled scalars in holographic conformal field theories.
Entanglement entropy of descendants of the vacuum are universal to all conformal field theories and in 
particular for holographic conformal field theories. 
Generalising the methods developed in \cite{Belin:2018juv,Belin:2019mlt,Belin:2021htw}, it should possible to verify the results of this section.

\subsection{Generalised free fields} \label{section4.1}

Generalised free fields are  operators whose correlation functions can  be evaluated by 
Wick contractions
\begin{eqnarray}\label{genffwick}
\langle \prod_{i =0}^{n-1} \o ( w_i) \o^* ( \hat w_i ) \rangle &=& 
\prod_{i= 0}^{n-1} \langle \o(w_i) \o^*(\hat w_i) \rangle + {\hbox{ all distinct permutations}}.
\end{eqnarray}
For example the four point function of generalised free fields  of 
weight $h$ is given by 
\begin{eqnarray} \label{genff4pt}
\langle \o( w_j) \o^*(\hat w_j) \o(w_k) \o^*( \hat w_k) \rangle &=& 
\frac{1}{ ( w_j - \hat w_j )^{2h} ( w_k - \hat w_k)^{2h} } 
+ \frac{1}{ ( w_j - w_k)^{2h} ( \hat w_j - \hat w_k)^{2h} } \nonumber  \\
& &+ \frac{1}{ ( w_ i -\hat w _k)^{2h} ( \hat w_i - w_k )^{2h} } .
\end{eqnarray}
Such a behaviour of correlators is expected in large $c$ holographic CFT's. 
Indeed minimally coupled scalar scalars in $AdS_3$ are to dual generalised free fields 
\cite{Belin:2018juv}. 
As shown recently, these 
correlation functions may get corrected in order $O(1/c)$ \cite{Belin:2021htw}. 
however for our purposes 
we will restrict our attention to the leading order in $c$. 
Writing the connected part of the four point function in  (\ref{genff4pt}) 
in terms of the cross ratio  $w$ defined in (\ref{crossratio})  as follows
\begin{eqnarray} \label{conngen4pt}
\langle \o( w_j) \o^*(\hat w_j) \o(w_k) \o^*( \hat w_k) \rangle _c
&=& \frac{1}{ ( w_j - \hat w_j )^{2h} ( w_k - \hat w_k)^{2h} }  (   w^{2h}  + \frac{w^{2h}}{ ( 1- w)^{2h} }) , 
\nonumber \\
&=& \frac{1}{ ( w_j - \hat w_j )^{2h} ( w_k - \hat w_k)^{2h} } \Big(    2 w^{2h} + \cdots \Big).
\end{eqnarray}
From the second line of the above equation, we can read out the lowest lying primary that 
appears in the OPE has dimension $h_p = 2h$ with structure constant $ C_{\o\o\o_p} C^{\o\o\o_p} = 2$. 
This primary corresponds to the composite  $[\o\o]$ which is a usually referred to as
a  double trace operator in holography. 
With this information of the four point function, it is easy to write down the  entanglement entropy 
of a generalised free field in the short distance approximation.   Using (\ref{eeleadsub}) we obtain
\begin{eqnarray} \label{eegenfie}
\hat S( \rho_{\o} ) = 2h ( 1- \pi x \cot\pi x)  -  
( \sin\pi x )^{4h}  \frac{\Gamma ( \frac{3}{2} ) \Gamma( 2h + 1) }{ \Gamma ( 2h + \frac{3}{2} )}. 
\end{eqnarray}
There is no contribution  from the stress tensor exchange term which is proportional to $1/c$, 
this is expected since the Wick contraction rules for generalized free fields in (\ref{genffwick}) 
is true only to the leading order in $c$.

Let us  use the result for the dressing factor for the descendants  $\partial^l \o$ on generalised free fields. 
Since $h_p  = 2h$,   from (\ref{gendressf}) we see it is 
only the $k=0$ term in the sum contributes. This results in 
the following dressing factor for the descendants of generalised free fields. 
\begin{eqnarray}
D_{\partial^l \o} ( h, h_p) &=&  \frac{\Gamma( 2h + l ) }{ l ! \Gamma (2h) }
= \frac{{\cal N}_l }{ ( l!)^2}, \qquad  \qquad {\cal N}_l =   \frac{ l! \Gamma( 2h + l ) }{  \Gamma (2h) }.
\end{eqnarray}
Here ${\cal N}_l$ is the norm of a descendant at level $l$. 
The dressing factor just depends on the conformal dimension and the level and it is related the 
norm in an interesting way. 
Thus the entanglement entropy of the global descendants of  generalised free fields is given by 
\begin{eqnarray} \label{eegenfie1}
\hat S( \rho_{\partial^l \o} ) = 2(h+l)  ( 1- \pi x \cot\pi x)  -  
( \sin\pi x )^{4h}  \frac{\Gamma ( \frac{3}{2} ) \Gamma( 2h + 1) }{ \Gamma ( 2h + \frac{3}{2} )} 
\times \left(   \frac{\Gamma( 2h + l ) }{ l ! \Gamma (2h) } \right)^2 .
\end{eqnarray}
As a cross check of the above result, we have directly used the four point function of generalized free 
fields 
 in (\ref{genff4pt})  and then evaluated the corresponding four point function for the global descendents. 
This just involves the knowledge of the two point function of the descendants since all generalized 
free field correlators are obtained from two point function. Evaluating the four point function 
of global descendants in this manner for the sub-leading corrections to the entanglement entropy
also results in (\ref{eegenfie1}). 

Global descendants are obtained from primaries by the action of $L_{-1}$  which corresponds to 
an isometry 
in $AdS_3$. Therefore, this dressing factor can in principle be seen in holography using 
the methods of \cite{Belin:2018juv}. 
It will be interesting to perform this check.   In \cite{Belin:2018juv}  to obtain  the entanglement 
entropy of generalised fields 
(\ref{eegenfie}) from the bulk, 
both the Ryu-Takayanagi minimal surface as well as the bulk entanglement entropy 
across this surface contributed to the the sub-leading term.
Thus deriving (\ref{eegenfie1}) from the bulk will teach us 
how the action of symmetry  $L_{-1}$ will 
affect the  Ryu-Takayanagi minimal surface as well as the bulk entanglement 
entropy across this surface.   

There is one more observation regarding dressing factor of descendants of generalised 
free fields worth pointing out. 
 In (\ref{dominus2}) and (\ref{dominus3})  we evaluated the dressing factors of  the Virasoro descendants 
$\o^{(-2)}, \o^{(-3)}$ . All these go to unity as $c\rightarrow \infty$. 
Then from the expression of relative entropy in (\ref{relenfd1}) we see  that 
 for generalised free fields,   relative entropy among  Virasoro descendants
as well as primaries vanish at the leading order.

\subsection{Descendants of the vacuum}

Evaluating entanglement entropy of descendants of vacuum is interesting. 
Since stress tensor correlators depend only the central charge the results are universal 
and depend only the central charge $c$ of the conformal field theory. 
Consider the state $L_{-2}|0\rangle$ which corresponds to the operator $T(z)$, the stress tensor. 
To evaluate the leading contributions to the short interval entanglement entropy of this state, 
we first evaluate the two point function of the stress tensor on a wedge of the uniformized
surface. 
\begin{eqnarray} \label{stress2pt}
\langle w\circ T(w_j) \, \hat w T( \hat w_j)  \rangle 
= \frac{2}{c}  \Big\langle
(  B_j ^2 T( w_j ) +  \frac{c}{12} S( w_j, z) |_{z\rightarrow 0} ) 
(  \hat B_j^2   T( w_j ) +   \frac{c}{12} S( \hat w_j, \hat z) |_{\hat z\rightarrow 0} ) 
\Big\rangle.  \nonumber \\
\end{eqnarray}
where the Schwarzian is given by 
\begin{equation}
S( w, z) =   \frac{w''' }{w'} - \frac{3}{2} \Big( \frac{ w''}{w'}\Big)^2 .
\end{equation}
In (\ref{stress2pt}) we have used the fact that the stress tensor is a quasi-primary and transforms as 
\begin{eqnarray} \label{cftstress}
\langle w\circ T( w (z) ) = ( w')^2 T( w(z) )  + S(w, z)  .
\end{eqnarray}
as well as the fact that its  expectation value  on the plane vanishes by translational 
invariance. 
The Schwarzian evaluated for the uniformization map is independent of the position $j$ is given by 
\begin{eqnarray}
\left. S( w_j, z) \right|_{z\rightarrow 0}  =  2 e^{- 2\pi i x} \frac{n^2 -1}{n} (  \sin\pi x)^2 ,
\qquad 
\left. S( \hat w_j, \hat z) \right|_{z\rightarrow 0}  =  2 e^{2\pi i x} \frac{n^2 -1}{n} (  \sin\pi x)^2 .
\nonumber \\
\end{eqnarray}
As expected the Schwarzian is proportional to $(n-1)$. 
Using this result for the Schwarzian  we obtain  we obtain 
\begin{eqnarray} \label{2ptstressu}
\langle w\circ T(w_j) \hat w \circ T( \hat w_j)  \rangle  &=& \Big( \frac{ \sin\pi x }{ n\sin\frac{\pi x}{n} }\Big)^4
+ \frac{c ( n^2 -1)^2 }{18 n^4 }  ( \sin\pi x)^4 , \\ \nonumber
&=&   \Big( \frac{ \sin\pi x }{ n\sin\frac{\pi x}{n} }\Big)^4 + O( (n-1)^2 ) .
\end{eqnarray} 
Thus the two point function of the stress tensor on the wedge of the uniformization plane 
is that of a primary field of weight $2$ to $O(n-1)$. This is the similar property satisfied by the 
conformal descendants seen in section \ref{dessubl}

Now let us examine the 4-point function of the stress tensor on the 
uniformized plane in which  a pair of operators are  inserted on 
two   distinct wedges.  This correlator is given by 
\begin{equation}\label{4ptstressu}
{\cal F}_4 = \frac{4}{c^2} \Big\langle w\circ T( w_j) \hat w T(\hat w_j) 
w\circ T( w_k) \hat w T(\hat w_j) \Big \rangle_c,
\end{equation}
where we have  divided by the norm of the stress tensor. 
The conformal transformation of the stress tensor  (\ref{cftstress}) involves the Schwarzian.
However since the Schwarzian is proportional to $(n-1)$ and we need  the correlator
in (\ref{4ptstressu}) to $O(n-1)$,  the contributions arise only from terms only to at most linear order in the 
Schwarzian. 
\begin{eqnarray}
{\cal F}_4 &=& \frac{4}{c^2} (B_{j} \hat B_j B_k \hat B_k )^2 \langle T( w_j) T( \hat w_j) T( w_k) T( \hat w_k ) 
\rangle_c  \\ \nonumber
&& +\frac{1}{3 c} S( w_j, z)|_{z\rightarrow 0}  \hat B_j B_k \hat B_k 
\langle  T( \hat w_j) T( w_k) T( \hat w_k ) +  ( 3\;\;  \hbox{permutations})  \\ \nonumber
&& + O( (n-1)^2 ) .
\end{eqnarray}
The second term arises from the single insertion of the Schwarzian. 
Substituting for the $4$ point and $3$ point function of the stress tensors we obtain
\begin{eqnarray}
{\cal F}_4 &=&   \Big( \frac{\sin \pi x}{ n\sin \frac{\pi x}{n} } \Big)^4 
\left[   w^4 + \frac{ w^4}{  ( 1- w)^4}    + \frac{8}{c}  w^2 \frac{ 1- w - w^2}{ ( 1- w)^2 }
\right]  \\ \nonumber
&& -\frac{2 ( n^2 - 1)}{ 3n^8} 
\frac{ \Big(  2 - \cos \frac{ 2( j - k + x) \pi }{n} -  \cos \frac{ 2( j - k 2 x)\pi }{n} \Big)}{ \Big( 
\sin \frac{(j -k)\pi}{n}  \sin  \frac{(j -k +x )\pi}{n}   \sin  \frac{(j -k xx )\pi}{n}  \Big)^2} \times
\frac{ (\sin \pi x)^8 }{ ( \sin \frac{\pi x }{n} )^2} \\ \nonumber
& & + O( (n-1)^2 ) 
\end{eqnarray}
where $w$ is the cross ratio (\ref{crossratio}).  We have removed the $1$  in the four point function 
of the stress tensor since this term is already accounted for in the pairwise contraction in each wedge. 
Though the second term is manifestly proportional to $n-1$, the sum over the wedge which is 
necessary to evaluate the entanglement entropy renders it $O(n-1)^2$. 
To see this expand the second term in $x$ 
\begin{eqnarray}
{\cal F}_4  &=&   \Big( \frac{\sin \pi x}{ n\sin \frac{\pi x}{n} } \Big)^4 
\Big[  w^4 + \frac{ w^4}{  ( 1- w)^4}    + \frac{8}{c}  w^2 \frac{ 1- w - w^2}{ ( 1- w)^2 } \Big] \\ \nonumber
&&  - \frac{8}{3} ( n^2 -1)  \frac{ (\sin \pi x )^8}{ (  \sin \frac{\pi x}{n} )^2 (  \sin \frac {( j - k) \pi }{n})^4} \left[ 1+ 
\left( \frac{3}{ (\sin \frac {( j - k) \pi }{n})^2    }- 2 \right) \frac{\pi^2 x^2}{n^2}   + O(x^4)  \right]
\\ \nonumber
&& + O( (n-1)^2) .
\end{eqnarray} 
From (\ref{cardysum}), 
we see that the sum over  the wedges $j, k$ will result in another factor of $(n-1)$ and 
therefore the contribution  of the second term can also be neglected for the purposes of evaluating the entanglement entropy. 
In the first line there are terms to the order $c^{0}$,  on comparision with (\ref{conngen4pt}) we see that 
this part of the correlator behaves as a generalized free field, while the terms at order $c^{-1}$
start at $w^2$ as expected for the stress tensor exchange. 
Now keeping the leading terms in the the short interval expansion and substituting for the cross ratio
we obtain
\begin{eqnarray} \label{4ptstressu1}
 {\cal F}_4  &=&  \Big( \frac{\sin \pi x}{ n\sin \frac{\pi x}{n} } \Big)^4  \left[  \frac{8}{c} 
 \Big( \frac{\sin\frac{\pi x}{n} }{ \sin \frac{ (j - k )\pi }{n} }  \Big)^4  + O ( x^5)  \right. 
 \\ \nonumber
 &&\qquad\qquad\qquad  \left. 
 + 2   \Big( \frac{\sin\frac{\pi x}{n}}{ \sin \frac{ (j - k )\pi }{n}}  \Big)^8    + O( x^9 ) \right]
 + O (  (n-1)^2 ) .
\end{eqnarray}
Thus the first term arises due to the stress tensor exchange and and the second term 
is similar to that of a generalized free field  with $C_{TT\o_p} C^{\o_p}_{\, TT} = 2$ and 
$h_p = 2h = 4$.  The composite operator involved in this exchange is the  bilinear of the stress tensor,
$\o_p = [TT]$.
Using the two point function on the wedge in (\ref{2ptstressu}) and the 
four point function (\ref{4ptstressu1}) we  use (\ref{2nlevel1d}) to obtain the leading 
contributions to the single interval  entanglement entropy of the 
state $L_{-2} |0\rangle$ 
\begin{equation} \label{eegrav1}
\hat S( \rho_{L_{-2}} ) =  4
( 1- \pi x \cot\pi x)  -  \frac{ 32}{ 15c}  \sin^4 \pi x - \frac{128}{315} \sin^8 \pi x
+ \cdots.
\end{equation}
In large $c$ holographic theories we expect to see only the 3rd term as the sub-leading 
correction. 
It will be interesting to verify this using the methods of \cite{Belin:2018juv}. 
This will involved the study of entanglement entropy of  gravitons which has largely been 
unexplored. 

Our derivation shows that for the purposes of  entanglement entropy  the stress tensor 
behaves as a primary of weight $h$ and the lightest state  $h_p=4$ corresponding to the composite
$\o_p = [TT]$ with 
$C_{TT\o_p} C^{\o_p}_{\, TT} = 2$. 
Therefore from (\ref{gendressf}) we can evaluate the dressing factors for the descendant 
$L_{-( l+2) } |0\rangle$ which corresponds to the operator $\partial^l T$. 
\begin{eqnarray}
D_{\partial^l T} ( 2, 4 ) =  \frac{ ( l +3) ( l + 2) ( l + 1) }{ 3!}, \qquad\qquad 
D_{\partial^l T} ( 2, 2 ) = \frac{ l + 2}{ 2} .
\end{eqnarray}
Using these expressions for the dressing factors we obtain the leading entanglement entropy 
of the descendants of the vacuum  $L_{-( l+2) } |0\rangle$
\begin{eqnarray} \label{eegrav2}
\hat S( \rho_{L_{-(l+2)}|0\rangle } ) &=&  2( l+2)  ( 1- \pi x \cot\pi x)  - \frac{ 8 ( l + 2)^2 }{ 15c} \sin^4 \pi x
\\ \nonumber
&& 
 - \Big(  \frac{ ( l +3) ( l + 2) ( l + 1) }{ 3!} \Big)^2 \frac{128}{315}  \sin^8 \pi x
+ \cdots.
\end{eqnarray}
Again for holographic conformal field theories it is only the 3rd term that contributes at the 
sub-leading order.

Since the dual to the stress tensor is the graviton, 
it should be possible to verify equations (\ref{eegrav1}), (\ref{eegrav2}) by generalizing the methods 
of \cite{Belin:2018juv, Belin:2019mlt} 
 to the graviton in the bulk.   This is interesting since it will give us some handle
on defining entanglement entropy for gravitons.

\section{Conclusions}\label{section5}

In this paper we have studied the single interval entanglement entropy of descendants in 
2 dimensional  conformal field theories. 
We have found simplifications in the evaluation of  contributions to the short interval 
expansion of entanglement entropy and relative entropies of descendants. 
One of the observations in the paper is that the leading contribution to the short 
interval expansion of a descendant at level $l$ is identical to that of a primary with 
weight $h+l$. This came about since the two point function of the descendant on 
the same wedge in the uniformized plane behaved as that of a primary of weight $h+l$ to 
$O((n-1)$ and deviated only at $O((n-1)^2)$ where $n$ is the replica index. 
It will be interesting to establish this property to all descendants as well as provide an
algorithm to evaluate the dressing factor to all descendants using the methods 
of \cite{Gaberdiel:1994fs}. 
We have seen that  the dressing factor can be thought of a ratio of generalised norm
to the usual norm of the descendant. The operator methods developed in \cite{Gaberdiel:1994fs} to  efficiently 
obtain conformal transformation of descendants would be useful to  obtain this deformed norm.

The general  question that motivated this study is how entanglement or relative entropy is
affected by the action of symmetries of a theory and how can this effect be seen in holography. 
When the conformal field theory  is deformed by a symmetry,  this question has been studied earlier in 
\cite{Datta:2014ska,Datta:2014uxa,Datta:2014zpa,Chowdhury:2019lfe}.
 Indeed when the theory is deformed by introducing higher spin fields, the  leading 
corrections to entanglement entropy proved to be universal  \cite{Datta:2014uxa} and provided a test 
for the Wilson line proposal of \cite{deBoer:2013gz,Ammon:2013hba}. 
In this paper we studied the action of the Virasoro symmetry on the entanglement of low lying 
states of the conformal field theory. For global descendants of 
generalised free fields and descendants of 
the vacuum the modification to the leading contributions to the single interval entanglement 
entropy are universal and are given by \ref{eegenfie1}, \ref{eegrav2}. 
It should be possible to verify these results by generalizing the methods of \cite{Belin:2018juv,Belin:2019mlt,Belin:2021htw} in the bulk. 
This would show how the Ryu-Takanayagi surface as well as bulk entanglement entropy 
across this surface are modified by the action of the isometries. It will also help us 
understand how to evaluate entanglement entropies of gravitons. 

\acknowledgments

We wish to thank Shouvik Datta for discussions during the initial phase of this project.

\appendix
\section{$2n$-point function of $\partial e^{il X}$ on the uniformised plane. } \label{appen1}

We are interested in the correlator  (\ref{expcor}) given by 
\begin{eqnarray}\label{expcor2}
{\cal C}_{2n} &=& \langle \prod_{k =0}^{n-1}  w\circ ( \partial e^{il X ( w_k) } )\,  \hat w \circ (\partial e^{ - i l X( w_k) )} 
\rangle , \\ \nonumber
&=& 
 \left. \prod_{k =0}^{n-1} ( B_k \hat B_k )^{\frac{l^2}{2} }  \left[   \frac{l^2}{2} A + B_k \partial_{u_k}  \right]
\left[  \frac{l^2}{2} \hat A + \hat B_k \partial_{\hat u_k}  \right] f( u, \hat u ) \right|_{ (u, \hat u) = (w , \hat w )}.
\end{eqnarray} 
where $f(u, \hat u)$ is defined in (\ref{taccor}), $B_k, \hat B_k$ defined in  (\ref{slope}), and 
 $A,  \hat A $ are defined in  (\ref{defA}) and (\ref{defhatA}). 
To un-clutter the expressions let us define the derivatives 
\begin{equation}
D_k = B_k \partial_{u_k}, \qquad \qquad \hat D_k = \hat B_k \partial_{\hat u_k}.
\end{equation}
Note that here is no summation over $k$. 
Opening out the expression  for the correlator in (\ref{expcor2}), we obtain 
\begin{eqnarray} \label{expcor1}
{\cal C}_{2n} &=&
\left( \prod_{k =0}^{n-1} ( B_k \hat B_k)^{\frac{l^2}{2}}  \right)  \\ \nonumber
  & & \times  \sum_{p=0}^{2n} \sum_{\stackrel{ i =0, i \leq n}{ 0\leq  p- i  \leq n }}^p  ( \frac{l^2}{2} )^{2n- p}
   A^{n-i}  \hat A^{ n-p+i} 
  \sum_{ \stackrel{ \{ k_1, \cdots k_i\} }  {  \{\hat k_1, \cdots \hat k_{p- i }\} }}
  [ D_{k_1} \cdots D_{k_i} ] [ \hat  D_{\hat k_1} \cdots \hat D_{\hat k_{p-i} } ] f(u, \hat u ).
\end{eqnarray}
The sum over $\{ k_1, \cdots k_i \}$ runs over $i$ integers in (\ref{expcor1}). None of these integers are equal and 
  they all take values from $0 $ to $n-1$. Similarly for the the set $\{  \hat k_1, \cdots \hat k_{p-i} \}$
   Also note that we can have maximum $2n$ derivatives. However, the  intermediate 
  $i$ cannot exceed $n$, and $(p-i)$ cannot exceed $n$.
  
  We now need to evaluate the derivatives acting on the function $f(u, \hat u)$. 
  To write down a general formula, it is first convenient to defined the notion of 
  `contraction'  between the  functions $g, h$ which arise due to action  the derivatives of the function $f(u, \hat u )$ 
  as seen in (\ref{definigh}) 
  \begin{eqnarray}
  g_k(u, \hat u )  &=& l^2 \left[ \sum_{j =0 j\neq k }^{n-1}   \frac{1}{ u_k - u_j } - \sum_{j =0 }^{n-1} \frac{1}{ u_k - \hat u_j} 
\right],  \\ \nonumber
h_k(u, \hat u ) &=& l^2  \left[ \sum_{j =0 j\neq k }^{n-1}   \frac{1}{ \hat u_k - \hat u_j } - 
\sum_{j =0 }^{n-1} \frac{1}{ \hat u_k -  u_j} 
\right].
  \end{eqnarray}
  Consider the product of  $g_{k_1} g_{k_2}$, we suppress 
  the dependence on $(u, \hat u)$ to un-clutter the expression. 
  Let us define the `contraction' as follows 
  \begin{eqnarray}\label{contract1}
\wick{\c{g}_{k_1} \c{g}_{k_2}    } = \frac{l^2  }{ ( u_{k_1} - u_{k_2} )^2}  .
 \end{eqnarray}
 Here $k_1\neq k_2$. 
 Using this definition we see that 
 \begin{eqnarray}
  \partial_{u_{k_2}}  g_{k_1} =  \partial_{u_{k_1}} g_{k_2} &=& 
   \frac{l^2  }{ ( \hat u_{k_1} - \hat u_{k_2} )^2} 
  =   \wick{ \c{g}_{k_1 } \c{g}_{k_2}  } , \qquad k_1 \neq k_2 .
 \end{eqnarray}
Similarly let us define  the contraction between two $h$'s as 
 \begin{equation} \label{contract2}
  \wick{ \c{h}_{k_1}  \c{h}_{k_2}  }  = \frac{l^2  }{  ( \hat u _{k_1} - \hat u_{k_2} )^2 } . 
  \end{equation}
 We see that 
 \begin{eqnarray}
  \partial_{\hat u_{k_2}}  h_{k_1} =  \partial_{\hat u_{k_1}}  g_{k_2} &=& 
   \frac{l^2  }{ ( \hat u_{k_1} - \hat u_{k_2} )^2} ,
  =   \wick{\c{h}_{k_1}  \c{h}_{k_2} } , \qquad k_1 \neq k_2 .
 \end{eqnarray}
 Finally the contraction between $g$ and $h$ is defined as 
 \begin{equation} \label{contract3}
 \wick{\c{g}_{k_1}  \c{h}_{k_2} } =  - \frac{l^2}{ ( u_{k_1}  - \hat u_{k_2}  )^2 }.
 \end{equation}
 This definition of the  contraction is related to the derivative 
 \begin{eqnarray} 
 \partial_{\hat u_{k_2}} g_{k_1} = \partial_{u_{k_1}} h_{k_1} =   - \frac{l^2}{ ( u_{k_1}  - \hat u_{k_2}  )^2 } =   \wick{ \c{g}_{k_1}  \c{h}_{k_2} } , \qquad  k_1 \neq k_2 .
\end{eqnarray}
Note that  higher order derivatives with respect to $u_{k_3}$ or $\hat u_{k_3}$  with $k_3\neq k_1, k_2$ on 
 $g, h$ vanish. 
 The rules for these contractions are similar to Wick contractions, for instance if one has a single 
 contraction on the product 
 \begin{equation}
 [g_{k_1} g_{k_2}  h_{k_3}  ]_{(1)} =  \wick{\c{g}_{k_1} \c{g}_{k_2} }   h_{k_3} + 
 +  \wick{ \c{g}_{k_1} g_{k_2} \c{g}_{k_3} }  + g_{k_1} \wick{\c{g}_{k_2} \c{h}_{k_3} } .
 \end{equation}
 The subscript in the square bracket of the LHS of the equation indicates the number of  contractions 
 performed. 
 One can consider more  contractions, for example the terms involving $2$ contractions on the 
 product 
 \begin{eqnarray}
 [g_{k_1} g_{k_2}  h_{k_3} h_{k_4} ]_{(2)}  =  \wick{ \c1{g}_{k_1} \c1{g}_{k_2} \c2{h}_{k_3} \c2{h}_{k_4}} +
 \wick{ \c1{g}_{k_1}  \c2{g}_{k_2}  \c1{h}_{k_3} \c2{h}_{k_4} } 
 + \wick{ \c2{g}_{k_1} \c1{g}_{k_2}  \c1{h}_{k_3} \c2{h}_{k_4}}.
\end{eqnarray}
Using this notation the derivatives on the function $f$ can be written as 
\begin{eqnarray}\label{partialder}
\partial_{u_{k_1} }  \cdots  \partial_{u_{k_i}}  
\partial_{\hat u_{\hat k_1} } \cdots \partial_{  \hat u_{\hat k_{p-i} }}  f && =
f g_{k_1} \cdots g_{k_i}   h_{\hat k_1} \cdots h_{\hat k_{p-i}} \\ \nonumber
& & + f \sum _{c= 1}^{[\frac{p}{2} ]} [g_{k_1} \cdots g_{k_i}   h_{\hat k_1} \cdots h_{\hat k_{p-i}} ]_{(c)},
\end{eqnarray}
where $[\frac{p}{2}] $ is the greatest integer less than or equal to $\frac{p}{2}$. 
This expression for the derivative can seen to be true by using the Leibnitz property of derivatives 
together with the fact none of the integers in the set $\{k_1\cdots  k_i\}$  and  $\{\hat k_1\cdots  \hat k_{p-i}\}$ 
are identical. 
As a simple check we can consider $3$ derivatives  acting on $f$. 
\begin{eqnarray}\label{3derv}
\partial_{u_{k_1} } \partial_{u_{k_2} } \partial_{\hat u_{\hat k_1}} f 
&=& fg_{k_1} g_{k_2} h_{\hat k_1}  + f\wick{ \c{g}_{k_1} \c{g}_{k_2}} h_{\hat k_1}  
 + f\wick{ \c{g}_{k_1} g_{k_2} \c{h}_{\hat k_1} }  + f g_{k_1} \wick{ \c{g}_{k_2}   h_{\hat k_1} },  \\ \nonumber
&=& fg_{k_1} g_{k_2} h_{\hat k_1} + f h_{\hat k_1}  \frac{l^2}{ ( u_{k_1} - u_{k_2} )^2 } 
- f g_{k_2} \frac{l^2}{ ( u_{k_1} - \hat u_{\hat k_1})^2} 
- f g_{k_1}   \frac{l^2}{ ( u_{k_2} - \hat u_{\hat k_1})^2} , \\ \nonumber
\partial_{u_{k_1} } \partial_{\hat u_{\hat k_2}}\partial_{\hat u_{\hat k_2}}f &=& f g_{k_1} h_{\hat k_1} h_{\hat k_2} +
f \wick{  \c{g}_{k_1} \c{h}_{\hat k_1} h_{\hat k_2} } + 
f \wick{ \c{g}_{k_1} h_{\hat k_1} \c{h}_{\hat k_2} }
+ f\wick{ g_{k_1} \c{h}_{\hat k_1} \c{h}_{\hat k_2} }, \\ \nonumber
&=& f g_{k_1} h_{\hat k_1} h_{\hat k_2}  - f h_{\hat k_2}   \frac{l^2}{ ( u_{k_1} - \hat u_{\hat k_1} )^2} 
- f h_{\hat k_1}   \frac{l^2}{ ( u_{k_1} - \hat u_{\hat k_2} )^2} 
+ f g_{k_1} \frac{l^2}{ ( \hat u_{\hat k_1} - \hat u_{\hat k_2} )^2}.
\end{eqnarray} 
Similarly the action of $4$ derivatives  is given by 
\begin{eqnarray}\label{4derv}
&& \partial_{u_{k_1} } \partial_{u_{k_2} } \partial_{\hat u_{\hat k_1}}  \partial_{\hat u_{\hat k_2}} f 
=  fg_{k_1} g_{k_2} h_{\hat k_1} h_{\hat k_2}  +  f [g_{k_1} g_{k_2} h_{\hat k_1} h_{\hat k_2}]_{(1)} 
+  f [g_{k_1} g_{k_2} h_{\hat k_1} h_{\hat k_2}]_{(2)} , \\ \nonumber
&=&   fg_{k_1} g_{k_2} h_{\hat k_1} h_{\hat k_2} \\ \nonumber
&+& f \wick{ \c{g}_{k_1} \c{g}_{k_2} h_{\hat k_1} h_{\hat k_2} }
+ f\wick{ \c{g}_{k_1} g_{k_2} \c{h}_{\hat k_1} h_{\hat k_2} }
+ f\wick{ \c{g}_{k_1} g_{k_2} h_{\hat k_1} \c{h}_{\hat k_2} }
+ f\wick{ g_{k_1} \c{g}_{k_2} \c{h}_{\hat k_1} h_{\hat k_2} }
+ f\wick{ g_{k_1} \c{g}_{k_2} h_{\hat k_1} \c{h}_{\hat k_2} }
+ f\wick{ g_{k_1} g_{k_2} \c{h}_{\hat k_1} \c{h}_{\hat k_2} }
\\ \nonumber
&+&  f  \wick{ \c1{g}_{k_1} \c1{g}_{k_2} \c2{h}_{\hat k_1}\c2{h}_{\hat k_2} }
+ f  \wick{ \c1{g}_{k_1} \c2{g}_{k_2} \c1{h}_{\hat k_1} \c2{h}_{\hat k_2} }
+ f  \wick{ \c2{g}_{k_1} \c1{g}_{k_2} \c1{h}_{\hat k_1} \c2{h}_{\hat k_2} }, 
\end{eqnarray}
Substituting the expressions for the contraction in (\ref{contract1}), (\ref{contract2}), (\ref{contract3}) we obtain 
\begin{eqnarray}
&& \partial_{u_{k_1} } \partial_{u_{k_2} } \partial_{\hat u_{\hat k_1}}  \partial_{\hat u_{\hat k_2}} f 
=  g_{k_1} g_{k_2} h_{\hat k_1} h_{\hat k_2} f, \\ \nonumber
&&+ f \frac{l^2}{ ( u_{k_1} - u_{k_2} )^2}  h_{\hat k_1} h_{\hat k_2}
- f   \frac{l^2}{ ( u_{k_1} - \hat u_{\hat k_1} )^2}  g_{ k_2} h_{\hat k_2}
- f   \frac{l^2}{ ( u_{k_1} - \hat u_{\hat k_2} )^2}  g_{ k_2} h_{\hat k_1} \\ \nonumber
& &- f   \frac{l^2}{ ( u_{k_2} - \hat u_{\hat k_1} )^2}  g_{ k_1} h_{\hat k_2}
- f   \frac{l^2}{ ( u_{k_2} - \hat u_{\hat k_2} )^2}  g_{ k_1} h_{\hat k_1}
+ f  \frac{l^2}{ ( \hat u_{\hat k_1} -\hat  u_{\hat k_2} )^2}  g_{ k_1} g_{ k_2}
\\ \nonumber
&&+ f  \frac{l^2}{ ( u_{k_1} - u_{k_2} )^2}   \frac{l^2}{ ( \hat u_{\hat k_1} -\hat  u_{\hat k_2} )^2}
+  f \frac{l^2}{ ( u_{k_1} - \hat u_{k_1} )^2}   \frac{l^2}{ (  u_{ k_2} -\hat  u_{\hat k_2} )^2}
+ f \frac{l^2}{ ( u_{k_1} -\hat u_{\hat k_2} )^2}   \frac{l^2}{ (  u_{ k_2} -\hat  u_{\hat k_1} )^2}.
\end{eqnarray}

The derivatives in $u_k$ and $\hat u_{\hat k} $
 that occur in (\ref{expcor1}) are always paired with the corresponding $B_k$ and $\hat B_{\hat k}$ respectively. 
 These can be simplified with the help of  the identities (\ref{derrel1}) and (\ref{derrel2}), which we  write down here 
 again 
 \begin{eqnarray} \label{derrel11}
B_k\partial_{u_k } f(u, \hat u) |_{(u, \hat u)  = ( w, \hat w ) } 
&=& g_k (u, \hat u) f( u, \hat u)   |_{(u, \hat u)  = ( w, \hat w ) } , \\ \nonumber
&=& -  \frac{l^2}{2}  f ( w, \hat w )  A, \\ \nonumber
\hat B_k\partial_{\hat u_k } f(u, \hat u) |_{(u, \hat u) = ( w, \hat w) } &=&  
g_k(u, \hat u) f( u, \hat u)  |_{(u, \hat u) = ( w, \hat  w) } , \\ \nonumber
&= &- \frac{l^2}{2} f( w, \hat w) \hat A.
\end{eqnarray}
Note that the above equations do not involve a sum over $k$. 
This relation helps us simplify the various terms which involve the derivatives.

 Let us simplify the derivatives that occur in (\ref{expcor1}), we have 
 \begin{eqnarray} \label{ddkeq}
 D_{u_{k_1}} \cdots D_{u_{k_i}} \hat D_{\hat u_{\hat k_1}}\cdots D_{\hat u_{\hat k_{p -i}} } f|_{( u, \hat u )= (w, \hat w) }
 &=&  (-1)^p  ( \frac{ l^2}{2}  )^p A^i \hat A^{ p -i}  f , \\ \nonumber
&&  + \sum_{c=1}^{[\frac{p}{2}]} T_{(c)} ( k_1, \cdots k_i; \hat k_1, \cdots \hat k_{p-i}) .
 \end{eqnarray}
 Here we have used the relations in (\ref{derrel11}) to simplify the first term on the LHS of 
 (\ref{partialder}) and written it in terms of only $A$' and $\hat A$'s. 
 $T_{(c)}$ is  given by 
 \begin{eqnarray}\label{deftc}
 T_{(c)}   ( k_1, \cdots k_i; \hat k_1, \cdots \hat k_{p-i}) 
    =  f B_{k_1} \cdots  B_{k_i} \hat B_{\hat k_1} \cdots \hat B_{\hat k_{p-i}}
   [g_{k_1} \cdots g_{k_i}   h_{\hat k_1} \cdots h_{\hat k_{p-i}} ]_{(c)}. \nonumber \\
   \end{eqnarray}
   Now again  using the identities (\ref{derrel11}) we see that the $B_k$ and $\hat B_{\hat k}$ which 
   are not associated with the contractions among $g_k, h_k$  can be reduced to $A$ and $\hat A$ respectively. 
   What remains are the $B_k, \hat B_{\hat k}$ associated with the contractions. 
   For example 
   \begin{eqnarray}
   T_{(1)} ( k_1, k_2; \hat k_1)  = - \frac{l^2}{2} \hat A \frac{ B_{k_1} B_{k_2} l^2}{ ( u_{k_1} - u_{k_2} )^2  }  f
   - \frac{l^2}{2} A \left(  - \frac{  B_{k_1} \hat B_{k_1} l^2}{ ( u_{k_1} - \hat u_{\hat k_1} )^2} 
   - \frac{  B_{k_2} \hat B_{k_1} l^2}{ ( u_{k_2} - \hat u_{\hat k_1} )^2}  \right)    f .\nonumber \\
   \end{eqnarray}
   Similarly 
   \begin{eqnarray}
& &    T_{(2)} ( k_1, k_2;\hat k_1, \hat k_2) =  \\ \nonumber
& & \quad   f \left( 
      \frac{ B_{k_1} B_{k_2}  l^2 }{ ( u_{k_1} - u_{k_2} )^2}  
       \frac{B_{\hat k_1} B_{\hat k_2} l^2} { ( \hat u_{\hat k_1} -\hat  u_{\hat k_2} )^2}
+   \frac{B_{k_{1}} \hat B_{\hat k_1} l^2 }{ ( u_{k_1} - \hat u_{k_1} )^2}  
 \frac{ B_{k_2} \hat B_{\hat k_2} l^2}{ (  u_{ k_2} -\hat  u_{\hat k_2} )^2}
+  \frac{ B_{k_1} \hat B_{\hat k_2} l^2}{ ( u_{k_1} -\hat u_{\hat k_2} )^2}  
 \frac{B_{k_2} \hat B_{\hat k_1} l^2}{ (  u_{ k_2} -\hat  u_{\hat k_1} )^2}
 \right) .
 \end{eqnarray}
 On examining the terms that occur in ( \ref{ddkeq}), we see that they are organised in terms of 
 powers of $l^2$.  
 We see that a pair of $B$'s  are associated with $l^2$, while a single $A$ or $\hat A$ is associated with
 $\frac{l^2}{2}$.   Note also the presence of $(-1)$ associated with $A$ or $\hat A$, that arises due to the 
 use of the identity (\ref{derrel11}).

 With all these observations, let us label the terms which occur in the correlator according to the 
 pair of $B_k, \hat B_{\hat k }$, that occur. 
 First consider the term with no $B$ or $\hat B$. 
 These terms arise from the first term in the expansion of the derivative in (\ref{ddkeq}). 
 Note that from (\ref{expcor1}) and (\ref{ddkeq}), they all occur with the coefficient $( \frac{l^2}{2} )^{2n}$. 
 \begin{eqnarray}
 \left. {\cal C}_{2n}  \right|_{ ( \frac{l^2}{2} )^{2n}} 
 &=& 
\left( \prod_{k =0}^{n-1} ( B_k \hat B_k)^{\frac{l^2}{2}}  \right)  f(u, \hat u) |_{(w, \hat w) }
 ( A^n \hat A^n) ( \frac{l^2}{2} )^{2n}  \\ \nonumber
  & & \times  \sum_{p=0}^{2n} \sum_{\stackrel{ i =0, i \leq n}{ 0\leq  p- i  \leq n }}^p 
  \sum_{ \stackrel{ \{ k_1, \cdots k_i\} }  {  \{\hat k_1, \cdots \hat k_{p- i }\} }} (-1)^p, \\ \nonumber
  &=&  ( A^n \hat A^n) ( \frac{l^2}{2} )^{2n}   \sum_{p=0}^{2n} \sum_{\stackrel{ i =0, i \leq n}{ 0\leq  p- i  \leq n }}^p 
  (-1)^p \binom{n}{i} \binom{n}{p-i} , \\ \nonumber
  &=& ( A^n \hat A^n) ( \frac{l^2}{2} )^{2n}   \sum_{p=0}^{2n}  \binom{2n}{p}  (-1)^p= 0.
  \end{eqnarray}
  In the first step we have used (\ref{iden5}), (\ref{derrel11}). Then we have replaced the 
  the sums over $\{ k_1, \cdots k_i\} ,  \{\hat k_1, \cdots \hat k_{p- i } \} $ by the number of terms. 
 To obtain the last step  we have used the combinatorial  identity 
  \begin{equation}
  \sum_{\stackrel{ i =0, i \leq n}{ 0\leq  p- i  \leq n }}^p   \binom{n}{i} \binom{n}{p-i}  =  \binom{2n}{p} .
  \end{equation}
  This same set of steps can be used to demonstrate that any term  
  containing $A$ or $\hat A$ vanishes. 
  Before going to the general case, let us illustrate it for the term containing a 
  single contraction   $B_0B_1l^2$ in the correlator which is proportional to $( \frac{l^2}{2} )^{2n- 2} l^2$.
  This term in the correlator is given by 
  \begin{eqnarray}
  && \left. {\cal C}_{2n}  \right|_{ ( \frac{l^2}{2} )^{2n -2} l^2 } 
    = 
\left( \prod_{k =0}^{n-1} ( B_k \hat B_k)^{\frac{l^2}{2}}  \right)  f(u, \hat u)  \frac{ B_0 B_1 l^2 }{ ( u_0 - u_1)^2}  \times \\ \nonumber
& &
\left.   \sum_{p=0}^{2n-2} \sum_{\stackrel{ i =0, i \leq n-2}{ 0\leq  p- i  \leq n }}^p 
( \frac{l^2}{2} )^{2n-p} A^{n- i} \hat A^{ n-p +i } 
  \sum_{ \stackrel{ \{ k_1, \cdots k_i\} , k\neq k_0, k_1  }  {  \{\hat k_1, \cdots \hat k_{p- i }\} }} 
  B_{k_1} g_{k_1} \cdots B_{k_i} g_{k_i}    \hat B_{k_1} h_{\hat k_1} \cdots \hat B_{\hat k_{ p-i}} \hat h_{ \hat k_{p-i} } 
  \right|_{( u, \hat u) = ( w, \hat w) }, \\ \nonumber
  & & \qquad  = \frac{ B_0 B_1 l^2 }{ ( w_0 - w_1)^2} ( \frac{l^2}{2} )^{2n-2}  A^{n- 2} \hat A^{ n }  
   \sum_{ \stackrel{ \{ k_1, \cdots k_i\} , k\neq 0, 1  }  {  \{\hat k_1, \cdots \hat k_{p- i }\} }} (-1)^p , 
   \\ \nonumber
   &&  \qquad = \frac{ B_0 B_1 l^2 }{ ( w_0 - w_1)^2} ( \frac{l^2}{2} )^{2n-2}  A^{n- 2} \hat A^{ n } 
   \sum_{p=0}^{2n-2} \sum_{\stackrel{ i =0, i \leq n-2}{ 0\leq  p- i  \leq n }}^p ( -1)^p   \binom{n-2}{i } \binom{n}{p-i}  
   ,  \\ \nonumber
   && \qquad = \frac{ B_0 B_1 l^2 }{ ( w_0 - w_1)^2} ( \frac{l^2}{2} )^{2n-2}  A^{n- 2} \hat A^{ n } 
     \sum_{p=0}^{2n-2}  (-1)^p  \binom{2n-2}{ p} , \\ \nonumber 
     && \qquad = 0.
  \end{eqnarray}
  The first two lines of the above equation contains all the terms  containing 
  $B_0 B_1$ that occur from summing the  various derivatives in (\ref{expcor1}). 
  Note that this  specific single contraction terms are present in $T_{(1)}$ as defined in (\ref{deftc}). 
  We then use  (\ref{iden5}), (\ref{derrel11}) and follow the same steps as before.  
  When one sums over the set of  integers $\{k_1, \cdots k_i \}$, none of these integers 
  can contain $\{0, 1\}$ since $g_0, g_1$ are contracted.  This is denoted in the subscript of the 
  sum in the third line as a result, the number of these set of integers  are given by $\binom{n-2}{i }$. 
  
  Now let us move to  the more general case. 
  Consider the term which arises from $a$  pairs of contractions involving $g$'s, 
  $b$ pair of contractions involving $h$'s and $c$ contractions between $g$ and $h$. 
  We also have $2( a  +b + c)  < 2n $.  Such a contraction will be present in $T_{(a + b+c)}$. 
  Among this type of contractions we focus on the term which contains
  the following products of $B, \hat B$. 
  \begin{eqnarray} \label{defcb}
  {\cal B} ( u, \bar u) &=& 
  \left[ \frac{ B_{0} B_{1} l^2 }{ ( u_0 - u_1)^2} \cdots   \frac{ B_{2a-2} B_{2a-1} l^2 }{ ( u_{2a-2} - u_{2a -1})^2}  \right]
  \times \left[ 
  \frac{ \hat B_{0} \hat B_{1} l^2 }{ ( u_0 - u_1)^2} \cdots   
  \frac{ \hat B_{2b-2} \hat B_{2b-1} l^2 }{ ( \hat u_{2b-2} - \hat u_{2b-1})^2}   \right]  \nonumber \\
  && \times \left[ 
   \frac{  -B_{2a} \hat B_{2b} l^2 }{ ( u_{2a}  - u_{2b} )^2} \cdots   
  \frac{ - \hat B_{2a+c -1} \hat B_{2b+c-1 } l^2 }{ ( \hat u_{2a + c-1 } - \hat u_{2b + c-1 }^2}  \right].
  \end{eqnarray}
  Let us now collect all the terms in the derivatives of the correlator (\ref{expcor1}) which contains 
  this specific contraction. 
\begin{eqnarray} 
  &&  \left. {\cal C}_{2n}  \right|_{ ( \frac{l^2}{2} )^{2( n- a- b-c) } l^{ 2( a+ b + c)  }} = 
    \left( \prod_{k =0}^{n-1} ( B_k \hat B_k)^{\frac{l^2}{2}}  \right)  f(u, \hat u)  {\cal B}( u, \bar u) 
    \\ \nonumber
    & & \times 
     \sum_{p=0}^{2(n-a-b-c)} \sum_{\stackrel{ i =0, i \leq n-2a -c}{ 0\leq  p- i  \leq n- 2b- c}}^p 
( \frac{l^2}{2} )^{2n-p} A^{n- i} \hat A^{ n-p +i } 
     \sum_{ \stackrel{ \{ k_1, \cdots k_i\} , k\notin  \{ 0, 1, \cdots 2a+c-1\} , }  {  \{\hat k_1, \cdots \hat k_{p- i }\}, 
    k\notin  \{ 0, 1, \cdots  2b+c-1 \}   }}  \Bigg(  \\ \nonumber
    &&\qquad\qquad\qquad\qquad\qquad\qquad\qquad\qquad  \left.   
  B_{k_1} g_{k_1} \cdots B_{k_i} g_{k_i}    \hat B_{k_1} h_{\hat k_1} \cdots \hat B_{\hat k_{ p-i}} \hat h_{ \hat k_{p-i} } 
  \Bigg) 
  \right|_{( u, \hat u) = ( w, \hat w) }, \\ \nonumber
  &=& {\cal B} (w, \hat w) ( \frac{l^2}{2} )^{2( n- a- b-c)}  A^{n- 2a- c} \hat A^{ n - 2b -c } 
     \sum_{ \stackrel{ \{ k_1, \cdots k_i\} , k\notin  \{ 0, 1, \cdots 2a+c-1\} , }  {  \{\hat k_1, \cdots \hat k_{p- i }\}, 
    k\notin  \{ 0, 1,\cdots  2b+c-1 \}   }}  (-1)^p , \\ \nonumber
    &=& {\cal B} (w, \hat w) ( \frac{l^2}{2} )^{2( n- a- b-c)}   A^{n- 2a- c} \hat A^{ n - 2b -c } 
    \sum_{p =0}^{ 2( n- a-b-c) } \sum_{\stackrel{ i =0, i \leq n - 2a - c}{ 0 \leq p- i \leq n- 2b -c} } ( -1)^p 
    \binom{ n - 2a - c}{ i } \binom{n- 2b-c}{p-i} ,
    \\ \nonumber
    &=&  {\cal B} (w, \hat w)   ( \frac{l^2}{2} )^{2( n- a- b-c)}   A^{n- 2a- c} \hat A^{ n - 2b -c }
      \sum_{p =0}^{ 2( n- a-b-c) }  \binom{ 2( n - a-b-c)}{ p } (-1)^p ,
      \\ \nonumber
      &=& 0.
  \end{eqnarray}
  The steps are as discussed before for the single contraction. 
  Note that it is clear from this derivation that 
  for any contraction of this form, not necessary the special one chosen in 
  (\ref{defcb})  vanishes as long as $2( n-a-b-c)>0$. 
  Thus  the  only non-zero element in the correlator is given in which all the 
  $n$ $g$'s and  $n$ $h$'s are contracted. Such a term occurs only once in 
  (\ref{expcor1}) and it occurs in the derivative or order $2n$. 
  Therefore the correlator of interest is given by 
  \begin{eqnarray}
  {\cal C}_{2n} &=& l^{2n} (-1)^n \left( \prod_{k=0}^{n-1}  ( B_k \hat B_k) \right)  {\cal J} , \\ \nonumber
  {\cal J } &=&   \left( 
  \prod_{k = 0}^{n-1} \frac{1}{ ( w_k - \hat w_k) } + {\rm distinct \;  permutations} \right) .
  \end{eqnarray}
  The $(-1)^n$ arises due to the contractions between $g$ and $h$ (\ref{contract3}). 
  ${\cal J}$ is identical to the correlator  that one would obtain by considering the $2n$ point functions
  of $U(1)$ currents on the uniformised plane and the $B, \hat B$'s are slopes of the conformal transformations
  (\ref{u1}) and (\ref{u2}).  Note that it is clear from the steps leading to this result,  the simplification 
  in the $2n$ point function occurs   only when the descendants are placed on the uniformized plane. 
  This concludes the derivation of the $2n$ point function of the descendant $\partial e^{il X}$ on the 
  uniformized plane.

\section{Properties of descendants} \label{appen2}

In this appendix we summarise as well as derive the properties of conformal descendants till  level 3. 

\subsection{Conformal transformations of descendants}

Transformations of descendants obtained by the action of derivatives on primaries are easily derived. Here
is the list till level-3

\paragraph{Level 1}

   \begin{equation}\label{transdelO}
        \begin{split}
            w\circ \partial \o( w(z) )=\left(\frac{\partial w}{\partial z}\right)^{h+1}\left[h\left(\frac{\partial w}{\partial z}\right)^{-2}\left(\frac{\partial^2 w}{\partial^2 z}\right)+\partial_{w}\right]\o(w (z) ).
        \end{split}
    \end{equation}
   For the purposes of this paper we would need the transformations under the conformal maps (\ref{u1}), 
   (\ref{u2}) where these maps take $z\rightarrow w_k,  \hat z\rightarrow \hat w_k$. 
   Using the definition of the slopes in (\ref{slope}) and  the second derivatives of these 
   transformations in (\ref{secder0}) and  (\ref{secder1}) , we can write  (\ref{transdelO}) as 
     \begin{eqnarray}
          w\circ \partial O(w_k)&=&B_k^{h+1}\left(h\frac{A}{B_k}+\partial_{w_k}\right) O(w_k),\\
     \hat w \circ \partial O(\hat w_k )&=&\hat B_k^{h+1}\left(h\frac{\hat A}{\hat B_k}+\partial_{\hat w_k}\right) O(\hat w_k).
   \end{eqnarray}
 
 \paragraph{Level 2}

The  second derivative of the operator transforms as 
   \begin{equation} \label{trans del2O}
    \begin{split}
     w\circ \partial^2 O( w(z) )=w'^{h+2}\left[h\left((h-1)\frac{w''^2}{w'^{4}}+\frac{w'''}{w'^3}\right)+(2h+1)\frac{w''}{w'^2}\partial_w+\partial_w^2\right]O(w).
    \end{split}
\end{equation}
Let us introduce the third derivatives of the conformal maps to the uniformized plane by 
\begin{eqnarray}
  &&\left. \frac{d^3 w}{ d z^3} \right|_{w_k}  =B_k(A^2+F), \qquad 
     \left. \frac{d^3 \hat w}{ d z^3} \right|_{\hat w_k}     = \hat B_k(\hat A^2+\hat F),\\
    &&   F=  (1 + e^{-4 \pi \mathrm{i}x}) + \frac{1}{n} (1 - e^{-4 \pi \mathrm{i}x}), \qquad
         \hat F =  (1 + e^{4 \pi \mathrm{i}x}) + \frac{1}{n} (1 - e^{4 \pi \mathrm{i}x}).
\end{eqnarray}
Then we can write the conformal transformations as
   \begin{equation}
    \begin{split}
      w\circ \partial^2 \o(w_k) &= B_{k}^{h+2}\left[h\left((h-1)\frac{A^2}{B_{k}^{2}}+\frac{A^2+F}{B_{k}^2}\right)+(2h+1)\frac{A}{B_k}\partial_{w_k}+\partial_{w^2_k}\right]  \o(w_k),\\
       \hat w \circ  \partial^2 \o (\hat w_k) &= \hat  B_{k}^{h+2}\left[h\left((h-1)\frac{\hat A^2}{\hat B_{k}^{2}}+\frac{\hat A^2+\hat F}{\hat B_{k}^2}\right)+(2h+1)\frac{\hat A}{\hat B_k}\partial_{\hat w_k} +\partial_{\hat w_k^2} \right]O(\hat w_k).
    \end{split}
\end{equation}

 \paragraph{Level 3} 
 The conformal transformation of the third derivative of the primary is given by 
 \begin{equation}\label{trans del3O}
    \begin{split}
        \partial_z^3O(z)=w'^{h+3}\Big[\partial_w^3 &+\frac{3(h+1)w''}{(w')^2}\partial_{w}^{2}+\frac{3h^2(w'')^2+(1+3h)w'w'''}{(w')^4}\partial_w\\
        &+\frac{h\left((h-1)w''((h-2)(w'')^2+3w'w''')+(w')^2w''''\right)}{(w')^6}\Big]O(w).
    \end{split}
\end{equation}
The fourth derivatives to the maps (\ref{u1}) , (\ref{u2}) to the uniformized plane is given by 
\begin{eqnarray}
 &&\left. \frac{d^4 w}{ d z^4} \right|_{w_k} 
  =B_k(A^3+3AF+2G), \qquad
    \left. \frac{d^3 \hat w}{ d z^3} \right|_{\hat w_k} 
    =\hat B_k(\hat A^3+3\hat A \hat F +2 \hat G),\\    
    && G= (1 + e^{-6 \pi \mathrm{i}x}) + \frac{1}{n} (1 - e^{-6 \pi \mathrm{i}x}), \qquad 
         \hat G=-(1 + e^{6 \pi \mathrm{i}x}) - \frac{1}{n} (1 - e^{6 \pi \mathrm{i}x}).
\end{eqnarray}
Ue use these expression to write the conformal transformation as 
   \begin{equation}\label{del3O transformation}
    \begin{split}
        w\circ \partial^3 \o(w_k)&=B_{k}^{h+3}\Big[\frac{h(2G+hA(3F+hA^2)) }{B_{k}^{3}}+\frac{(3h+1)F+(1+3h+3h^2)A^2}{B_{k}^{2}}\partial_{w_k}\\
        &\hspace{1.5cm}+\frac{3(h+1)A}{B_k}\partial_{w_k}^{2}+\partial_{w_k}^{3}\Big]\o(w_k),\\
        \hat w \circ\partial^3 \o(\infty_k)&=\hat B_{k}^{h+3}\Big[\frac{h(2\hat G+h \hat A (3\hat F+h \hat A^2)) }{\hat B_{k}^{3}}+\frac{(3h+1)\hat F +(1+3h+3h^2) \hat A^2}{\hat B_{k}^{2}}\partial_{\hat w_k}\\
        &\hspace{1.5cm}+\frac{3(h+1)\hat A }{\hat B_k}\partial_{\hat w_k}^{2}+\partial_{\hat w _k}^{3}\Big] \o (\hat w_k) . \end{split}
\end{equation}

Let us now obtain the conformal transformation of Virasoro descendants defined by 
\begin{equation}\label{Okdefiniton}
    \begin{split}
        \o^{(-k)}(z)=\oint_z \frac{d\tilde{z}}{(\tilde z-z)^{k-1}}T(\tilde z)\o(z).
    \end{split}
\end{equation}
Here the contour is around  $z$. 
 Consider the conformal transformation $w(z)$, we can use the conformal transformation of 
 the stress tensor and the primary to write the above equation as 
\begin{equation} \label{omtrans}
    \begin{split}
        \o^{(-k)}(z)=\oint_e \frac{d\tilde{w}}{(\tilde{z}-z)^{k-1}}\frac{d\tilde{z}}{d\tilde{w}}\left(\left(\frac{d\tilde{w}}{d\tilde{z}}\right)^2 T(\tilde w)+\frac{c}{12}S[\tilde w,\tilde z]\right)\left(\frac{dw}{dz}\right)^h \o(w)
    \end{split}
\end{equation}
where $\tilde w = w( \tilde z) $ and the contour is around $w$. 
The Schwarzian is defined as 
\begin{equation}
    S[w,z]=\frac{2w''' w' -3(w'')^2}{2(w')^2},\hspace{1cm} w'=\frac{dw}{dz}.
\end{equation}
The next step is to write the  denominator $\tilde z - z$ in terms of $\tilde w - w$ and also write the 
slope $\frac{d \tilde z}{d \tilde w}$ in terms of the slope  $\frac{dw}{dz}$. 
This can be done  by inverting the Taylor series of  the function  $\tilde w ( \tilde z) $ around  $z$
which is given by 
\begin{equation}
    \tilde w=w(z)+(\tilde z-z)w'+\frac{1}{2!}(\tilde z-z)^2 w''+\frac{1}{3!}(\tilde z-z)^3 w'''+\frac{1}{4!}(\tilde z-z)^4 w''''+\cdots.
\end{equation}
Using this we can obtain 
\begin{equation}\label{dz}
    \begin{split}
        \tilde z-z &=\frac{1}{w'}(\tilde w-w)-\frac{1}{2}\frac{w''}{(w')^3}(\tilde w-w)^2+\frac{1}{(w')^5}\left(\frac{(w'')^2}{2}-\frac{1}{3!}w'''w'\right)(\tilde w-w)^3\\
        &\hspace{2.8cm}-\frac{1}{(w')^7}\left(\frac{15(w'')^3-10w'w''w'''+(w')^ 2w''''}{4!}\right)(\tilde w-w)^4+\cdots, \\
        \frac{d\tilde z}{d\tilde w}&=\frac{1}{w'}-\frac{w''}{(w')^3}(\tilde w-w)+\frac{3}{(w')^5}\left(\frac{(w'')^2}{2}-\frac{1}{3!}w'''w'\right)(\tilde w-w)^2\\
        &\hspace{2.8cm}-\frac{1}{6(w')^7}\left(15(w'')^3-10w'w''w'''+(w')^ 2w''''\right)(\tilde w-w)^3+\cdots.
    \end{split}
\end{equation}
Substituting these expressions in (\ref{omtrans}) and performing the contour integral with $k=2, 3$  we can obtain the
conformal transformation of  the Virasoro descendants. 

\subsubsection*{
$\o^{(-2)}$ }

The conformal transformation of the level-2 Virasoro descendant is given by 
\begin{equation}\label{transO2}
\begin{split}
  w\circ  \o^{(-2)}(w(z) )&=(w')^{h+2}\Big[\o^{(-2)}(w)+\frac{3w''}{2(w')^2}\o^{(-1)}(w)+\Big(\frac{3h(w'')^2}{4w'^4}+(4h+\frac{c}{2})\frac{S[w,z]}{6(w)'^2}\Big)\Big]\o(w), \\
    &=(w')^{h+2}\Big[\o^{(-2)}(w)+\frac{3w''}{2w'^2}\o^{(-1)}(w)\\
    &\hspace{2cm}+\frac{1}{12(w')^4}\Big((-3h-\frac{3c}{2})(w'')^2+2(4h+\frac{c}{2})w'''w'\Big)\o(w)\Big].
\end{split}\end{equation}
This agrees with the transformation given in \cite{Caputa:2015tua} which uses the operator approach of \cite{Gaberdiel:1994fs} to obtain
conformal transformations of descendants. 
For the maps in (\ref{u1}) and (\ref{u2}), this transformation can be written as 
\begin{equation}{\label{O2transformation}}
    \begin{split}
       w\circ \o^{(-2)}(w_k)&=B_{k}^{h+2}\Big[\o^{(-2)}(w_k)+\frac{3A}{2B_k}\o^{(-1)}(w_k)+\frac{1}{12}\Big((5h-\frac{c}{2})\frac{A^2}{B_{k}^{2}}+2(4h+\frac{c}{2})\frac{F}{B_{k}^{2}}\Big)\o(w_k)\Big],\\
         \hat w \circ \o^{(-2)}(\hat w_k)&= \hat B_{k}^{h+2}\Big[\o^{(-2)}(\hat w_k)+\frac{3\hat A}{2 \hat 
         B_k}\o^{(-1)}( \hat w_k)+\frac{1}{12}\Big((5h-\frac{c}{2})\frac{ \hat A^2}{\hat B_{k}^{2}}+2(4h+\frac{c}{2})
         \frac{\hat F}{\hat B_{k}^{2}}\Big)\o ( \hat w_k)\Big].\\
    \end{split}
\end{equation}

\subsubsection*{
$\o^{(-3)}$ }

Similarly the transformation of the level 3 descendant is given by 
\begin{equation}
    \begin{split}
       w \circ  \o^{(-3)}(  w(z) )&=(w')^{h+3}\Big[\o^{(-3)}(w)+\frac{2w''}{(w')^2}\o^{(-2)}(w)+\left(\frac{(w'')^2}{4(w')^4}+\frac{5}{6}\frac{w'''}{(w')^3}\right)\o^{(-1)}(w)\\
        &\hspace{1cm}+
        \Big(\frac{3w'w''''-2w''w'''}{12(w')^5}-\frac{c}{12}\frac{6 (w'')^3 - 6 w' w'' w''' + (w')^2 w''''}{(w')^8}\Big)\o(w)\Big].
    \end{split}
\end{equation}
Restricting the conformal transformation to the uniformisation map we can write the above transformation
as 
\begin{equation} \label{ominus3trans}
    \begin{split}
       w\circ  \o^{(-3)} (w_k ) &= B_{k}^{(
  h + 3)} \Big[\o^{(-3)} (w_k) + 
    2 \frac{A}{B_k} \o^{(-2)}(w_k)  + \frac{13 A^2 + 10 F}{12 B_{k}^{2}}
      \o^{(-1)}(w_k)  \\
      &\hspace{2cm}+ \left(\frac{h (A^3 + 7 A F + 6 G)}{12 B_{k}^{3}} + 
      \frac{ c}{12} \frac{(-A F + 2 G)}{B_{k}^{3}}\right) \o(w_k)\Big],\\
      \hat w \circ \o^{(-3)} (\hat w_k) &= \hat B_{k}^{(
  h + 3)} \Big[\o^{(-3)} (\hat w_k) + 
    2 \frac{\hat A}{\hat B_k} \o^{(-2)}( \hat w_k)  + \frac{13 \hat A^2 + 10 \hat F}{12 \hat B_{k}^{2}}
     O^{(-1)}(\hat w_k)  \\
      &\hspace{2cm}+ \left(\frac{h ( \hat A^3 + 7 \hat A \hat F + 6 \hat G)}{12 \hat B_{k}^{3}} + 
      \frac{ c}{12} \frac{(- \hat A \hat F + 2 \hat G)}{\hat B_{k}^{3}}\right) \o(\hat w_k)\Big].
    \end{split}
\end{equation}
The  conformal transformations of the last state at level $3$  is given by 
  \begin{equation}
    \begin{split}
        w\circ \partial \mathcal O^{(-2)}(w_k)&=B_{k}^{(h+3)}\Big[\partial_{w_k} \mathcal O^{(-2)}(w_k)+\frac{A (h+2)}{B_k}\mathcal O^{(-2)}(w_k)\\
        &\hspace{2cm}+\Big\{\frac{3 A}{2 B_k}\partial_{w_k}^{2} +\frac{A^2 (-c+46 h+36)+2 F (c+8 h+18)}{24 B_{k}^{2}}\partial_{w_k}\\
        &\hspace{2cm}+\frac{1}{24B_{k}^{3}} \Big(2 A^2 (c + 8 h) + 2 F (c + 8 h) - A^3 (2 (8 - 5 h) h + c (2 + h))\\
        &\hspace{3.5cm}+2 A F (c (-4 + h) + 2 h (-7 + 4 h))\Big)\Big\}\mathcal O(w_k)\Big],\\
       \hat w \circ \partial \mathcal O^{(-2)}(\hat w )&=B_{\hat k}^{(h+3)}\Big[\partial_{\hat w_{ k}}\mathcal  O^{(-2)}(\hat w_{ k})+\frac{\hat A (h+2)}{\hat B_{ k}}\mathcal O^{(-2)}(\hat w_{ k})\\
        &\hspace{2cm}\Big\{+\frac{3 \hat A}{2 \hat B_{k}}\partial_{\hat w_{ k}}^{2}+\frac{\hat A^2 (-c+46 h+36)+2 \hat F (c+8 h+18)}{24 \hat B_{ k}^{2}}\partial_{\hat w_{ k}}\\
        &\hspace{2cm}+\frac{1}{24\hat B_{k}^{3}} \Big(2 \hat A^2 (c + 8 h) + 2 \hat F (c + 8 h) - \hat A^3 (2 (8 - 5 h) h + c (2 + h))\\
        &\hspace{3.5cm}+2 \hat A \hat F(c (-4 + h) + 2 h (-7 + 4 h))\Big)\Big\}\mathcal O(\hat w_k)\Big].
       \end{split}
\end{equation}

\subsection{Correlators with  descendants}

Correlators with derivatives of primaries are straightforward to obtain from the corresponding 
correlators of the primaries. 
To evaluate correlators involving Virasoro descendants we would need the OPE with the 
stress tensor. 
To evaluate the OPE, we first use the definition (\ref{Okdefiniton}) and use the $TT$ OPE to 
obtain
\begin{equation}
    \begin{split}
        &T(w)\oint_z \frac{dz'}{(z'-z)^{k-1}}T(z')\o(z)\\&=\oint_z \frac{dz'}{(z'-z)^{k-1}}\left(\frac{2T(z')}{w-z'}+\frac{\partial_{z'}T(z')}{w-z'}+\frac{c}{2}\frac{1}{(w-z')^4}+:T(w)T(z'):\right)\o(z).
    \end{split}
\end{equation}
We can then use the OPE of the stress tensor with the operator $\o$. 
For $k=2$ we obtain 
\begin{equation} \label{tominus2}
    \begin{split}
        T(w)\o^{-2}(z)\sim \frac{(\frac{c}{2}+4h)\o(z)}{(w-z)^4}+\frac{3\partial_z \o(z)}{(w-z)^3}+\frac{(h+2)\o^{(-2)}(z)}{(w-z)^2}+\frac{\partial_z \o^{(-2)}}{(w-z)}.
    \end{split}
\end{equation}
Similarly for $k=3$ we get
\begin{equation}
    \begin{split}
        T(w)\o^{-3}(z)\sim \frac{2(c+3h)\o(z)}{(w-z)^5}+\frac{5\partial_z \o(z)}{(w-z)^4}+\frac{4\o^{(-2)}(z)}{(w-z)^3}+\frac{(h+3)\partial \o^{(-3)}}{(w-z)^2}+\frac{\partial_z \o^{(-3)}}{(w-z)}.
    \end{split}
\end{equation}

Let us illustrate how to use these OPE's  in a couple of correlators. 
Consider 
\begin{equation}
    \begin{split}
        \langle \o(z_1) \o^{(-2)}(z_2)\rangle=\langle \o(z_1)\oint_{z_2}\frac{dz'}{z'-z_2}T(z')\o(z_2)\rangle.
    \end{split}
\end{equation}
Here by the definition of $\o^{(-2)}$ the contour is around $z_2$. By deforming the contour around infinity 
we obtain
\begin{equation}
    \begin{split}
        \langle \o(z_1) \o^{(-2)}(z_2)\rangle=-\langle\oint_{z_1}\frac{dz'}{z'-z_2}T(z')\o(z_1)\o(z_2)\rangle.
    \end{split}
\end{equation}
Note that now the contour is around $z_1$. Using the OPE of the stress tensor with $\o(z_1)$ we obtain
\begin{equation} \label{oominus}
    \begin{split}
        \langle \o(z_1)\o^{(-2)}(z_2)\rangle&=-\langle \int \frac{dz'}{z'-z_2}\left(\frac{h\o(z_1)}{(z'-z_1)^2}+\frac{\partial_{z_1}\o(z_1)}{(z'-z_2)}\right)\o(z_2) \rangle,\\
        &=\frac{h}{(z_1-z_2)^2}\langle \o(z_1)\o(z_2)\rangle-
        \frac{1}{(z_1-z_2)}\partial_{z_1}\langle \o(z_1)\o(z_2)\rangle,\\
        &=\frac{3h}{(z_1-z_2)^{2h+2}}.
    \end{split}
\end{equation}
Consider another example, the two  point function of $\o^{(-2)}$ and $\o^{(-3)}$. 
\begin{equation}
    \begin{split}
        &\langle \o^{(-2)}(z_1) \o^{(-3)}(z_2)\rangle,\\
        &=\langle \o^{(-2)}(z_1)\oint_{z_2}\frac{dz'}{(z'-z_2)^2}T(z')\o(z_2)\rangle,\\
        &=-\langle\oint_{z_1}\frac{dz'}{(z'-z_2)^2}T(z')\o^{(-2)}(z_1)\o(z_2)\rangle,\\
        &=-\langle\oint_{z_1}\frac{dz'}{(z'-z_2)^2}\Big[\frac{(\frac{c}{2}+4h)\o(z_1)}{(z'-z_1)^4}+\frac{3\partial_{z_1}\o(z_1)}{(z'-z_1)^3}+\frac{(h+2)\o^{(-2)}(z_1)}{(z'-z_1)^2}+\frac{\partial_{z_1}\o^{(-2)}(z_1)}{(z'-z_1)}\Big]\o(z_2)\rangle, \\
%        &=-\Big[\Big\{\frac{-4(\frac{c}{2}+4h)}{(z_1-z_2)^5}+\frac{9\partial_{z_1}}{(z_1-z_2)^4}\Big\}\langle \o(z_1)\o(z_2)\rangle\\
%        &\hspace{.5cm}+\Big\{-\frac{2}{(z_1-z_2)^3}(h+2)+\frac{\partial_{z_1}}{(z_1-z_2)^2}\Big\}\langle \o^{(-2)}(z_1)\o(z_2)\rangle\Big]\\
        &=\frac{2c+52h+12h^2}{(z_1-z_2)^{2h+5}}.
    \end{split}
\end{equation}
In the first step we used the definition of $\o^{(-3)}$, then deformed the
contour so that it is around $z_1$. 
We then applied the OPE (\ref{tominus2}),  performed the contour integrals 
and also used the two point function (\ref{oominus}).  

Using similar methods we can obtain the following two point functions. 
\begin{equation}\label{2ptlist}
    \begin{split}
         \langle \o(z_1) \o^{-1}(z_2)\rangle&=\frac{2h}{(z_1 -z_2 )^{2h+1}},\\
          \langle \o(z_1) \o^{-2}(z_2)\rangle&=\frac{3h}{(z_1-z_2)^{2h +2}},\\
          \langle \o(z_1)\o^{-3}(z_2)\rangle&=\frac{4h}{(z_1-z_2)^{2h+3}},\\ 
         %   \langle \o^{-1}(z_1) \o^{-1}(z_2)\rangle&=\frac{-2h(2h+1)}{(z_1 - z_2)^{2h +2}}\\
            % \langle \o^{-1}(z_1) \o^{-2}(z_2)\rangle&=\frac{-(3h)(2h+2)}{(z_1-z_2)^{2h +3}}\\
            %  \langle \o^{-1}(z_1) \o^{-3}(z_2)\rangle&=\frac{-(4h)(3+2h)}{(z_1-z_2)^{2h+4}}\\
            % \langle O^{-2}(w_j) O(w'_j)\rangle&=\frac{(3h-h_p)}{(w_j-w'_j)^{2h-h_p+2}}\\
            %  \langle O^{-2}(w_j) O^{-1}(w'_j)\rangle&=\frac{(3h-h_p)(2h-h_p+2)}{(w_j-w'_j)^{2h-h_p+3}}\\
               \langle \o^{-2}(z_1) \o^{-2}(z_2)\rangle&=\frac{ \frac{c}{2} + h(9h +22 )  }{(z_1 - z_2)^{2h+4}},\\
               \langle \o^{-2}(z_1)\o^{-3}(z_2)\rangle&=\frac{ 2c+ h( 12 h +52  )}{(z_1 - z_2)^{2h +5}},\\
              % \langle O^{-3}(w_j)O(w'_j)\rangle&=-\frac{(4h-h_p)}{(w_j-w'j)^{2h-h_p+3}}\\ 
            %   \langle  O^{-3}(w_j) O^{-1}(w'_j)\rangle&=\frac{(4h-h_p)(-3-2h+h_p)}{(w_j-w'j)^{2h-h_p+4}}\\ 
               %\langle O^{-3}(w_j)O^{-2}(w'_j)\rangle&=-\frac{(2c+12h^2+h(52-7h_p)+(-15+h_p)h_p)}{(w_j-w'j)^{2h-h_p+5}}\\
               \langle \o^{-3}(z_1)\o^{-3}(z_2)\rangle&=\frac{-10c-2h(71+8h)}{(z_1-z_2 )^{2h +6}}.\\
    \end{split}
\end{equation}
All the other 2-point functions required in this paper can be obtained by the action of derivatives on 
the above correlators.

\bibliographystyle{JHEP}
\bibliography{references}

\end{document}